# Hints and Principles for Computer System Design

Butler Lampson
May 14, 2021

## Abstract

This new long version of my 1983 paper suggests the goals you might have for your system—Simple, Timely, Efficient, Adaptable, Dependable, Yummy (STEADY)—and techniques for achieving them—Approximate, Incremental, Divide & Conquer (AID). It also gives some principles for system design that are more than just hints, and many examples of how to apply the ideas.

## Table of Contents

# 1. Introduction

*There are three rules for writing a novel. Unfortunately, no one knows what they are.* —Somerset Maugham[Q51]

*You got to be careful if you don't know where you're going, because you might not get there.* —Yogi Berra[Q8]

*If I have seen farther, it is by standing on the shoulders of giants.* —Bernard of Chartres[Q7]

*Shakespeare wrote better poetry for not knowing too much; Milton … knew too much finally for the good of his poetry.* —A.N. Whitehead[Q93]

Designing a computer system is very different from designing an algorithm:
- The external interface (the requirements) is more complicated, fuzzy and changeable.
- The system has much more internal structure, and hence many internal interfaces.
- The measure of success is much less clear.

The designers usually find themselves floundering in a sea of possibilities, unclear about how one choice will limit their freedom to make other choices, or affect the size and performance of the entire system. There probably isn't a 'best' way to build it, or even any major part of it; what's important is to avoid a terrible way, and to have clear division of responsibilities among the parts.

I have designed and built a number of computer systems, some that succeeded and some that didn't. I have also used and studied many other systems, both successes and failures. From this experience come some general hints for designing good ones. Most are part of the folk wisdom of experienced designers, but, "It is not sufficiently considered that men more often need to be reminded than informed."[Q35] There are also some principles (about abstraction and modules) that almost always apply, and some oppositions that suggest different ways to look at things.

I've organized the hints along three axes, corresponding to three time-honored questions, with a catchy summary: STEADY by AID with ART.

| What? | **Goals** | STEADY | — **S**imple, **T**imely, **E**fficient, **A**daptable, **D**ependable, **Y**ummy |
| How? | **Techniques** | by AID | —**A**pproximate, **I**ncremental, **D**ivide & Conquer |
| When, who? | **Process** | with ART | —**A**rchitecture, **A**utomate, **R**eview, **T**echniques, **T**est |

There are a lot of hints, but here are the most important ones:
- Keep it simple.
- Write a spec.
- Design with independent modules.
- Exploit the ABCs of efficiency.
- Treat state as both being and becoming.
- Use eventual consistency to keep data local.

These are just hints. They are not novel (with a few exceptions), foolproof recipes, guaranteed to work, precisely formulated laws of system design or operation, consistent, or always appropriate and approved by all the leading experts. Skip over the ones you find wrong, useless or boring.

The paper begins with a list of oppositions (simple vs. rich, imperative vs. declarative, etc.), which can help you decide on priorities and structure for a system. §2 presents the principles: abstraction, specs, code, modularity and the value of a point of view. In §3 each goal gets a section on the techniques that support it, followed by one for incremental techniques that didn't fit under a goal. "Efficient" gets by far the most space, followed by "dependable", because locality and concurrency fall naturally under the first and redundancy under the second, and these three are fundamental to today's systems. Finally there's a short nontechnical §4 on process, and a discussion of each opposition in §5. Throughout, short slogans capture the most important points without any nuance, and quotations give a sometimes cynical commentary on the text.

There are lots of examples to illustrate specific points; I've tried to choose ones that are well-known or well-described online. Look for an example that matches your problem; it can help you find a good technique. I've also told some longer stories, marked » and in smaller type. Many things fit in several places, so there are many cross-reference links. A term of art is in *italics* the first time it's used, and it's a good starting point for a web search; so are the names of techniques and examples. I've put in explicit references when I think a search won't find what you need.

I'm afraid that what I've written is rather dense—you'll need to think carefully about many of the points to get the most out of them —why I said it that way, what's wrong with obvious alternatives, how it connects to other points. And I've omitted many details and caveats that you can find in the literature. Otherwise there would be a thousand pages, though.

Most of what I have to say is about software—the hardware usually comes off the shelf. If your system does need new hardware design, many of these ideas still apply.

For many years I saw no reason to rewrite or extend my 1983 paper on hints for system design.[R59] It said what I knew about personal distributed computing, operating systems, languages, networking, databases and fault tolerance, and things hadn't changed that much from the 1970s. But since the mid-1990s the Internet, the World Wide Web, search engines, mobile phones, social media, electronic commerce, malware, phishing, robots and the Internet of Things have become part of everyday life, and concurrency and scaling are now dominant themes in systems.

Then I could fit nearly everything I knew into a reasonable number of pages, but today computing is much more diverse and I know a lot more; this paper is unreasonably long. I couldn't find a single way to organize it, so I've taken several different perspectives and put in links (like this) to help you find what you need if you read it online, as well as an index.

This is not a review article; the work I cite is the work I know about, not necessarily the earliest or the best. I've given some references to material that expands on the ideas or examples, but usually only when it would be hard to find with a web search.

A shorter version is here, half the size of this one but still 50% longer than the 1983 paper.

## 1.1 Oppositions and slogans

*I've looked at life from both sides now.* —Joni Mitchell[Q54]

It often helps to think about design in terms of the opposition between two (or three) extremes. These are not irreconcilable alternatives, but the endpoints of a range of possibilities. Here are some useful ones, each with a few slogans that hint at its (sometimes contradictory) essence. They are ordered by the first goal or technique they serve, and discussed in §5.

| Goal | Opposition | Slogans |
|---|---|---|
| Principles | Spec ↔ code | { Have a spec. Get it right. Keep it clean. <br> Don't hide power. Leave it to the client. |
| Simple | Simple ↔ rich, fine ↔ features, general ↔ specialized | { KISS: Keep It Simple, Stupid. Don't generalize. <br> Do one thing well. Don't hide power. <br> Make it fast. Use brute force. |
| | Perfect ↔ adequate, exact ↔ tolerant | { Good enough. Worse is better. <br> Flaky, springy parts. |
| | Spec ↔ code | { Keep secrets. Free the implementer. <br> Good fences make good neighbors. <br> Embrace nondeterminism. Abstractions leak. |
| | Imperative ↔ functional ↔ declarative | Make it atomic. Use math. Say what you *want*. |
| | Immutable ↔ append-only ↔ mutable | Make it stay put. |
| Timely | Precise ↔ approximate software | Get it right. Make it cool. |
| Efficient | | { ABCs. Use theory. Latency vs. bandwidth. <br> $S^3$: shard, stream or struggle. |
| | Dynamic ↔ static | Stay loose. Pin it down. Split resources. |
| | Indirect ↔ inline | Take a detour, see the world. Use what you know. |
| | Time ↔ space | Cache answers. Keep data small and close. |
| | Lazy ↔ eager ↔ speculative | Put it off. Take a flyer. |
| | Centralized ↔ distributed, share ↔ copy | Do it again. Make copies. Reach consensus. |
| Adaptable | Fixed ↔ evolving, monolithic ↔ extensible | { The only constant is change. <br> Make it extensible. Flaky, springy parts. |
| | Evolution ↔ revolution | Stay calm. Ride the curve. Seize the moment. |
| | Policy ↔ mechanism | Change your mind. |
| Dependable | Consistent ↔ available ↔ partition-tolerant | Safety first. Always ready. Good enough. |
| | Generate ↔ check | Trust but verify. |
| | Persistent ↔ volatile | Don't forget. Start clean. |
| Yummy | Simple ↔ rich, fine ↔ features | KISS: Keep It Simple, Stupid. |
| Incremental | Being ↔ becoming | How did we get here? Don't copy, share. |
| | Iterative ↔ recursive, array ↔ tree | Keep doing it. A part is like the whole. |
| | Recompute ↔ adjust | Take small steps. |
| Process | | Build on a platform. Keep interfaces stable. |

# 2. Principles

*The quest for precision, in words or concepts or meanings, is a wild goose chase.* —Karl Popper[Q66]

The ideas in this section are not just hints, they are the basic mental tools for system design. But if you are math-averse and section 2.1 puts you off, just skip it.

## 2.1 Abstraction—Have a spec

*The purpose of abstraction is not to be vague, but to create a new semantic level in which one can be absolutely precise.* —Edsger Dijkstra[Q21]

*Without a specification, a system cannot be wrong, it can only be surprising.* —Gary McGraw[Q52]

*If you're not writing a program, don't use a programming language.* —Leslie Lamport[Q47]

*The hardest part ... is arriving at a complete and consistent specification, and much of the [work is] the debugging of the specification.* —Fred Brooks[Q9]

*Architecture is the things about a computer that a machine language programmer must understand to write a correct (timing independent) program for the machine.* (paraphrased) —Gene Amdahl et al[Q2]

Spec ↔ code $\begin{cases} \text{Keep secrets. Free the implementer.} \\ \text{Good fences make good neighbors.} \\ \text{Embrace nondeterminism. Abstractions leak.} \end{cases}$

*Abstraction* is the most important idea in computing. It's the way to make things simple enough that your limited brain can get the machine to do what you want, even though the details of what it does are too complicated for you to track: many, many steps and many, many bits of data. The idea is to have a *specification* for the computer system that tells its clients

– *what*: everything they need to know to use the system,

– but not *how*: nothing about how it works inside—the *code* ("implementation" is too long).

The spec is normally much smaller and clearer than the code, and it decouples the client from the details of the code, so that (a) the client's life is simpler and (b) changes in the code don't affect the client. An abstraction is better if it's simpler and clearer; it's good enough if your limited brain can use it effectively.

A system's *state* is the values of its variables. The spec describes a client's view of the state using basic notions from mathematics, usually relations (sets of ordered pairs) and their special cases: sets, sequences, tuples, functions, and graphs. This is the *abstract* state. For example, a file system spec describes a file as a sequence (array or list) of bytes (a sequence or tuple is a function from $0..length - 1$ to its elements, in this case bytes). Internally the code has data blocks, index blocks, buffer caches, storage allocators, crash recovery, etc., but none of this appears in the spec. The spec *hides* the complexity of the code from the client and keeps *secret* the details that the client shouldn't depend on because they are irrelevant and might change. Almost always the spec is much simpler, so the client's life is much easier. If it's not, you are probably doing something

wrong—usually, giving details that should be hidden from the client, or describing something that's too small.

The spec also describes the *actions* that read and change the state; a file has read, write, and set-length actions. The state and actions define a *state machine* or *transition system*. An action $a$ is just a set of possible transitions or *steps* from a pre-state $s$ to a post-state $s'$, so it too can be described by a relation, a predicate $a(s, s')$ on pairs of states that is true exactly when a step from $s$ to $s'$ is one of the action's steps. There are many notations (usually called programming languages) for writing down these relations easily and clearly, but first-order logic underlies all of them. Example: x:=y is short for the predicate $x' = y$ **and** $(\forall\, v\ \textbf{except}\ x \mid v' = v)$; the value of x is now y and the other variables stay the same. There might be more than one possible next state if an action is nondeterministic, or none if it's blocked. A *behavior* of the system (also called a *history* or *trace*) is just a sequence of steps that the system could take.

An action $a$ is *enabled* in state $s$ if there's at least one next state $s'$ such that $a(s, s')$, and *blocked* if it isn't enabled (presumably some other action is enabled, so the system can do something). In a sequential program, the atomic action at program counter $p$ is $a$ **and** $(s.PC = p)$ **and** $(s'.PC = p')$: it's only enabled when the PC is $p$, and the next PC is $p'$.

The appendix explains how to give the meaning of a whole program as a logical formula.

Why use math in a spec? For clarity and precision. English prose comments are a substitute for a precise description of the state and the actions. They are a good place to start, certainly much better than nothing, and such comments can also help the developer's intuition about the system. It's surprisingly hard to make an English spec complete and correct, though. When you try to translate prose into math you usually find that you overlooked many details, and that you don't even have the right vocabulary to express them clearly and concisely. The computer won't overlook any details.

The actions are as important to the spec as the state, though sometimes the choice of state makes the actions so obvious that they don't seem important. But usually the actions should follow from what the abstraction is supposed to be good for. A simple example is a key-value store, where the state $s$ is a partial function from keys to values—a set of (key, value) pairs with no two keys the same—and (with $s.\text{dom}$ for the domain of $s$) the actions are

$read(key) = $ **return** $s(key)$,
$write(key, value) = s(key) \coloneqq value$ if it's not read-only,
$enumerate(k_1, k_2) = $ **return** $\{key \in s.dom \mid k_1 \leq key \leq k_2\}$ if keys are ordered, and
$search(query) = $ **return** $\{k \in s.dom \mid query(k, s(k))\}$ for pairs that satisfy a query.

Many systems have this interface under different names: OS file directories, JavaScript objects, PostScript dictionaries, document stores, property lists, arrays (with keys in $0..n$), … Nesting gets you JSON or XML.

A spec can be very partial, describing only some aspects of the behavior; then it's often called a *property*. For example, it might just specify "no segfaults" by allowing any step that isn't a segfault. Type safety and memory safety are properties. As well as being partial, a spec can be *nondeterministic*: any of a set of results is acceptable; for example, a timing spec such as "Less than 200 ms". And often details *should* be left open: eventual consistency just says that an update will surely be visible by the end of the next sync (if there is one); it tells you nothing else about when. Language specs are often nondeterministic, for example when function arguments that may have side effects can be evaluated in any order (of course the compiled code is deterministic).

The code should *satisfy* (meet, conform to) the spec. This means that every visible behavior of the code is a visible behavior of the spec: code behaviors are a *subset* of spec behaviors. The "visible" is important; typically the code has internal state that's invisible, and often the spec does too. A partial spec usually has less visible state. This doesn't mean that the code does *everything* the spec allows. In particular, the spec is often nondeterministic where the code takes a single path.

Satisfying is subset, hence it's transitive: if $C \subseteq R$ and $R \subseteq S$ then $C \subseteq S$. So you can get from spec to code in several stages, putting in more details at each stage. This is called *stepwise refinement*. You usually stop the formal development once you have correct code for the tricky parts, even if it's still far from executable,[R78] because the bugs that you add in getting from there to something you can ship are much less tricky.

"Hyperproperties" that constrain sets of possible behaviors such as non-interference (the result doesn't depend on the value of a secret) are beyond the scope of this sketch.

Finding good abstractions is the most important part of designing a system. A language gives you some built-in abstractions: strings, arrays, dictionaries, functions. These are useful, but they are less important than the abstractions in the platform you are building on, such as files, networking, relational data, vectors and matrices, etc. And those in turn are less important than the abstractions specific to the application, such as calendars, protein structure or robot motion.

Which comes first, the spec or the code? In theory the spec should come first, since it reflects what you want done; this is called top-down design, and the code is a *refinement* of the spec. In practice they evolve together, because you can't tell what the spec should be until you see how it affects the code and the system's customers. The first ideas for a spec are usually either too ambitious or too close to the code, providing both more and less than the customers need. Thus the process for getting the spec is distinct from the way to think about the system once you have it.

### 2.1.1 *Safety and liveness*

Any spec that just says what visible behaviors are okay is the conjunction of two parts:
- A *safety* spec, which says what the code *may* do, or equivalently, that nothing bad ever happens. If the code violates a safety spec the bad thing happens in a finite number of steps. Code that does nothing satisfies every safety spec.
- A *liveness* spec, which says what the code *must* do, or equivalently, that something good eventually happens, usually that it's *fair*: every action allowed by safety eventually happens. No finite behavior can violate liveness, because the good thing could happen later. If the spec includes real time then there's a clock that ticks (a liveness spec), so the code always does something.

For a non-interactive program such as *sort*(*a*:**seq**) **returns** *sa* (one that just takes an input and produces a result), the spec is just the relation between the pre-state and the post-state. The traces are of length two, and safety and liveness are called partial correctness and termination.

Usually safety is what's important, because "eventually" is not very useful; you care about getting a result within two seconds, and that's a safety property (violated after two seconds). But a liveness proof often works by counting steps, loop iterations, function calls or whatever, and so contains a proof of a time bound. Sometimes safety of the code state implies liveness of the spec.

### 2.1.2 *Operations*

The description above is written entirely in terms of state changes. Usually it's easier to describe visible behavior as sequences of *operations* the client can invoke and results that the system returns. For example, the spec for a first-in first-out buffer has `put(x)` and `get()` **returns** `y` operations (the parameter and result values are part of the operation), and its internal state is a sequence of items that have been `put` and not yet retrieved by `get`. A state machine with an operation for each transition (either visible or internal) is also called a *labelled transition system.*

Describing the visible behavior as a sequence of operations seems very different from describing it as a sequence of states. Operations (function calls) are natural in a programming language, and changes to state (registers and memory) are natural in hardware. But these two views are equivalent; what reconciles them is the idea of a *calling sequence*, a sequence of state changes that correspond to invoking an operation and getting a result. There are many possible calling sequences:

- Executing a machine instruction; the details are buried in the CPU hardware.
- Putting some arguments into specific registers, executing a branch-and-link instruction that saves the PC in a register, and getting the result from a specific register.
- Pushing some arguments onto a stack, executing a call instruction that pushes the PC onto the stack, and getting the result from the top of the stack.
- Marshaling arguments into a packet, sending it to an RPC server, and waiting for a result packet.

Usually an operating system chooses a PC-saving sequence and a runtime chooses an RPC scheme, so you can forget about the details and describe the visible behavior in terms of operations.

## 2.2   Writing a spec—KISS: Keep It Simple, Stupid.

*Seek simplicity, and distrust it.* —A.N. Whitehead[Q95]
*Reality is that which, when you stop believing in it, doesn't go away.* —Philip K. Dick[Q19]
*What is written without effort is in general read without pleasure.* —Samuel Johnson[Q74]
*The problem ... is not* precise *language. The problem is* clear *language.* —Richard Feynman[Q27]

*It is impossible to speak in such a way that you cannot be misunderstood: there will always be some who misunderstand you.* —Karl Popper[Q66]

How should you go about writing a spec? There are two steps:

(1) Write down the state of the spec (the **abstract** state).

You have to know the state to even get started, and finding the simplest and clearest abstract state is *always* worth the effort. It's hard, because you have to shake loose from the details of the code you have in mind and think about what your clients really need. The mental tools you need for this are the elementary discrete math of relations, and a good understanding of the clients.

Often people say that the abstract state is not real or that the spec is an illusion; only the RAM bytes, disk blocks and machine instructions are real. I can't understand this; a physicist will say that only the quantum mechanics of electrons in silicon is real. What they probably mean is that the spec doesn't actually describe the behaviors of the system. This can happen in several ways:

- It can be wrong: the code does things the spec doesn't allow. This is a bug (in either the spec or the code) that should be fixed.
- It can omit important details: how accurate a sine routine is or what happens if there's a failure.
- It can omit unimportant details by being leaky. This is a matter of judgment.

For the file system example, the spec state has files $F$, directories $D$, and inode numbers $N$. A file is $F = $ **seq** Byte (a function $0..l-1 \to$ Byte) and a directory is a (partial) function $D = Name \to N$. The state is a function $s = N \to (F$ **or** $D)$ that gives the current contents of the nodes. The $D$'s must organize the nodes into a graph where the $F$'s are leaf nodes and the $D$'s form a tree or DAG rooted in $s(0)$; an invariant on the state says this.

(2) Write down the spec actions: how each action depends on the state and changes the state. You may need English comments to guide the developer's intuition.

Now you have everything the client needs to know. If you haven't done this much, you probably can't do a decent job of documenting for the client.

For example, here are three of the file system actions (somewhat simplified). Notation: $f$.dom is the domain of $f$ ($0..l-1$ if $l$ is $f$'s length), $PN = $ **seq** $Name$ is a path name; for **var** see here. Note that this spec is declarative, without any loops.

$read(f, i) = $ **if** $i \in f$.dom **then return** $f_i$

$write(f, i, b) = $ **var** $f'$ | $\quad f'$.dom $= f$.dom $\cup\; 0..i-1$
$\qquad\qquad\qquad\quad$ **and** $\forall j \in f'$.dom | $f'_j = ($**if** $j = i$ **then** $b$ **elseif** $j \in f$.dom **then** $f_i$ **else** $0)$
$\qquad\quad$ **in** $f := f'$

**let** $isName(pn, n) = \exists\, path$: **seq** $N$ | $\qquad$ // $pn$ is a path name for node $n$
$\qquad path_0 = 0$ **and** $path_{pn.\text{size}} = n$ **and** $\forall i \in pn$.dom | $s(path_i)(pn_i) = path_{i+1}$ **in**
$open(pn) = $ **var** $n$ | $isName(pn, n)$ **in return** $n$

The invariant says there's at least one path to each node, but at most $k$ paths (hence no cycles) and one to a $D$: $\forall n \in s.\text{dom} \mid \exists\, k \mid \bigl(0 < \{pn \mid isName(pn, n)\}.\text{size} \leq (\textbf{if } s(n) \in D \textbf{ then } 1 \textbf{ else } k)\bigr)$.

Writing down the actions precisely is a lot more work than writing down the state, as the example suggests, and you probably need a good notation (language) to do it clearly and concisely. The lecture notes for my MIT course 6.826, Principles of Computer Systems, have many realistic (if abstract) examples worked out in some detail, unfortunately using a made-up language.[R63] A spec that captures most of the observable behavior of actual code for TCP and sockets, written in this state-and-nondeterministic-actions style, illustrates the state of the art in 2018.[R11]

A spec should be simple, it should be complete enough, and it should admit code that is small and fast enough. Good specs are hard, because each spec is a small programming language with its own types and built-in operations, and language design is hard. Also, the spec mustn't promise more than the code can deliver—not the best possible code, but the code you can actually write. In Dennis Ritchie's words: "In spite of these limitations, the stream I/O system works well. Its aim was to improve design rather than to add features, in the belief that with proper design, the features come cheaply. This approach is arduous, but continues to succeed."[R87]

There is nothing special about concurrency, except that it makes the code (and perhaps the spec) nondeterministic: the current state doesn't determine the next step, which could come from any thread that isn't blocked. Likewise there is nothing special about failures. A crash or the misbehavior of a component is just another action. Crashes cause trouble because they may destroy state that you would prefer to keep, and because they add nondeterminism that's not under your control. But these are facts of life that you have to deal with, not flaws in the method.

If there is concurrency, the file system operations often are not *atomic* actions that make a single state change; indeed, in Windows and Unix none of them is atomic. Instead there is a *start* action that collects the arguments and an *end* action that returns the result, and other actions that affect the result can occur in between; this is a bit like eventual consistency. Such non-atomic operations are difficult to reason about, but their code can be much cheaper to run.

The spec describes *all* the possible visible behaviors of the system in terms of the visible abstract state (which includes the arguments and results of operations that the client invokes). Sometimes people talk about a "functional spec", a vague notion based on the wrong idea that the system has no internal state or actions, and that there's nothing to say about its behavior beyond the results that its actions return. But reliability, failures, resource consumption, latency and system configuration can all be visible state that a spec can describe.

Is the spec correct? In other words, is it actually what you want the system to do? There's no way to *prove* that it is; what standard would you judge it against? Sometimes desired properties follow from the spec, such as a bound on resource consumption or the ability to read out and restore persistent state, but the best bet is to keep it simple. A 20 page spec won't be correct.

Specs need modularity too; a part that's repeated, perhaps with some parameters changed, is like a routine in a program. Give it a meaningful name and write the body just once.

It's tempting to write a spec axiomatically to make it as independent of the code as possible, but experience says that this doesn't work, because it's too hard to get the axioms right.

2.2.1 *Leaky specs and bad specs*

Specs are usually incomplete or *leaky*; they don't describe all the behaviors of the system that are visible to the client. Most notably, specs often don't say much about the resources they consume, especially running time. This may be okay, depending on what the client needs, but for a system that interacts with the physical world (which includes users) predictable response time is important, especially for interactions that should lead to smooth motion such as scrolling on a touch device. Also, it can be disastrous when the code for a frequently used abstraction changes and an application suddenly runs much more slowly. Ideally there will be an approximate expected or worst-case timing spec, written as a function of the cost of primitives that the code invokes: cache misses, remote procedure calls, network file accesses, etc. Often, however, the guarantee is probabilistic and the best you can do is to monitor it and report unexpected delays. Don't ignore the problem; that will end in tears. There are similar issues for other resources such as RAM, storage space or network bandwidth. If the code depends on external resources that it doesn't control and that can fail or perform badly, such as networking, the spec must say this. For example, network file systems have caused many problems when actions such as `stat` that were assumed to be fast become very slow or never finish at all.

The whole point of a spec is to suppress irrelevant detail, but leaky specs can also be a problem for security, since observable state changes that are not part of the spec such as timing, power consumption, cache allocation, acoustic or electromagnetic radiation, etc. can reveal secrets; these are called *side channels* in cryptography, where they have been studied extensively. In general, abstractions don't keep secrets very well. If you care about side channels, don't share resources.

Sometimes the spec needs to be leaky, in the sense that it exposes some internal secrets, to give clients the access they need to run fast.[R26] Increasing the number of assumptions that one part of a system makes about another may allow less work to be done, perhaps a lot less.

Being leaky is not necessarily a bad thing, and in general it's unavoidable. But there are other properties of a spec that *are* usually bad:

- *Complexity* is hard for the client to understand, and hard to code. It often comes from being overambitious, ignoring the state-and-actions recipe, or exposing the code's secrets.
- *Brittleness* makes the spec depend on details of the environment that are likely to change, or on details of how it is called that are easy to get wrong.
- *Errors* or *failures* in the code mean that the code won't satisfy the spec, unless the spec gives it a way to report them. A common example is a synchronous API that makes the code look local, fast and reliable even though it's really remote, slow and flaky.
- Similarly, *contention* or overload may keep the code from meeting the spec if there's no way to report these problems or set priorities.

- *De facto specs*, in either function or performance, happen when the code has properties that clients come to depend on even though they are not in the spec. Unless you can change the clients, you are now stuck. Sometimes virtualization can fix these problems.

»Bug-for-bug. Code is supposed to keep secrets; it shouldn't expose properties of the system that are not in the spec and that you might want to change, because clients may come to depend on them. A sad example of this was a fairly popular app for Windows 3 (perhaps it had 10 million users). The app had a routine that did something minor to a string, maybe removing all the punctuation. Because of a bug, in one place it called the routine with a bogus string pointer that happened to point to the garbage area off the end of the stack. The routine would run, make enough changes to the garbage to satisfy it, and return; no one noticed the string that didn't get processed. When Windows 95 came along, however, enough things were different that the bogus pointer had a different value, and Windows 95 has more memory protection, so now the bogus call caused a segfault, crashing the app. The fix to keep the app working, in the days before online updates? A special check for this app and this call in the segfault routine, which did whatever was necessary to satisfy the app. This is what it means to be bug-for-bug compatible.

»TPM 1.1. In the late 1990s Paul England and I at Microsoft invented the mechanisms that ended up in the Trusted Platform Module (TPM). By 2002 the TPM had been standardized, and later I wanted to know some details about how it worked. So I downloaded the spec[R107] and started to read it, but I got stuck right away. Even though I actually invented these mechanisms, I couldn't understand the language of the standard.

### 2.2.2 *Executable specs*

Another kind of spec is *executable*: the machine can run it fast enough for its clients to actually use it, perhaps not to do useful work but at least to see whether they like its functionality.[R53] This has some advantages:

– You can try it out, often a better way than thinking about it to learn whether you like it.
– You can use it as an oracle for testing real code.[R85] This is tricky if the spec is nondeterministic, because the execution has to explore all the possible paths.[R11]
– Perhaps you can evolve it into code that's good enough to ship.

If it's a spec for a module in a bigger system and it's fast enough (perhaps running on a supercomputer), you can run that system before there's real code for the module.

An executable spec also has some drawbacks:

– It may not be simple and clear enough to be useful as a spec; you need powerful primitives and a strong will to keep from putting in too many details.
– Nondeterminism is hard, and a single choice in the spec may constrain the code too much.
– It can't use the full power of mathematics. For example, it can't say, "There exists a path through this network such that …"

A related idea is a *reference implementation*. Sometimes this means an executable spec, but more often it means practical but unoptimized code, intended to make it clear that the spec itself is practical, and often to guide implementers about what to do.

### 2.3   Writing the code: Correctness—Get it right

*Smart companies know reliable software is not cost effective. It's much cheaper to release buggy software and fix the 5-10% of bugs people complain about.* (paraphrased)   —Bruce Schneier[Q72]

*For simple mechanisms, it is often easier to describe how they work than what they do, while for more complicated mechanisms, it is usually the other way around.* —John von Neumann[Q89]

*Debugging is twice as hard as writing a program in the first place. So if you're as clever as you can be when you write it, how will you ever debug it?* —Brian Kernighan[Q44]

*Testing can show the presence of bugs, but never their absence.* —Edsger Dijkstra[Q22]

*Beware of bugs in the above code; I have only proved it correct, not tried it.* —Donald Knuth[Q45]

Most of this paper is about how to write the code. But is the code correct? In other words, does it satisfy the spec? (You don't have a spec? Then the question makes no sense.) In theory this question has a yes-or-no answer. If

- the spec is a predicate that describes every allowed or required action (step) of the system,
- the code precisely specifies every action that the system takes, and
- you know which parts of the state are *visible* to the client,

then correctness is a theorem: "Every visible code behavior is a spec behavior," either true or false. This section explains how to prove this theorem; even though it's seldom worthwhile to complete this proof, you can find bugs and get insight into why the code works by writing down the abstraction function and invariants described in step (3) below.

If the theorem is true, a surprising fact is that it has a *simulation proof*: there is an *abstraction function* $f$ from the code state to the spec state that matches each code action with a spec action that has the same visible effect: every code action $c \to c'$ from a reachable state $c$ has a matching spec action $f(c) \to f(c')$; it's skip (no change) if the action doesn't change any visible state.

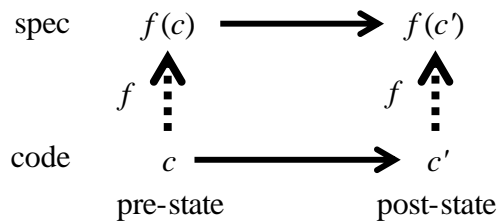

**Fig. 1**: Inductive step for a simulation proof

Figure 1 is the inductive step in the proof that every visible code behavior is a spec behavior. The basis is that $f$ takes any initial state of the code into an initial state of the spec.

For a non-interactive system such as $sort(a{:}\mathbf{seq})$ **returns** $sa$ the figure is the whole story, since the spec has only two states. The code has lots of internal states as it reorders the sequence, but they all map to the spec's pre-state. For the file system spec action $write(f, i, b)$ (ignoring the complications of crashes) the code does

- a long sequence of internal actions that simulate skips in the spec, to bring into RAM the index block and data block for byte $i$ of the file, allocating new blocks if necessary,
- followed by a visible action to update byte $i$ to $b$ that simulates the $write$ action,
- followed by more internal code actions to write the changed blocks back to disk.

A later $read(f, i)$ will return $b$. The abstraction function says that if $j$ is $i$ mod the disk block size and byte $i$ of $f$ is in disk block $b$ (a function of the inode and index blocks for $f$), then $f(i) = b_j$, unless $b$ is in the buffer cache at address $c$, in which case $f(i) = c_j$.

You might need to add *history variables* to the code to write the abstraction function (or alternatively, define an *abstraction relation*). These are variables that keep track of spec state that the code doesn't need; they aren't allowed to affect the original code state. For example, the spec for a simple statistics system might allow you to add data points one at a time, and then obtain the mean of the whole set. If the code computes the mean incrementally and doesn't store the set, you need to add the set as a history variable. A few specs also need *prophecy variables*,[R1] but you're unlikely to run into this. I know only three real-life examples; see MIT handout 8,[R63] pages 8-16.

Following this script, once you have the spec (steps (1) and (2) above) and the code state and actions, there are two more steps to connect them:

(3) Find an abstraction function (or relation) from code to spec state. You'll need some of the code state, but a lot of it is there for performance and won't show up in the abstraction function.

At the same time, find the *invariants* on the code state, that is, define the states that the code can reach; the proof only needs to deal with actions from reachable states. For example, code that has a sorted array has an invariant that says so, and you need it to show that the code actions that use binary search to look up a key in the array actually work.

The reachable states are not declarative because they are defined by the closure of the $Next$ relation, which has lots of steps. An *inductive invariant* is strong enough that you can prove it by induction: it's true initially and if it's true before any action it remains true afterward. This makes (some superset of) the reachable states declarative. The superset needs to be small enough (the invariant strong enough) that it implies the property you want.

(4) Finally, do the actual simulation proof that every code action preserves the visible behavior and the invariants.

Step (4) requires reasoning about *every* action in the code from *every* reachable code state, so it's by far the most work. Step (3) requires *understanding* why the code works, and it usually uncovers lots of bugs. Unfortunately, the only way to be sure that you've done it right is to do step (4), which is usually not worthwhile. With modern proof systems it's possible to do a machine-checked simulation proof for systems with thousands of lines of code, and to redo the proof fairly easily after changing the code.[R45] But it's still a lot of work, which only pays off as research, or when correctness is very important. Writing a spec is always worthwhile, though, because it decouples the client from the code.

An alternative is *model checking*: exploring a subset of the code's state space systematically, looking for behaviors that violate the spec. This doesn't give any guarantee of correctness (unless there are so few behaviors that the checker can try them all), but it finds lots of bugs. In fact, many people's experience is that it finds most of the bugs.[R78,R35] And it's much less work than a simulation proof, since the checking itself is automatic and only costs CPU time. Often you have to modify the code, though, to coax the model checker into exploring the right parts of the state space.

Testing, model checking, and proof are all much easier when a big system is decomposed into well-specified modules with simple abstract states, because you only have to consider the code state of one module at a time, which is much smaller than the code state of the entire system.

You might also worry about whether the spec is correct.

### 2.3.1 *Types*

Types are a way to express some facts about your code that the machine can understand and check, in particular some stylized preconditions and postconditions. The idea is that
- a value $v$ of type $T$ has an extra `type` field whose value is $T$,
- if $R$ is a routine with type $T \rightarrow U$, its argument must have type $T$ (the precondition): $R(v)$ is an error unless $v.\text{type} = T$ (or more generally, $v.\text{type}$ is a subtype of $T$),
- $R$'s result has type $U$ (the postcondition).

With dynamic types the type field is there at runtime (most often it's called a class) and a call of $R$ checks the precondition. In a static system `type` is a "ghost" field not present at runtime, because the compiler knows the type of every expression $e$ and guarantees that if $e$ has type $T$, then $e.\text{type} = T$. In other words, the spec (the type declarations) says that the value of $e$ has type $T$, and the compiler proves it. So the compiler can also check the preconditions.

Why are static types good? For the same reason that static checking in general is good: the compiler can try to prove theorems about your program, and if it fails you have found a bug early, when it's cheap to fix. Most of the theorems are not very interesting, since they just say that arguments have the right types; sometimes the type system can figure out (infer) the types as well, as in Haskell or TypeScript. But the first draft of a program almost always has lots of errors, most pretty obvious, so type checking finds lots of bugs fast when it can't prove its trivial theorems.[R83]

A type carries with it a table of methods (distinguished by their names), routines that operate on data of that type, or a named function carries its overloaded methods (distinguished by the types of their arguments). Statically the compiler knows the types and looks up the method; dynamically the lookup is at runtime. All languages use one of these schemes, and they do it statically at least for built-in methods like assignment or adding numbers.

### 2.3.2 *Languages*

What programming language should you use? There is no universal answer to this question, but here are some things to think about:
- How hard is it to write your program so that the language guarantees a safe, bulletproof abstract state, in which a variable always has the expected type and only an explicit write can change its value? Usually this means strong typing and garbage collection. Java is safe in this sense,

C++ is not (unless you hide unsafe features behind a safe abstraction),[R101] and JavaScript is in between. If the abstract state isn't bulletproof, debugging is much harder.
- Is the language well matched to your problem domain? Is it easy to say the things that you say frequently? Is it *possible* to say all the things that you need to say?
- What static checking does the compiler do? A bug found at compile time is much easier to fix.
- How hard is it to make your program efficient enough, and to measure how it uses resources?

## 2.4 Modules and interfaces—Keep it clean. Keep basic interfaces stable.

*Interfaces should follow the principle of least astonishment. —Geoffrey James*[Q35]

The only known way to build a large system is to reinforce abstraction with divide and conquer: break the system down into independent abstractions called *modules*. I'll call the running code of a module a *service*; sometimes people call it an object. The spec for a module does two things:
- it *simplifies* the client's life by hiding the complexity of the code (see above), and
- it *decouples* the client from the code, so that the two can evolve independently.

Thus many people can work on the system productively in parallel without needing to talk to each other. Since a spec embodies assumptions that are shared by more than one part of a system, and sometimes by a great many parts, changing it is costly.

It's common to call the spec of a module its *interface*, and I'll do this too. Unfortunately, in common usage an interface is a very incomplete spec that a compiler or loader can process, giving just the data types and the names and (if you're lucky) the parameters of the operations, rather than what the actions do with the state. Even a good description of the state is often missing.

A really successful interface is like an hourglass: the spec is the narrow neck, with many clients above and many codes below; it can live for decades. Examples: CPU ISAs (instruction set architectures such as x86 and ARM), file systems (Posix), databases (SQL), programming languages (C, C++, JavaScript). In networking, interfaces are especially important because there's no authority to coordinate simultaneous changes in a module and its clients. Examples: ethernet, reliable messages (TCP), names for Internet services (DNS), email addresses, end-to-end security (TLS), web pages (HTTP and HTML). Ousterhout's book on software design[R81] gives many smaller examples, emphasizing how important it is to make the spec much smaller and simpler than the code.

A popular view of decoupling is that the spec is a *contract* between the client and the service:
- The client agrees to depend only on the behavior described by the spec; in return the client only has to know what's in the spec, and can count on the code to actually behave as the spec says it should.
- The service agrees to obey the spec; in return it can do anything it likes internally, and it doesn't have to deliver anything not laid down in the spec.

Many people shun "spec" in favor of "contract", perhaps because they think specs are too formal.

Module interfaces serve much the same purpose as contracts between business entities: to reduce transaction costs by simplifying and standardizing communication. But as with firms, you sometimes have to look inside because you can't understand the spec, it's incomplete, or you just don't believe that the code will actually satisfy it.

A module boundary doesn't just decouple its *code* from the clients; it can decouple its *execution* and resource consumption as well. If the interface is asynchronous neither side waits for the other, so the service can keep running no matter what the clients are doing, and vice versa. And it can manage the way it consumes storage and other resources independently of its clients. Thus the service is an autonomous agent. How does this show up in the spec? A complete spec doesn't just say enough about the service's internal state to say what results it returns; the spec also describes how it consumes any resources it shares with its clients (including real time). An autonomous service doesn't share resources (except time), so its spec is simpler and a system that uses it is more dependable and easier to change. Distributed transactions are an interesting example.

### 2.4.1 *Classes and objects*

*Mathematics is the art of giving the same name to different things.* —Henri Poincaré[Q63]

A very popular variation on modules attaches the spec and code to a data item, usually called an *object*, that has hidden private state and public methods that can read and modify that state. Think of it as an independent computer. Programs organized this way are called *object-oriented*, and the style usually has a lot of support from the programming language. You package the specs for a set of routines called *methods* with the same type of first argument into a single spec, here called a *classspec* (it's called a type class in Haskell, a protocol in Smalltalk, an interface in Java, a concept or abstract base class in C++ and Python). The code for the classspec is a *class*, a dictionary that maps each method name to its code. A class with enough methods can satisfy more than one spec. An object of the right type that has the class attached is an *instance* of the class (and of the classspec). With static typing you can put the class in the type instead of the object.

For example, the classspec `Ordered T` might have methods `eq` and `lt` that take two values of type `T`, return a `Bool`, and satisfy the axioms for a partial order. If `x` is an instance of `Ordered T`, then `x.eq(y)` calls the `eq` method (and perhaps so does the prettier `x==y`); to run it, look up `eq` in `x`'s class to get the method's code and call it with `(x,y)` as its arguments. If the compiler knows the class it can do the lookup. This is not magic, and these ideas can work even with no language support. In C, for example, the method lookup is explicit: an object `x` is a pointer to a `struct` with a `class` field that points to the dictionary, which is another `struct` with an `eq` field that points to the method code, and instead of `x.eq(y)` you write `x->class->eq(x,y)`.

Adding methods to a class makes a *subclass*, which *inherits* the superclass methods (and likewise for a classspec); thus `Ordered T` is a subclass of an `Equal T` class that has only the `eq` method. An instance of `Ordered T` is also an instance of `Equal T`. The subclass provides code for the added methods, and it can replace or *override* the superclass methods as well. The subclass should satisfy the classspec of the superclass, which is all that its clients can count on. Then there'll be no surprises when a subclass object is used in code that expects the superclass. An easy way to ensure this is to not override the superclass methods, and to keep the added methods from changing the private data of the superclass. The `final` modifier in Java enforces this, but inheritance in general does not. It's very easy to break the superclass abstraction, because
- usually the spec is very incomplete and
- actually proving correctness is beyond the state of the art, so
- at most you get a guarantee that the method names and types agree.

There are two distinct ideas here: hiding (abstraction) and overloading (using the same name for more than one thing).

- The class is doing the usual job of *abstraction*, describing the object's behavior and hiding its code from clients.
- The class is providing *overloading* for its methods, making it easy to invoke them using names local to the class, but the same as the names of methods in other classes with the same spec or a closely related one.

These two things work together when the different overloaded methods do indeed satisfy the same spec, but can be a rich source of errors when they don't, since there's no way to get the machine to tell you. This may be okay when the same team owns both superclass and subclass, but it's very dangerous if the superclass has many independent clients, since there's no single party ensuring that the subclass satisfies its spec. Overloading can be pushed a lot farther, to automatically select the right method for each combination of argument types from a sea of routines, as in Julia[R10 §4] for numeric computing, or in C++[R101 §C] more generally. Like subclassing, this works well when the methods satisfy the same spec, and it works best for abstractions where mathematicians have done the work, such as + for integers, vectors of reals, sparse matrices of complex numbers, etc... Overloading leads to chaos when the methods don't satisfy the same spec. Beware.

The `Ordered` and `Equal` examples are unusual because unlike most objects, they have no internal state (data). Typically the code state of an object is a set of named fields, for example, a `Point` class with `x` and `y` fields. The state can be immutable, so that `pt1.add(pt2)` returns a new object, or it can change, so that `file.write(i,b)` replaces byte `i` of the file with `b`. Since an object is a module with an abstract state, it's not really doing its job if the code state is the same as the spec state. In particular, it's a bad sign if there is a `setx` method for field `x`; this object looks like an expensive record or struct, without any abstraction. At the opposite extreme, the idea in Simula and Smalltalk was that an object is a small computer, with its methods as the I/O system.

Many languages embed classes in different and confusing ways. It helps to organize them into two main categories.

- **Object-based**: the object hosts the class, as in Smalltalk or Java. Of course to know what methods it makes sense to call, the programmer needs to know what classpecs an expression's class implements. In Smalltalk the programmer has to keep track of this (with a "method not understood" error if they get it wrong); in Java the expression's type tells you statically, but the object's class might be a subclass of the type so the compiler doesn't know the method's actual code (unless it's `final`). In C++ virtual functions are in the object.

- **Type-based**: a type hosts the class, as in Haskell or for built-in methods like "+" in most languages. For example, `Integer` is a host for `Ordered` given suitable code: `integerEq` for `==` and `integerLT` for `<`. A type can host several classes if it has code for all their methods. Because the method code is part of a type, the compiler knows it unless the type is a parameter; in that case the class dictionary is a runtime parameter. In C++ methods (called members) that are not virtual are in the type.

2.4.2 *Layers and platforms*

A system usually has lots of modules, and when a module's spec changes you need to know who depends on it. To make this easier, put related modules into a *layer*, a single unit that a team or vendor can ship and a client can understand. The layer only exposes chosen interfaces, and a lower layer is not allowed to call a routine in a higher layer. So a layer is a big module, normally a client of its *host*, a single layer below it, with one or more layers as its clients above it. Thus the term layer makes some sense, although the turtle example below has some exceptions. A layer can be a client of several hosts; for example, it's usually a client of the CPU hardware (or the language virtual machine) as well as of the layer just below itself.

| | Clients | |
|---|---|---|
| Peers | **YOU** | Peers |
| | Host | |

Layers are good for decoupling, but they are not free. Unless you're very careful, there's a significant cost for each level of abstraction. Usually this cost is worth paying, but if performance is important it's prudent to measure it. There are two ways to reduce it: make it cheaper to go from one layer to another, or bypass some layers (making the system a lot more complicated and hard to maintain). The ideas in § 2.4.4 on open systems can also help.

Usually you build a system on a *platform*, a big layer that serves a wider range of clients and comes from a different organization. Common platforms are a browser (the interface is a document object model accessed through JavaScript) or a database system (the interface is SQL), built on an operating system platform (Windows or Linux; the interface is kernel and library calls) built on a hardware platform (Intel x86 or ARM; the interface is the ISA). The ISA is generally the lowest layer in such a picture, but it's turtles all the way down: the hardware is built on gates and memory cells, which are built on transistors, which are built on electrons obeying the laws of quantum mechanics. Here is a 2020 example with all the turtles:

| Layer | Example |
|---|---|
| application | Gmail |
| web framework | Django |
| database     browser | BigTable    Chrome |
| operating system | Windows 10 |
| virtual machine | VMware |
| ISA | X86 |
| CPU hardware | AMD Ryzen 7 2700X |
| gates     memory | TSMC 7 nm     Micron MT40A16G4 |
| transistors | 7 nm finFET    LPDDR4X-4266 |
| quantum mechanics | electrons |

»Celebrity beeps. Microsoft Windows makes sounds, generically called beeps, to notify the user about various conditions. Some of these conditions happen at low levels of the system. At one point Windows engineering introduced layers to control dependencies, about 50 of them. Then someone had the idea of celebrity beeps, by analogy with celebrity ringtones. But of course a celebrity beep is a valuable property that needs digital rights management, which is done at level 45. This meant that code to beep at level 10 was calling up to level 45. When the checkin tool rejected this, the developers were baffled—they couldn't understand what they were doing wrong.

### 2.4.3 Components

*Reusing pieces of code is like picking off sentences from other people's stories and trying to make a magazine article.* —Bob Frankston[Q28]

*It's harder to read code than to write it.* —Joel Spolsky[Q75]

A module that is engineered to be reused in several systems is called a *component*. Obviously it's better to find a component that does what you need than to build it yourself (don't reinvent the wheel), but there are some pitfalls:

- You need to *understand* its spec, including its performance.
- You need to be confident that its code actually *satisfies* the spec and will be maintained.
- If it doesn't quite do everything that you want, you have to *fill in the gaps*.
- Your environment must satisfy the *assumptions* the component makes: how it allocates resources, how it handles initialization, exceptions and failures, how it's configured and customized, and the interfaces it depends on.

Building a truly reusable component costs several times as much as building a module that does a good job in one system, and usually there's no business model that can pay this cost. So an advertised component probably won't meet your needs for a reliable, maintainable system, though it could still be fine if dependability is not critical (for example, for approximate software).

There are two ways to keep from falling into one of these pitfalls:

- Copy and paste the module's code into your system and make whatever changes you find necessary. This is usually the right thing to do for a small component, because it avoids the problems listed above. The drawback is that it's hard to keep up with bug fixes or improvements.
- Stick to the very large components usually called platforms. There will only be a few of them to learn about, they encapsulate a lot of hard engineering work, and they stay around for a long time because they have a viable business model (since it's impractical to write your own database or browser).[R61] A well-maintained library can also be a source of safe components that are smaller than a whole platform.

### 2.4.4 *Open systems*—Don't hide power. Leave it to the client.

The point of an abstraction is to hide how the code is doing its work, but it shouldn't prevent a client from using all the power of its host. An abstraction can preempt decisions that its clients could make; for example, its way of buffering I/O might keep a device from running at its full bandwidth. If it's an ordinary module, a client can always hack into it, but that's not an option if it's an operating system that isolates its clients, or if you want to keep taking bug fixes. The alternative is careful design that doesn't hide power, but gives clients access to all the underlying performance. The Internet's *UDP protocol* is an example; unlike reliable TCP, it gives clients direct access to the basic unreliable, best-efforts packet delivery, which is critical for real-time applications like voice. *Scheduler activations* are less convenient than threads, but give the client control over scheduling and context switching. *Exokernels* carry this idea further to a *library OS* that absorbs most of the code of an OS so that the client can change it if necessary. Similarly, the fact that locks and condition variables don't let you control the scheduling of waiting threads, often cited as a drawback, is actually an advantage, since it leaves the client free to provide the scheduling it needs, perhaps using several locks or conditions as the interface to the basic thread scheduler.

Another way to expose an abstraction's power (and also to make it extensible) is to make it programmable, either by *callbacks* to client-supplied functions or by programs written in an application-specific instruction set. A successful abstraction will have many clients depending on all the details of this interface, so choose it carefully. There are many examples of programmability:
- The SQL query language, a functional instruction set.
- JavaScript embedded in data: webpages, documents, database records, etc.
- Display lists and more elaborate programs for GPUs.
- Programmable network interfaces (NICs); leaving it to the client is very important here.[R52]
- Software-defined networking.
- Patching of binaries, or of code written in other languages.

Binary patching was first done in the Informer, a tool for instrumenting an OS kernel; it checked the proposed machine code patch for safety.[R33] Now there are tools to instrument and optimize binaries when the source is unavailable,[R99,R71] to instrument network code as in the Berkeley Packet Filter, and to do software fault isolation.[R112]

Client-supplied programs start out being used for very specific purposes, but often evolve to be more general-purpose until you have a full-blown computer; then the cycle can start again.[R76]

Hiding secrets also doesn't mean that the *code* should be secret; the success of open source systems like Linux, Apache, gcc, EMACS and llvm shows the value of having lots of people reading the code and contributing. This is especially important for security, since security through obscurity doesn't work well when attacks can come from anywhere and there are powerful analysis tools. Many eyes are not a substitute for thorough testing or code verification, though.

2.4.5 *Robustness*—Flaky, springy parts.

*Principle of robustness: Be conservative in what you do, be liberal in what you accept from others.*
    —Jon Postel[Q67]

How strong should a spec be? A stronger spec promises more to the client, but is harder to code. A looser (weaker) spec promises less and is easier to code. If you want to build a robust system, though, it's often better to be more conservative as a client and more liberal as a service. A conservative client tries to rely only on the essential features of the spec (by guessing what they are, or by studying other successful clients). If the service turns out to be flaky (that is, doesn't implement the whole spec correctly) the client will still work. A liberal service tries to be springy in anticipating the client's mistakes and catering to them. The resulting system is more likely to work in spite of inevitable bugs and misunderstandings. This applies in spades to standards, which usually have a lot of unnecessary features that don't really work.

### 2.4.6 Standards

*Q: What do you get when you cross a mobster with an international standard?*

*A: Someone who makes you an offer that you can't understand.* —Paul Mockapetris[Q55]

Specs that are either widely accepted, or agreed upon by an accepted process, are called *standards*. Sometimes they are very successful: ethernet, the IBM PC, USB, TCP/IP, HTML, C, C++, JavaScript, PostScript, PDF, RSA and Linux are obvious examples. It's notable that all of these were originally designed by one person or a small group, not by an accepted process. In fact, most standards that start out in a standards committee end up on the scrapheap, because these committees are political animals that tend to drive away good engineers and to take the path of least resistance by including everyone's ideas. Examples: OSI networking,[R88] IPsec network security, IPv6, XML, UML, ATM networking. *Caveat emptor.*

A standard that is created by a government to meet its own needs is especially likely to fail. The US Department of Defense has had at least two major computing standards failures, the Ada programming language and the Orange Book standard for multilevel security.

On the other hand, it's fine for a standard that starts out in an ad hoc form and succeeds to end up in a committee that takes responsibility for the boring task of evolving it in a conservative, backward-compatible way to meet new needs, and this is usually what happens. It's not fair to be too hard on failed standards either, since most new ideas do fail.

The important thing about standards is this: one that's been around for a while and is known to have good implementations gives you a stable base on which to build. But robustness means that it's wise to stick to the simplest and most heavily used parts; it's quite likely that the rest of it is poorly designed or implemented. And if there's no working code, stay away from it.

»X.509. Sometimes a badly designed and overly complex committee standard does take hold. A striking example is the X.509 standard for digitally signed certificates, the only part of OSI networking that has survived. It's such a mess that whenever I put two X.509 experts in a room and ask them a question about it, I get at least two answers.

»WEP. Sometimes a standard is messed up for political reasons by some members of the committee, either deliberately or because they have been misled. The Wired Equivalent Privacy (WEP) standard to support Wi-Fi security, for example, has flaws that were well known to experts on encryption protocols, but they were either not consulted or ignored.[R13]

## 2.5 Points of view

*A point of view is worth 80 points of IQ* —Alan Kay[Q39]

*Well, if you have two ways, one of them is real: what the machine executes. One of them is not. Only if one is massively more brief ... than the other is it worthwhile.* —Ken Thompson[Q83]

A good way of thinking about a system makes things easier, just as the center-of-mass coordinate system simplifies dynamics problems, or statistical mechanics summarizes the behavior of many particles in a few parameters such as temperature and pressure. It's not that one viewpoint is more correct than another, but that it's more convenient for some purpose. Many of the oppositions reflect this idea. Here are some examples of alternative points of view, discussed in more detail later:
- Being vs. becoming: the system state is the values of the variables (a map), or it's the sequence of actions that made it (a log).
- An interface adapter for compatibility is part of a component, adapting it to many environments, or part of the environment, making it hospitable to many components.
- Iterative vs. recursive, sequence vs. tree: you can do the same thing over and over, or you can divide a case into sub-cases and keep dividing until it's really simple.
- Declarative vs. imperative, solution vs. state machine vs. table, intensional vs. extensional: define a result by its properties or the equation it satisfies, or by the steps that achieve it, or by a table that lists the result for each input.
- Interpreter vs. compiler: the same program, but different primitive operations (x86 machine instructions, C statements, Java virtual machine instructions, Lisp functions or relational queries) get you different speed, size, or ease of change.
- Set $s$ (listing the elements) vs. predicate $p_s$ ($p_s(x)$ is true if $x \in s$). An important special case is relation $r$ (a set of ordered pairs) vs. predicate $p_r$ ($p_r(x, y) \equiv (x, y) \in r$) vs. function $f_r$ to the powerset ($f_r(x) = \{y \mid (x, y) \in r\}$ vs. directed graph $g$ ($r(x, y) \equiv (x, y) \in g.\text{edges}$).

### 2.5.1 *Notation*

*By relieving the brain of all unnecessary work, a good notation sets it free to concentrate on more advanced problems, and in effect increases the mental power of the race.* —A.N. Whitehead[Q94]

*The limits of my language are the limits of my world.* —Ludwig Wittgenstein[Q97]

*There are more useful systems developed in languages deemed awful than in languages praised for being beautiful—many more.* —Bjarne Stroustrup[Q77]

Notation is closely related to viewpoint, making something that's important easier to think about. Every system has at least some of its own notation: the datatypes and operations it defines, which are a domain-specific language (DSL) without its own syntax. A notation can also be general-purpose: a programming language like C or Python, or a library like the C++ standard template library. Or it can be for a domain: a DSL like the Unix shell (for sequential string processing) or Julia (for numerical computation), or a library like TensorFlow (for machine learning).

A notation consists of:
- *Vocabulary* for naming relevant objects and actions (`grep`, `awk`, `cat`, etc. for the shell). Generic terms make it easier for people: "sort" for different sorting methods, "tree" for partially ordered or recursive structures. In a spec, the foundation should be mathematics, most often relations.
- *Syntax* for stringing them together (in the shell, "|" for pipes, ">" for redirect, etc.). In a DSL, syntax is a way to make common things in the domain easy to write and read. By contrast, a library for a general-purpose language has to live with the syntax of the language, typically

method selection and function call, as in this example from Spark in Python: `df.select(multiply(col("x"), col("y")))`.

There are tools that make building a DSL easy, but a successful one will tend to evolve into being general-purpose, losing much of its simplicity and elegance.

In addition to being well matched to the domain, a DSL may be good because it's easy to optimize (as in SQL queries), or because it provides useful properties like type safety or predictable execution time even if it is compiled into a lower-level language like C or machine code where native programs lack these properties. Sometimes it works to embed a DSL into a general-purpose language, as with Linq (Language Integrated Query) in C#.

A language needs more machinery than a library, but there are advantages:
- It provides syntactic sugar to make the program shorter and easier to read.
- It can do more static checking; a library is limited to the host language's type checking of parameters and results.
- With a view of the whole program and knowledge of the primitives it can optimize better.

## 3. Goals and Techniques

### 3.1 Overview

The summary is STEADY by AID with ART: reach goals by using techniques with the process.

#### 3.1.1 *Goals—STEADY*

*[Data is not information,] Information is not knowledge, Knowledge is not wisdom, Wisdom is not truth, Truth is not beauty, Beauty is not love, Love is not music and Music is THE BEST —* Frank Zappa[Q98]

By goals I mean general properties that you want your system to have, not the problem it tries to solve. You probably want your system to be STEADY: **S**imple, **T**imely, **E**fficient, **A**daptable, **D**ependable, and **Y**ummy. Since you can't have all these good things at the same time, you need to decide which goals are most important to you; engineering is about trade-offs.

**Simple** should always be the leading goal, and abstraction is the best tool for making things simpler, but neither one is a panacea. There's no substitute for getting it right. Three other goals are much more important now than in the 1980s: Timely, Adaptable, and Yummy.

- **Timely** (to market) because cheap computer hardware means that both enterprises and consumers use computer systems in every aspect of daily life, and you can deploy a system as soon as the software is ready. You can order up hardware in the cloud in a few minutes, and your customer has a smartphone. If you can't deliver the system quickly, your competitor can.
- **Adaptable** because the Internet means that a system can go from a few dozen users to a few million in a few weeks. Also, user needs can change quickly, and for many applications it's much more important to be agile than to be correct.

- **Yummy**[Q69] because many systems are built to serve consumers, who are much less willing than organizations to work hard to learn a system, and much more interested in features, fashions and fads. Even for professionals, the web, social media and GitHub mean that it's easy for enthusiasm to build up in defiance of formal procurement processes.

| **Goals** | Simple | Timely | Efficient | Adaptable | Dependable | Yummy |
|---|---|---|---|---|---|---|
| *As questions* | Is it clean? | Is it ready? | Is it fast? | Can it evolve? | Does it work? | Will it sell? |
| *Alliterative* | Frugal | First | Fast | Flexible | Faithful | Fashionable |
| *As nouns* | Beauty | Time to market | Economy | Evolution | Fidelity | Elegance |
| *Antonyms* | Complex | Late | Wasteful | Rigid | Flaky | Crummy |

3.1.2 *Techniques—AI$^n$D*

Techniques are the ideas and tools that you use to build a system; knowing about them keeps you from reinventing the wheel. The most important ones are about abstraction and specs; those are principles, not just hints. Most of the rest fall under three major headings:

- **Approximate** rather than exact, perfect or optimal results are almost always good enough, and often much easier and cheaper to achieve. Loose rather than tight specs are more likely to be satisfied, especially when there are failures or changes. Lazy or speculative execution helps to match resources with needs.
- **Incremental** design has many aspects; often they begin with "i". The most important is to build the system out of *i*ndependent, *i*solated parts called modules with *i*nterfaces, that you can put together in different ways. Such parts are easier to get right, evolve and secure, and with *i*ndirection and virtualization you can reuse them in many different environments. *I*terating the design rather than deciding everything up front keeps you from getting too far out of touch with customers, and extensibility makes it easy for the system to evolve.
- **Divide and conquer** is the most important idea, especially in the form of abstractions with clean specs for imposing structure on your system. This is the only way to maintain control when the system gets too big for one person's head, now or later. Other aspects: making your system concurrent to exploit your hardware, redundant to handle failures, and recursive to re-use your work. The incremental techniques are other aspects of divide and conquer.

For each technique, many examples show how it's used and emphasize how widely applicable it is. A small number of ideas show up again and again, often concealed by the fact that people use different words for the same thing. The catalog below is both short and surprisingly complete.

I describe most of the techniques in the context of a goal, telling you for each one:
- What it is.
- Why it's good.
- How it can go wrong (when to avoid it; things to watch out for).

The examples can inspire you when you have a design problem; if you find one that's a good match for an important part of your problem, you can see what techniques it uses and how it uses them. A helpful example might be from a very different application domain than yours. For another source of inspiration, look at these links to important techniques:

> **Simple**: abstraction, action, extensible, interface, predictable, relation, spec.
> **Efficient**: algorithm, batch, cache, concurrent, lazy, local, shard, stream, summarize, translate.
> **Adaptable**: dynamic, index, indirect, scale, virtualize.
> **Dependable**: atomic, consensus, eventual, redundant, replicate, retry.
> **Incremental**: becoming, indirect, interface, recursive, tree.

## 3.2 Simple

*I'm sorry I wrote you such a long letter; I didn't have time to write a short one.* —Blaise Pascal[Q62]

*Everything should be made as simple as it can be, but not simpler.* —Albert Einstein[Q26]

*Simple things should be simple, complex things should be possible.* —Alan Kay[Q41]

*The onion principle: doing a simple task is simple, and if it's less simple, you peel one layer off the onion. The more layers you peel off, the more you cry.* (paraphrased) —Bjarne Stroustrup[Q79]

*Perfection is attained not when there is no longer anything to add, but when there is no longer anything to take away.* —Antoine de Saint-Exupéry[Q70]

*Beauty is our business.* —Edsger Dijkstra[Q23]

*Beauty is more important in computing than anywhere else in technology … because software is so complicated…. Beauty is the ultimate defense against complexity.* —David Gelernter[Q30]

| | | |
|---|---|---|
| Simple ↔ rich, general ↔ specialized [Y] | { | KISS: Keep It Simple, Stupid. Do one thing well. Don't generalize. Don't hide power. Leave it to the client. Make it fast. Use brute force. |
| Spec ↔ code [P] | { | Keep secrets. Free the implementer. Good fences make good neighbors. Embrace nondeterminism. Abstractions leak. |
| Perfect ↔ adequate, exact ↔ tolerant [TD] | | Good enough. Worse is better. Flaky, springy parts. |
| Immutable ↔ append-only ↔ mutable | | Make it stay put. |
| Imperative ↔ functional ↔ declarative [E] | | Make it atomic. Say what you *want*. |

The main thing is to keep the spec simple and to divide the system into modules with simple specs, points that I've already discussed. This section is about keeping the code simple.

### 3.2.1 *Do one thing well*

*Figure out how to solve one really tricky sticky problem and then leave the rest of the system straightforward and boring. I call this the "rocket science" pattern.* —Terry Crowley[Q15]

*There are some insurmountable opportunities around.* —Don Mitchell[Q53]

*Work expands so as to fill the time available for its completion.* —C. Northcote Parkinson[Q60]

Design your system around a small number of *key* modules with simple specs and predictably good performance. If you're lucky you can get these modules from your platform or from a library. If not, you have to build them yourself, but your goal should be the same. Finding this system design and building the key modules is hard work, but it's rewarded throughout the system's life because you can concentrate on the customers' needs; the rest of the code is easy to change, since it won't need any real cleverness. A successful key module will grow over time, improving performance with better algorithms and adding a few features, but building on a solid foundation. Make it fast rather than general or powerful, because then the client can program the function it wants. Slow, powerful operations force the client who doesn't want the power to pay more for the basic function. Usually it turns out that the powerful operation is not the right one. Well-known examples are CISC vs. RISC instruction sets and guaranteed vs. best-efforts packet delivery.

A wide range of examples illustrate this idea:

- The inode structure in a file system represents variable-length byte strings efficiently, even very large ones. Many variations fit in: variable-length extents (ranges of disk blocks) to keep the index small, sharing parts of the byte string for copy-on-write, logs for crash recovery.
- The Unix version 6 operating system separates file directories from inodes, and uses shell programs to connect applications through byte streams.
- The basic Internet protocols (TCP and UDP) provide reliable and best-efforts communication among billions of nodes.
- The BitBlt or RasterOp interface's simplicity and generality make it the standard for raster display applications.
- The original versions of HTTP and HTML provide basic facilities for making fairly pretty web pages and linking them. The current versions, of course, have overflowed these bounds.
- The eventually consistent hierarchical name space of DNS is the basis of Internet naming, for the web, email, and many other things. It maps a path name such as `csail.mit.edu` into a set of small "records", each with a type and (small) value, usually an IP address.
- Google's Chubby is a highly fault-tolerant, consistent lock service and low-volume store that's the root of many large distributed data structures. It scales to hundreds of thousands of clients.
- Relational databases structure very large amounts of data as tables with named columns (called attributes). Query (called selection) returns rows satisfying a predicate and composition (called join) combines tables *A* and *B* into a table that has all of *A*'s and *B*'s columns, and a row for each pair of rows from *A* and *B* that are equal on the columns that they share. A production database system is a poor example of doing one thing well, though, because of the many extensions to this simple interface.
- Domain Specific Languages (DSLs) such as Ruby or Matlab are generic examples of this idea. Domain Specific Architectures[R47] [§7] extend it to hardware such as GPUs, network controllers and encryption engines; the end of Moore's Law makes them more valuable.[R65]

Often a module that succeeds in doing one thing well becomes more elaborate and does several things. This is okay, as long as it continues to do its original job well. If you extend it too much, though, you'll end up with a mess. Only good judgment can protect you from this.

»Word Line Services. The Line Services component of Microsoft Word lays out lines of text in a way that looks good and respects the typographical conventions of hundreds of languages, including different fonts and sizes, ligatures, right-to-left writing, computed fields, mathematical typesetting, and many other complications. It also has to be fast enough to lay out a whole page in a fraction of a second. It was retrofitted into Word by Eliyezer Kohen and his team, an amazing achievement.[R91]

»DEC Alpha vs. Pentium Pro. The experience of Digital Equipment Corporation's Alpha microprocessors shows that keeping things simple doesn't always work. To have any chance of success, an incompatible CPU had to have at least twice the performance of the competing Intel x86 chip. The Alpha designers were confident that they could achieve this, since the Alpha had a much simpler architecture, designed from scratch for high performance and to meet modern needs rather than incrementally over 20 years. But they failed, even though both the design and implementation were good. Brilliant engineering and a very large design effort gave Intel's Pentium Pro much better performance than anyone had imagined was possible.[R22]

»DEC Cluster Interconnect (CI). Special purpose hardware sometimes succeeds and takes hold—GPUs and encryption engines are examples—but usually it fails because it doesn't get updated as often as general-purpose CPUs. This happened in the '80s with the network interface to DEC's high-speed local area network, the Cluster Interconnect. It off-loaded the work of sending and receiving packets and reliable messages from the VAX 780 CPU, delivering several times as much bandwidth. But it was roughly as complex as the 780 CPU, and when faster CPUs came along the CI NIC became a bottleneck rather than an accelerator. Here is a 2019 take on this.[R52] GPUs have avoided this fate because there's a large gaming market that keeps being willing to pay for more performance.

»Windows Vista. The successor to Microsoft's very successful Windows XP system was Windows Vista, six years in development. Vista was planned to have four major innovations:
- A new file system called WinFS with many features from databases, aspiring to give applications uniform access to structured and semi-structured data as well as to old-fashioned byte stream files.
- A new graphics system called Avalon (later released as Windows Presentation Foundation), an incompatible replacement for the entire Windows graphics stack, up to and including the browser's rendering system.
- A new networking system called Indigo (later released as Windows Communication Foundation) designed to support distributed computing where services have remote clients.
- Extensive use of "managed code" written in Microsoft's C# language for much better security, and using garbage collection for memory management.

All four failed and had to be removed from Vista.[R28] The first three were "universal" goals: to create powerful storage, display canvas and distributed computing infrastructure good enough to underlie all applications. They failed because they did not deliver enough value, added a lot of generality and complexity, were incompatible with the large base of existing applications, and were unpredictably slow. Lesson: beware of universal goals.

Managed code failed in Vista for a more interesting reason, after quite a lot of code had been written. C# was designed as a replacement for both C++ and Visual Basic; since the latter had a much larger user base, the needs of those users had priority as C# was developed. They write applications that are often complex but relatively light-duty, and they don't worry much about exception handling, resource exhaustion or multiple versions, all very important to Windows systems programmers. By the time this mismatch became so obvious that even the responsible managers could no longer ignore it, there was not enough time to fix the C# implementation, and the managed code in Vista had to be rewritten in C++.

### 3.2.2 *Brute force*

*Entities should not be multiplied beyond necessity.* —William of Occam[Q58]

Computers are fast, and specialized hardware is even faster—take advantage of this. Exhaustive search (perhaps only up to some "depth") is a simple brute force technique. Its cost is $O(n)$, and often $n$ is not too big, so always consider it first. Examples: `grep` over a file, model checking, program synthesis (if there's a short program that solves the problem), many optimization problems, and a host of attacks on security measures such as password guessing, defeating address space layout randomization (ASLR), and searching for encryption keys. It's also the only way to query a database if you don't have an index. It works best when you have locality.

*Broadcast* is another example of brute force. It is to routing as exhaustive search is to indexing, and likewise scales badly. In networking you often need a broadcast to get started, for example when an ethernet node needs to find an Internet router, but this doesn't need to scale. A third example is *polling* for pending work, instead of notification; it's simple, and it's efficient if you can tolerate enough latency (that is, poll infrequently enough) that you usually find work to do.

### 3.2.3 *Reduction*

*Less is more.* —Robert Browning, *Andrea del Sarto*

Reduction is fundamental in theoretical computer science: solving a problem using one that's already solved. It's also used in systems, but in two different ways. The good way is reduction to a much simpler problem that is much easier to solve correctly and efficiently. For example:

- Redo logging reduces making an arbitrary update atomic in spite of crashes to atomically appending update records to a log, and then to just atomically writing a commit record after logging all the updates. To do this when writes may be done out of order or if the log is replicated, add a hash of all the updates to the commit record, and treat the record as a commit only if the log matches the hash.
- Similarly, making an arbitrary update atomic in spite of concurrency reduces to holding locks on data that the update touches, which in turn reduces to just acquiring a lock with an atomic test-and-set or compare-and-swap instruction.
- Encryption reduces keeping stored or transmitted data secret to just keeping the keys secret.
- Certificates digitally signed by a trusted authority reduce learning another party's key to learning just the authority's key.

The second kind of reduction is more dangerous: reducing a problem to an already solved problem that is not much simpler, and may indeed be more complex. For example, to make a speech recognizer for the words "yes", "no" and "cancel", call an existing general recognizer for English words. The resulting code is larger and slower than a custom recognizer for those three words, but much easier to write. Many powerful modules already exist, and using a fraction of their power to solve your problem is often good engineering. But it can be wasteful of computing resources; this is not always bad, but it's easy to lose track of how much is being wasted.

## 3.3 Timely

Precise vs. approximate software         [D]  Get it right. Make it cool. Shipping is a feature.
Perfect ↔ adequate, exact ↔ tolerant [SD] Good enough. Worse is better. Flaky, springy parts.

Building a timely system (one that ships soon enough to meet your time-to-market needs) means making painful choices to give up features and dependability. If it's extensible you can add features later; adding dependability is harder. It's easier to make approximate software timely.

»The web. Perhaps the biggest reason the web is successful is that it doesn't have to work. The model is that the user will try again, switch to an alternative service, or come back tomorrow. It's quite rare to find a web service that is precise. For example, there's no spec for a search engine, since you can't write code for "deliver links to the 10 web pages that best match the customer's intent", and indeed engines are ruthless about ignoring parts of the Internet in order to deliver results faster.

» Uncoordinated software. A more surprising example comes from a major retail web site, where the software is developed as hundreds of modules, each one hosted on the retailer's internal servers. Each module is developed by a small team that has complete control over the specs and code: they define the interface, write the code, deploy it on the servers, fix bugs and make changes as they see fit. Any module can call any other module over the internal network. There is no integration testing or release control. Not surprisingly, it's common that a module fails to deliver expected or timely results; this means that its caller must be programmed defensively. Retail customers may notice that some of the web pages they see are incomplete or wrong, but this doesn't matter much as long as it doesn't happen too often. It's more important to be able to change the software and add features quickly—the only page that really must be correct is the one with the "Place Your Order" button. Of course, credit card processing uses precise software.

»Web design. In 2010 I went to an O'Reilly "un-conference" where one of the events was a workshop chaired by Tim Berners-Lee on what could have been done differently in the initial design of the web. About a dozen ideas were proposed, but I thought that in every case Tim was right to reject the idea. Why? Each idea would have made it more difficult to deploy the web widely, without enough compensating benefit. The most controversial and most important decision was to allow broken links. The only idea I know that doesn't make deployment harder is that every page and every link should include a large random number. If search engines indexed this number, broken links would be a thing of the past. But web-wide search engines didn't seem practical in 1990.[R80]

## 3.4 Efficient—ABCs. Use theory. Latency vs. bandwidth. $S^3$: shard, stream or struggle.

*An efficient program is an exercise in logical brinksmanship.* (paraphrased) —Edsger Dijkstra[Q24]

*The cheapest, fastest, and most reliable components of a computer system are those that aren't there.* —Gordon Bell[Q6]

*An engineer can do for a dime what any fool can do for a dollar.* —Anonymous

| | | |
|---|---|---|
| Dynamic ↔ static | [A] | Stay loose. Pin it down. Shed load. Split resources. |
| Indirect ↔ inline | [I] | Take a detour, see the world. |
| Time ↔ space | | Cache answers. Keep data small and close. |
| Lazy ↔ eager ↔ speculative | | Put it off. Take a flyer. |
| Centralized ↔ distributed, share ↔ copy | [D] | Do it again. Make copies. Reach consensus. |
| Imperative ↔ functional ↔ declarative | [S] | Make it atomic. Say what you *want*. |

Efficiency is about doing things fast and cheaply. Most of what I have to say about it is in the ABCs below: **A**lgorithms, **A**pproximate, **B**atch, **C**ache, **C**oncurrent, **C**ommute, **S**hard/**S**tream. Bentley's book says more about these ideas and gives many others.[R8] But first some generalities.

### 3.4.1 *Before the ABCs*

*The greatest performance improvement of all is when a system goes from not-working to working.* —John Ousterhout[Q59]

*Two fundamental rules for program optimization: Rule 1: Don't do it. Rule 2: (for experts) Don't do it yet.* —Michael Jackson[Q34]

It used to be that machines were small and slow, and it was a struggle to get your problem to fit. Today machines are big and fast, and for many problems efficiency is not an issue; it's much more important to be timely, dependable and yummy. And well-optimized libraries for numerical computing, machine learning, databases, etc. often take care of efficiency. But there are still plenty of big problems: genomics, molecular dynamics, web search, social media graphs. And there are devices with limited energy that can't afford to execute too many instructions, and new real-time problems where the computer needs to keep up with the human or with the physical world, responding in just a few milliseconds.

It's tricky to write an efficient program, so don't do it unless you really need the performance. If a shell script is fast enough to solve your problem, by all means use a shell script.[R9] If you do optimize, remember the rule: first design, then code and debug, then measure, finally (if ever) optimize. In other words, make the code correct first and then make it fast. It's a good idea to keep the unoptimized code around as an oracle to test the optimized code against.

The resources you are trying to use efficiently are computing, storage, and communication. The dimensions are time and space: how long something takes (or how long a resource is tied up), and how many resources. For time the parameters are *bandwidth* (or throughput) and *latency* (or response time). Latency is the time to do the work (including communication) plus the time spent waiting for resources because of contention (queuing).

A system design needs to consider efficiency as well as simplicity and functionality, even though it shouldn't involve detailed optimization. To evaluate a design idea, start by working out roughly how much latency, bandwidth and storage it consumes to deliver the performance you need. Then ask whether with optimistic assumptions (including plausible optimizations), you can afford that much. If not, that idea is no good; if so, go on to a more detailed analysis of the possible bottlenecks, and of how sensitive the cost is to the parameters of the platform and workload.

If you can divide the work into independent parts, you can use concurrency to trade more resources (more bandwidth) for less latency. With enough parts the only limit to this is the budget, as cloud services for search, email, etc. demonstrate. Likewise, a cache lets you trade locality and bandwidth for latency: if you use a fraction $f$ of the data, you need $1/f$ times as much bandwidth to get everything into the cache. The best case is a stream of data when you don't need real time response; then you only need bandwidth (plus enough buffering to cover variations in latency).

When performance of a module or application is bad or unpredictable, you have incurred *performance debt*, a special case of technical debt. This takes several forms:
- It's unknown—you haven't measured it realistically.
- It's bad—worse than your intuition says it should be, or than what you need.
- It's fragile—it's okay now, but you don't have any process to keep it that way.

If performance is important to your clients, you need to fix all of these things.[R25]

Here is a summary of how to build a high-performance system today:[R69]

1. **Exploit modern hardware.** Wait-free data structures and delta updates are friendly to deep cache hierarchies and fast flash storage.
2. **Make critical paths short**, especially for hotspot data where Amdahl's Law applies.
3. **Minimize conflicts** with multi-version concurrency.
4. **Minimize data movement.** Data transfers are very costly; put data in its final resting place immediately, and keep it small.
5. **Exploit** batching to reduce the per item cost.

**Fast path and bottlenecks**

There are two basic ways to reduce latency: concurrency and *fast path*—do the common case fast, leaving the rare cases to be slow. For caching, the fast path is a cache hit. *Amdahl's Law* governs the performance of fast path: if the slow path has probability $p \ll 1$, the fast path takes time $f$, and the slow path takes time $s \gg f$, then the average time is $f + ps$. The *slowdown* from the slow path is $(f + ps)/f = 1 + p(s/f)$. Thus a RAM cache with $p = 1\%$ (99% hits) and $s/f = 100$ (1 ns to cache, 100 ns to RAM) is 2 × slower than a hit every time.

Amdahl invented his law to describe the limit on speedup from concurrency. Here the slow path is the part that must be done serially. The *speedup* from the concurrent fast path is $s/(f + ps) = 1/(f/s + p)$. With $n$-way concurrency $f/s = 1/n$, and for large $n$ this goes to 0 and the speedup is just $1/p$. If $p = 1\%$ (only 1% is serial), the maximum speedup is 100 ×, no matter how much concurrency there is. Whether you think of the result as a speedup or slowdown depends on your expectations. You can evade the law with an algorithm whose serial fraction shrinks as fast as the problem size grows.

Sometimes the fast path is hard to spot because its code is tangled up with various rare cases. When this happens you have to find the fast path code (by hand or by profiling) and then restructure the code to make it obvious, hence easier to improve.

You want the fast path to be the normal case, and often it's best to handle the worst case separately, because the requirements for the two are quite different:
- The normal case must be fast.
- The worst case must make some progress.
- Sometimes radically different strategies are appropriate in the normal and worst cases.

For example, Microsoft Word uses a piece table to keep track of edits. If there are lots of edits, the table gets big and editing slows down. Writing out a new version flattens it.

A variation on fast path is the distinction between *data plane* and *control plane*, common in networking but relevant in many other places. The data plane is the fast path, the part of the system through which lots of data flows; it's important to keep the latency and the cost per byte low. The control plane configures the data plane; usually neither bandwidth nor latency is critical. Outside of networking this is often called system management.

»Finding the fast path. The RPC that Mike Burrows, Mike Schroeder, and Sheng Yang Chiu implemented for the Firefly multiprocessor at DEC SRC in 1987 was unexpectedly slow. It took a line-by-line analysis to extract the fast path from the complicated setup, marshalling, error handling, congestion control and scheduling code and then to measure the cost of each step and tune it. The builders were very reluctant to do this, fearing that it would be too hard to maintain the resulting code, but I insisted and the code remained manageable.[R93] A 2018 example is the tradeoff between storage and CPU costs for database systems; if the code is properly structured, a fairly simple analysis reveals how the cost depends on a few properties of the workload and the underlying hardware.[R70]

Almost the opposite of a fast path is a *bottleneck*, the part of the system that consumes the most time (or other resources). Look for the bottleneck first. Usually you don't need to look any farther; it dominates the performance, and optimizing anything else wastes your time and adds complexity. Once you've found it, find a fast path that alleviates it. In other words, design your code to use it as little as possible, and measure and control how it's used. The most fundamental bottleneck is the speed of light, but that's usually not what's limiting.

**Predictable performance**

*That, Sir, is the good of counting. It brings everything to a certainty, which before floated in the mind indefinitely.* —Samuel Johnson[Q37]

*What you measure is what you'll get. Period.* —Dan Ariely[Q3]

Your guess about where the time is going is probably wrong. Measure before you optimize. If you depend on something unpredictable, measure it in the running system and either adapt to it, or at least report unexpected values so that developers or operations staff can tell what's going on.

It's often not enough for a spec to describe only the state that the program can name. Resources must be part of the state, including real time, and an action must say roughly (perhaps within a factor of two) what resources it consumes, and especially how long it takes. Ideally this won't depend on the environment or on parameters of the action, but often it does and you need to know how in order to use the action effectively. A module can control many aspects of its performance: internal data structures and algorithms, optimization, code quality, compression, etc. But the environment controls other aspects: latency and bandwidth to different levels of storage, between address spaces and between machines. This can change as the clients' demands or the underlying platform change, and a robust application must either adapt or report that it can't.[R29] Sometimes there are different ranges in which the dependence is different because the code runs into different bottlenecks (compute, storage, network). This can change with time as well. For example, encryption used to be considered slow, but today you can do an AES encryption of a cache line in 50 cycles while a miss takes 200.

Don't try to be precise; that's too hard. It's enough to know how to avoid disaster, as in paging, where you just need to keep the working set small enough.

A refinement of the worst or average case is tail latency, knowing not just the mean or median cost but also the maximum cost for 99% of the tasks, or 99.9%.[R31] To control tail latency you need quotas or admission control, or the freedom to shed load, the way the Internet drops packets.

»Shedding load. Bob Morris, with whom I shared an office in Berkeley in 1967 when he was on sabbatical from Bell Labs, used to say that a time-sharing system terminal needs a big red button that you push if you are dissatisfied with the service. The system responds by either improving the service or throwing you off. The idea is that you'll only push it when you think you'd be better off doing something else. An overloaded system should make this decision itself: either decent service or no service.

It's hard to deliver predictable latency when your host is unpredictable. With an interface that does a lot of optimization such as SQL queries, client code that doesn't get optimized often looks a lot like code that does. A remote procedure call (RPC) is especially dangerous, because it looks just like a local call. Other examples are the Internet, paging, and fancy graphics. These all have artificial similarity that isn't real: a common interface but surprisingly different performance. Network access in general is very unpredictable (except in a data center, where the environment is tightly managed) and you can't control it very well (except perhaps in a datacenter), so it's best to work only on local data (which might be stale) when responding to a user input, unless it's very obvious to the user that the network is involved, for example in a web search. This means that the UI should communicate asynchronously with anything that might be slow.

If there are actions with highly variable latency, such as reading something over a network, you have to plunge ahead rather than waiting for a response. Usually the response affects the state, so you need a way to postpone that effect. This requires being very clear about the semantics of unordered sets of operations, a form of eventual consistency. Web pages do this all the time, and word processors often do it for slow operations like hyphenation. You must also pay attention to scheduling and contention, issues that sequential programs handle automatically.

**Locality**—Keep data small and close

Because communication is expensive and memory hierarchies are deep, keep the data close to the computation. The L1 cache is the closest it can get, but you just need the data close enough that moving it to the computation doesn't slow things down too much. The two main strategies are:

- Keep the parts that run concurrently as independent as possible, to minimize communication
- Make the data smaller, so that more of it is local and there's *less* to communicate. Try to get by with a summary of the full dataset, or rearrange the data so that what's accessed a lot is small together and compressible, as a column-store database does.

Often it helps to process data in a stream: it uses the cache efficiently, especially with prefetching, and auxiliary data such as a summary or a mapping table will stay in the cache. To interact with a large object (such as a big document or a complex drawing), store it so that all the data needed for computing the display is together. Don't let the display depend on the whole object, or on everything up to the current point. For example, page breaks are your friend.

In a big fault-tolerant system individual nodes are constantly failing and recovering, and coming and going in response to the changing load, so that keeping data local means moving it frequently. This is why cloud systems separate persistent data storage from computation, keeping data in objects or blobs on specialized storage nodes; they make huge investments in network bandwidth and replicated storage. To make up for the fact that objects are slower and more expensive to access than local storage, clients build volatile caches in RAM or SSD, which is much faster and cheaper, and use log-based redo to handle failures; this is complicated, though.[R30]

Instead of moving the data to the computation, you can do the reverse if the data store can run code that a client provides. The most common examples are queries (including complex ones on data cubes) and stored procedures in relational databases, and map and reduce functions in a map-reduce system.

**Contention**

If there aren't enough resources to process the instantaneous load there will be *contention*, which shows up as *queuing* for access to a resource and increases the latency. It's hard to understand queuing in general, but the simplest case is important and easy: if a resource is busy (utilized) for $u$ seconds per second on average and tasks arrive randomly, then a task that uses it for a second will take $1 / (1 - u)$ seconds. For example, at $u = 90\%$ it takes 10 seconds—ouch! *Batch sampling* avoids this for a job with $m$ independent tasks competing for a pool of resources, when all the tasks need to complete so that you want to speed up the slowest one. It queues the whole job at $d > 1$ resources per task ($dm$ resources in total); $d = 2$ is usually good. A resource asks the job for a task when it becomes free. This yields near-optimal performance, and there is no centralized scheduling bottleneck.[R82]

The other simple fact about a single queue is Little's Law, $L = \lambda W$: $L$ is the number of requests being processed, $\lambda$ the throughput or bandwidth (the rate at which requests arrive and depart), and $W$ the latency or response time for a request; all three are averages. A request passes through one or more resources in the system, and if $W$ is not close to the sum of their latencies, at least one of them is congested. In networking, $L$ is the bandwidth-delay product, the amount of data in flight that hasn't been acknowledged but is still on the wire or buffered, either in a router or in the receiving node. If this is not close to twice the link delay times the link bandwidth, either a router or the receiver is congested.

One way to avoid contention is to break a resource into lots of pieces and choose one to use at random; if there are many more pieces than clients, contention is unlikely. VLB is a classic example.

**Translation**

Compiling often makes a system faster, moving from code that's better for people (easier to write, understand, and change) to code that's better for the machine (closer to machine instructions, friendlier to the cache, easier to optimize) or more widely implemented (compiling to JavaScript or C; anything can be a target). For a dynamic source language such as JavaScript, just-in-time (JIT) translation and trace scheduling usually yield good code for the common cases, falling back to an interpreter or to recompiling when new cases come up.

Anything can be a source as well. Translating from one machine language to another is common for backward compatibility. This can be *very* dynamic; Intel CPUs translate from clunky x86 instructions to micro-ops (internal RISC instructions) on the fly as they fetch the instructions.[R36]

Often you can "translate" to something simpler, the way a PostScript document is translated to PDF by simply running the PostScript program after redefining each imaging action to record itself, or a large piece of graphics geometry is stripped down to just the lines and triangles that are visible now on the physical screen.

### 3.4.2 *Algorithms*

*[In many areas] performance gains due to improvements in algorithms have vastly exceeded even the dramatic performance gains due to increased processor speed.* —PCAST[Q64]

*Fancy algorithms are slow when N is small, and N is usually small.* —Rob Pike[Q65]

*Once you succeed in writing the programs for these complicated algorithms, they usually run extremely fast. The computer doesn't need to understand the algorithm; its task is only to run the programs.* —Robert Tarjan[Q81]

*When in doubt, use brute force.* —Ken Thompson[Q84]

There's been a lot of work both on devising algorithms for important problems and on analyzing their performance. Typically the analysis bounds the running time $t(n)$ asymptotically as the problem size $n$ grows: $t(n) = O(n)$ means that there's a constant $k$ such that $t(n) \leq kn$ as $n \to \infty$, $t(n) = O(n \log n)$ means that $t(n) \leq kn \log n$, and so forth; $O(2^n)$ is *exponential*. Anything worse than $O(n \log n)$ is bad unless $n$ is sure to be small, but this is not the whole story.

- There can be a large fixed overhead (which is bad when $n$ is small), and $k$ can also be large.
- You might care about the average rather than the worst case.
- Your problem might be much easier than the hardest problem, or even than a randomly chosen problem.[R89] For instance, the simplex method for linear programming is exponential in the worst case, but it's always fast in practice.

It's usually best to stick to simple algorithms: a hash table for looking up a key, a B-tree for finding all the keys in a range, a DHT for strong fault tolerance. Books on algorithms[R24] tell you a lot more than you need to know. If you have to solve a harder problem from a well-studied domain such as numerical analysis or graph theory, look for a widely-used library.

If $n$ is really large (say the Facebook friends graph), look for a randomized *sublinear* algorithm with time $< O(n)$; for example, the median of a large set of size $n$ is very close to that of a random subset of size $\log n$. Randomization can keep other algorithms both simple and efficient, for example testing whether a number is prime, or reaching consensus with asynchronous communication. It helps with network scheduling too, as in Valiant Load Balancing, where routing each packet through a randomly chosen intermediate node doubles the bandwidth, but avoids contention at hotspots with high probability.[R108]

### 3.4.3 Approximate—Flaky, springy parts

*It is better to have an approximate answer to the right question than an exact answer to the wrong one.* —John Tukey[Q86]

*There is nothing so useless as doing efficiently what should not be done at all.* —Peter Drucker[Q25]

Very often you don't need an exact answer; a good enough approximation is fine. This might be "within 5% of the true answer" or "the chance of a wrong answer is less than 1%." If the "chance" in the latter is truly random and the runs are independent, doing it twice makes it .01%. Sometimes the answer is just a guess (a hint), which you need to validate by watching the running system.

You can approximate the *analysis* rather than the solution; this is "back of the envelope" analysis, and usually it's all you need. How to do it: find the few bottleneck operations that account for most of the cost, estimate the cost and the number of times you do each one, multiply and add. For example, for a program that does $10^{10}$ memory operations, has a cache hit rate of 95%, and runs on a machine with RAM access time of 100 ns, if memory access is the bottleneck it will take about $10^{10} \times .05 \times 100/10^9 = 50$ sec. To `grep` a 1 GB string from a remote file server on a machine with a 10 Gb/s ethernet connection will take at least $8 \times 10^9/10^{10} = .8$ sec. In each case there are lots of other things going on, but they shouldn't matter to the performance.[R70]

Another way to approximate a system's behavior is by its *time constants*, the time it takes for most of the effects of a change to propagate through the system. Examples:

- the round-trip time in a network,

- the mean time to repair a component,
- the temporal locality of data usage and the churn that occurs when there's too little locality,
- the hysteresis by which the effects of a change stick around for a while after it is reversed.

It often pays to *compress* data so that it's cheaper to store or transmit. LZW is a widely used general-purpose scheme that typically reduces the number of bits by $2 - 5 \times$. It's probably fast enough for your purposes, since it only takes about four machine instructions per byte. LZW is not an approximation, since it is lossless, returning the original data.

The most powerful compression produces a *summary* that is much smaller than the input. Unlike lossless compression, this cannot recover the original data.

- A *sketch* keeps the most important things about the input. Examples: a low resolution version of an image, a vector of hashes that maps similar documents to nearby points,[R18] a Bloom filter.
- A *Bloom filter* is a bit vector that summarizes a set for testing membership. If a value is in the set the filter will say so; if it's not, the filter will wrongly say that it is with some false positive probability $f$. With 10 filter bits per set element $f < .01$, with 20 filter bits $f < 10^{-4}$.[R74]
- *Sampling* a data set summarizes it with a much smaller set whose properties are good approximations to properties of the original. Often $\log n$ samples from a set of size $n$ are enough.
- *Principal component analysis* (PCA; it has many other names) takes a set of $n$ observations of $p$ properties and finds the linear combinations of the properties that account for as much of the variation as possible; that is, it summarizes all the properties by their most important features. Often hundreds or thousands of properties boil down to two or three.
- *Fitting* a line to a set of data points summarizes the data by a linear model, minimizing the mean squared error. Of course it generalizes to quadratic, exponential and other models and to other definitions of error.
- Similarly, *clustering* summarizes the data by a small number of points and radii.
- A *classifier* tells you some property of the input, for example, whether it's a picture of a kitten.
- A *Merkle tree* lets you prove that an item $i$ is in a set $s$ in $O(\log n)$ time and space. Make the set elements the leaves of a balanced tree, and summarize each subtree by a hash of its children. The root hash defines $s$; verifying $i \in s$ needs the hashes of the siblings between $i$ and the root.
- *Abstract interpretation* summarizes the dynamic behavior of a program by making it static, replacing each variable with one whose value only depends on where you are in the program, not on how you got there. It needs abstract versions of each data type's primitive operations, and a merge operation that combines the abstract values on two execution paths that join together. Other kinds of data flow analysis make the program static in similar ways.
- An application that shows views of a large data set needs to boil the data down to something that fits on the screen. If it's interactive, it needs a local "projection" of the data that is not too big, but lets the user select and navigate. This is usually a fundamental part of the design. A simple example is a word processor that shows the current page and a table of headings.

**Approximate behavior**

Another kind of approximation works on a program's *behavior* rather than its data.

- A hint is a value that might be what you want, but you need to check that it's valid; see below.
- Many numerical computations iterate until some *tolerance* is reached. Examples: relaxing a field (an array of values on a grid), truncating a series, searching for a minimum, training a neural network. The danger is that there will be many more iterations than you expected.
- *Multigrid* systems vary the tolerance, and perhaps the algorithm, at different grid resolutions. More generally, you can use different algorithms at different scales in a recursion. Quicksort, the best in-memory sorting algorithm, recursively divides the region it is working on in half, but once the region gets small enough it's faster to switch to bubble sort.
- Cryptographic protocols have a *security parameter* $\lambda$, which is the probability that an adversary can break the security by exhaustive search. A popular value is $\lambda = 2^{128}$: if a trial takes 1 ns and you run on a million machines, then a break takes $5 \times 10^{15}$ years.
- In *exponential backoff* an autonomous agent responds to an overload signal by decreasing its offered load (rate) by some factor. Examples: ethernet, Internet TCP, Wi-Fi, spin locks that wait before retrying. The right way to *increase* the offered load depends on details; for TCP it's additive (with some complications), so the simple rule is additive increase, multiplicative decrease (AIMD).[R20] In the Internet the overload signal is a packet that isn't acknowledged, but in a data center network[R52] it's more likely to be
  - an explicit congestion notification (ECN)—a congested switch turns on a bit in a packet and the receiver sends it back in an acknowledgment, or
  - an increase in the round-trip time, which data center NICs can measure to the microsecond.
- A randomized algorithm gives an answer with probability $p < 1$ of being wrong. If you don't have a check, there's a single right answer, and $p$ isn't small enough, repeat $n$ times and the chance of being wrong is $p^n$, as small as you like.
- Networks usually don't guarantee to deliver a packet, but simply provide "*best-efforts*" and may discard a packet if delivering it gets too hard. End-to-end protocols provide stronger guarantees at the price of bad worst-case behavior.
- Eventual consistency lets applications operate on stale data.
- Other uses of *stale data* are common. For example, the user interface of a browser often operates only on local data, to avoid getting hung up if the network is flaky. Word processors usually do hyphenation, figure placement and proofing in background, only updating the display every now and then.
- Most computing systems don't try to optimize their use of the underlying resources, because optimization is too hard when the load is unpredictable and bursty, as it usually is. Instead they simply try to *avoid disaster*: deadlock or total resource exhaustion.
- Natural user interfaces accept input from people in the form of speech, gestures, natural language etc. Machines have gotten pretty good at recognizing this kind of input, but they're certainly not perfect. To some extent the machine is *guessing* what the user wants, and it's important to make it

    clear what the machine's guess was,

    easy to undo any undesired effects, and

    possible for the user to do something besides just repeating the failed input.

- More dramatically, an "interim" system approximates a desired "final" system (that may never be finished). Examples: Network Address Translation instead of IPv6, SSL instead of IPsec.
- Agile software development approximates the system spec to get something running quickly for both developers and users to try out. Their reactions guide the evolution of the spec.

**Hints**

A hint (in the technical sense) is information that bypasses an expensive computation if it's correct; it's cheap to *check* that it's correct, and there's a (perhaps more expensive) *backup* path that will work if it's wrong. *Soft state* is another name for a hint, somewhat more general because it's okay, if inefficient, to act on the soft state even if it's wrong, so there's no need for a check. This is common in network routing, where end-to-end fault tolerance and rerouting make the unreliability acceptable, and eventually the soft state times out.

There are many examples of hints throughout the paper, but here are some general patterns:

- An approximate index points to an item in a large data set that contains a search term, or more generally that satisfies a query. To check the hint, follow the pointer and check that the item does satisfy the query. The backup is consulting a more expensive index, or an exhaustive search. Unfortunately, the check doesn't work if there might be more than one item and you want all of them. Web and desktop search work this way, without reliable notifications of changes, so they are very loosely coupled to the data and hence don't depend on its code.
- A *predictor* uses past history to guess something. A CPU predicts whether a conditional branch will be taken; the check is to wait for the condition, the backup is to undo any state changes that have happened since the wrong prediction.[R36] A method predictor for a class guesses that the method's code is the same as last time. More generally, JIT and trace scheduling predict that some property of a variable won't change.
- Routing hints tell you how to forward a packet or message. These are usually soft state. The backup is rerouting.

**Strengthening**

If you need some property $p$ to be true but it's hard to guarantee $p$ exactly, look for a stronger property $q$ (that is, $q \Rightarrow p$) that's easier to guarantee. For example:

- If a bit table marks free pages, it's enough to have "bit $i = 1 \Rightarrow$ page $i$ is free" instead of equality; freeing the page before setting the bit guarantees this. If a crash intervenes, a few free pages may get lost because they don't have the bit set.

- If actions $a$ and $b$ don't commute, their locks must conflict, but it's okay for the locks to conflict even if the actions do commute.
- If a thread wants to wait until some predicate $p$ has become true, and the code signals condition variable $c$ whenever it makes $p$ true, then it's sufficient to wait on $c$. There's no guarantee that $p$ is still true when the thread runs, so the waiter has to check this.
- In redo recovery, the log has the property that repeatedly redoing prefixes of it (which happens if there are crashes during recovery), followed by redoing the whole log, is equivalent to redoing the whole log once.
- In security, if principal $A$ speaks for principal $B$, that means you trust $A$ at least as much as $B$.

Related ideas are to strengthen a loop invariant to get a strong enough precondition for the loop body, or to weaken a spec in order to make the code's job easier. Thus the Internet only makes its best effort to deliver a packet rather than guaranteeing delivery; the receiver must acknowledge and the sender must retry to get a guarantee. This is an example of the end-to-end principle. On the other hand, strengthening a spec (reducing the number of different actions while preserving liveness) promises the client more and may make the code's job harder; of course the code itself is a strengthening of the spec. Note that adding operations the client can invoke is not strengthening, but extending the spec (for objects, this is subclassing), and you have to be careful that the new actions don't violate properties the client is depending on (break backward compatibility).

**Relaxation**

The idea of relaxation is to take a lot of steps concurrently that get the system closer to where you wanted to be, with very little centralized scheduling. Disruptions such as node failures may make things worse temporarily, but the relaxation will still make progress. It may end in a finite number of steps, or it may just get closer and closer to an exact solution, in which case you need a benefit (or loss) function and a tolerance to match against the cost of continuing. It's important to understand the set of initial states from which relaxation will succeed (ideally every state). Examples:

- Taking averages of the values at neighboring points solves the Laplace equation numerically on a grid.
- Eventual consistency relaxes the system toward a state in which all the updates are known everywhere. Gossip protocols are one way to code this.
- A distributed hash table (DHT) relaxes toward a state in which each entry is replicated the right number of times and the load is evenly balanced, even when nodes arrive or leave.[R100]
- A switched ethernet relaxes toward a state in which each switch knows how to forward to every active MAC address on the network. The Internet is similar.
- A self-stabilizing system relaxes from an arbitrary state to a state that satisfies some invariant, usually by repeatedly reducing some bounded measure of disorder until it reaches 0.

### 3.4.4 *Batch*—Take big gulps

Whenever the overhead for processing $b$ items is much less than $b$ times the overhead for a single item, batching items together will make things faster. If the batch cost is $s$, the cost per batched item is $f$ and the batch size is $b$, the total cost is $s + fb$ and the cost per item is $f + s/b$. This is just the fast path formula $f + ps$, with $p = 1/b$; bigger batches are like a smaller chance of taking the slow path. Batching increases bandwidth, at the cost of increased latency for the earlier elements in the batch. For example, when you pack a number of bytes into a packet, the first one arrives later by the time it takes to accumulate the rest of the bytes. Of course, if at least a packet's worth of bytes arrive at once the added latency will be 0.

Here are some examples of batching:

- Buffering many items in a stream (characters, lines, records, etc.) in memory. Usually it's much cheaper to get or put an item from the buffer than from the stream.
- A cache with a line size bigger than the size of the data requested by a load instruction. This works well when there is enough locality that later instructions will consume much of the rest of the line. Otherwise it wastes memory bandwidth and silicon in the cache.
- Blockwise linear algebra, a special case of optimizing the cache that distorts the order of processing the elements of a matrix so that all the data that gets fetched into the cache is used before it gets evicted.
- Minibatches for deep learning; each minibatch trains a set of weights that fits in the cache.
- Piggybacking acknowledgments or network metrics on packets going the other way.
- Group commit, packing the commit records for many transactions into one log record.
- Indexing, which pays a big cost upfront to build the index so that later queries will be fast.
- Coarse-granularity locks, which protect a whole subtree with a single lock. Usually there's a way to substitute finer-granularity locks if necessary to avoid contention.
- Mining rather than probing, collecting lots of data and using it to answer lots of queries, rather than just getting the data needed to answer one query. Internet search engines are the most dramatic example.
- Merkle trees, using hashing to authenticate that an item is a member of set $s$ at cost $O(\log |s|)$.
- Epochs, batching deletions or other changes to reduce syncing, as in read-copy-update,[R73] generational garbage collectors, and the Hekaton in-memory database.

Often one reason for batching is to gather up work and defer it until the machine is idle; examples are defragmentation and garbage collection. This can work well, but fails when the load is continuous, when the machine is usually off if it's idle, or when battery power consumption is important. In these cases steady-state performance is the goal, and maintenance costs need to be steadily amortized, avoiding a balloon payment.

The opposite of batching is *fragmenting*, artificially breaking up a big chunk of work into smaller pieces. This is good for load-balancing, especially when either the load or the service time is bursty. An important example is assigning work to a number of servers that run concurrently.

There should be a lot more fragments than servers and a bound on the cost of the biggest fragment, so that stragglers don't do too much harm.[R7] Fragmenting bounds the variation in latency, and it also reduces head-of-line blocking: small jobs stuck behind big ones. Of course the fragments can't be too small or the per-fragment overhead will kill you. Fragments in a network are called packets; another reason to limit their size is to limit the size of the buffers needed to receive them.

3.4.5 *Cache*

The idea of caching is to remember the result of a function evaluation $f(x)$, indexed by the function $f$ and its arguments $x$. This yields an overlay of the partial function defined by the cache on the base function $f$; you try the overlay first, and only run $f(x)$ if the cache is undefined at $x$. The best-known application is when $f(x)$ is "the contents of RAM location $x$"; CPUs implement this in hardware, using a fast on-chip memory that is much smaller than the RAM. File and database systems do the same in software, keeping disk pages in RAM. Web proxies, content distribution networks and device-local email stores are also caches. Most references will *hit* in the cache if there's enough locality and it's bigger than the *working set* of frequently referenced locations; otherwise the cache will *thrash*. You can think of a cache for such a storage system as lazy partial replication, done for speed rather than fault tolerance.

The software indexes of databases and search engines are equally important; here $f(x)$ is "the table rows or documents matching $x$": $x = (col, y)$ matches $\{row \mid \text{table}(row, col) = y\}$ and $x =$ string $s$ matches documents containing $s$. Without an index you have to scan the entire database to evaluate these functions. For a table, if range queries are important a sorted index can find $(col, y, z)$, matching $\{row \mid y \leq \text{table}(row, col) \leq z\}$. A coarse index that says "every record in this region has a `salary` field between \$40k and \$50k" takes much less space than a complete index on `salary`, and it lets you skip that region when you search for `salary = $30k`; the price is that you have to search the whole region to find \$45k salaries.

What results do you cache? *Historical* caching saves a result that was obtained because the program needed it. *Predictive* caching guesses that a result will be needed and precomputes it; in a RAM cache it's called prefetching, and in a database system it's called a materialized view. Both are forms of speculation, betting that a computed result will be used again, and that a result that hasn't been used yet will be used in the future. Usually the second bet is riskier, unless the resources it uses are free.

If $f(x)$ depends on the state as well as $x$, then when state changes cause $f(x)$ to change you must tolerate stale cache values, treat a cache hit as a hint and check it (often by checking a version number), or invalidate or update a cache entry. The last requires that the source of the change either
- sends a *notification* to any cache entries that depend on it, or
- broadcasts every state change, and the cache watches the broadcasts.

For a RAM cache a state change is another processor's write to an address in the cache, and the two techniques are called *directory* and *snooping*. A directory entry is much like a lock on the address, except that the entry doesn't release a read lock explicitly, but holds it until there's a conflicting write and then invalidates the cache entry and releases the lock automatically. If using a cache entry involves going through a page table entry that you control, you can invalidate it by setting the entry to trap.

A database index is part of the system and it's kept consistent: updated immediately when the data changes; this is complicated. A file or document index is usually an addon that is updated eventually; this is simple—it doesn't affect the code of the target system—but expensive, since you have to rescan all the data. It's like broadcast; it always works and it needs no help from the target. A web search engine must work this way, perhaps favoring the parts that change more often. Notifications in a change log, even a coarse one that just notes the region of a change, can make index updates much faster; many file systems keep this log.

Whether it's better to invalidate or update is also a speculation. A subtle example is a word processor like Bravo that caches information of the form, "For a display line at character $cp$ of the document, the layout depends on characters $[cp-1, cp+\Delta]$, it contains $l$ characters, and it has this display image." Any change in the dependent range invalidates this entry; it may also become useless, though not invalid, if there's no longer a line being displayed that starts at character $cp$.

Here are some other examples of caching a function:

- Network routing tables, which say what link to use to reach a destination address. These are soft state, updated lazily by a routing protocol such as BGP, OSPF, or ethernet switching.
- Shadow page tables in virtual machines, which cache values of the mapping $(guest\ VM, virtual\ address) \rightarrow host\ physical\ address$, the composition of $guest\ VA \rightarrow guest\ PA$ and $guest\ PA \rightarrow host\ PA$. If the value of either function changes the entry is invalid; this is tricky because the two functions are owned by different systems.[R3] This also works for other hardware resources such as interrupt tables.[R5]
- Materialized views in a database, which cache the table that's the result of a query in the hope that it will be reused many times.

A twist on storage caching for a database is spilling: you have enough RAM that *all* the data usually fits. Then slow memory (for a cloud application, object storage) is just there to handle the peaks in size. The catch is complexity: you can't just follow a pointer to access data, and you need something else to do while waiting for a miss. But a RAM-only system doesn't work at all if the database is bigger than your RAM.

3.4.6 *Concurrency*—S$^3$: shard, stream or struggle. Make it atomic.

Now that single-stream general-purpose processors are not getting faster because speeding up the clock makes the chip too hot,[R65] there are only three ways to speed up a computation: using fewer instructions or cache misses (by better algorithms or tighter code), specialized hardware, and concurrency. Only the latter is reasonably general-purpose, but it has two major problems:

- It's hard to reason about concurrent computations that make arbitrary state changes, because the concurrent steps can be interleaved in so many ways. Hence the S$^3$ slogan.

- To run fast, data must be either immutable or local, because when a remote variable changes, getting its current value is costly. Fast computations need P&L: parallelism and locality.

The other reason for concurrency is that part of the computation is slow. Disk accesses, network services, external physical devices, and user interaction take billions of processor cycles. This kind of concurrency doesn't have the second problem, but it still has the first. And when the slow part is done it has to get the attention of the fast part, usually by some form of *notification*: interrupt a running thread, wake up a waiting thread, post to a queue that some thread will eventually look at, or run a dispatcher thread that creates a new thread.

**Sharding** (also called striping or partitioning) is really easy concurrency that breaks the state into $n$ pieces that change *independently*. A single thread touches only one shard, so the steps of threads that touch different shards don't depend on the interleaving. A *key* determines which shard to use. Sharding is essential for scale-out, making an application able to handle an arbitrarily heavy load by running on arbitrarily many machines.

The simplest example is disk striping: a few bits of the address are the key that chooses the disk to store a given block, and all the disks read or write in parallel. Fancier is a sharded key-value store with ordered keys; $n - 1$ *pivot* values divide the keys into $n$ roughly equal chunks. To look up a key, use the pivot table to find its shard. Everything is concurrent, except for rebalancing the shards, which is trickier.

You can make the sharding flat, usually by hashing the key to find the shard as in DHTs,[R100] or hierarchical if there are natural groupings or subsets of keys, as with path names for files, DNS names, and Internet addresses. Hierarchy is good for change notifications, since it makes it easy to notify all the "containing" shards that might need to know, but not so good for load-balancing, since there may be hot spots that get much more than their share of traffic. The fix is to make the path names into ordered keys to pivot on, disregarding their structure and treating them as strings.

Often there's a *combining function* for results from several shards. A simple example is sampling, which just takes the union of a small subset from each shard; the union is serial, but it's typically processing $\log n$ out of $n$ items so this doesn't matter. Union generalizes to any linear (homomorphic) function between two monoids $f: M \to N$, a function that "preserves" operations: $f(a +_M b) = f(a) +_N f(b)$, and $f(0_M) = 0_N$. Each shard evaluates a linear function, and then a tree combines the results with $+_N$. The familiar Map and Reduce operators are linear.[R19]

A tricky example is sorting on $n$ processors: shard the input arbitrarily, sort each shard, and combine by merging the sorted shards. A straightforward merge requires processing all the data in one place, and Amdahl's Law will limit the performance. To avoid this, sample the data in parallel to make a pivot table;

send the data to the shards in parallel—each item in shard $i$ precedes any item in shard $i + 1$;

sort the data in parallel in each shard.

Concatenating the data from the $n$ shards in order, rather than merging it, yields a sorted result.[R7]

**Streaming** (sometimes called pipelining) is the other really easy kind of concurrency: divide the work for a single item into $k$ sequential steps, put one step on each processor, and pass work items along the chain. If it's systolic, that's even better. This scheme generalizes to *dataflow*, where the work flows through a DAG. The number of distinct processing steps limits concurrency. Use batching to reduce the per-item overhead. You can evaluate any expression that doesn't have side effects this way because it forms a tree, or a DAG if there are common sub-expressions. An important example is a database query.

Map-reduce combines these two techniques, alternating a sharded map phase with a combining reduce phase that also redistributes the data into shards that are good for the next phase. The redistribution is the reason for having phases rather than a single DAG. You can reuse the same machines for each phase, or stream the data through a DAG of machines. The combining phase of map-reduce illustrates that even when you have a lot of independence, concurrency requires communication, and that's often the bottleneck.

The fastest way to send data through a shared memory system is to "lend" the memory, just passing a pointer. If the goal is to model a message send, the data had better be immutable.

Concurrent tasks need to be scheduled.

**Constructs for concurrency**

There are many ways to get a concurrent program:
- An explicit `fork(r)` of a new thread that runs routine `r` concurrently, returning a *promise* `p`; then `await(p)` returns the result of `r`. Languages package this in many different ways.
- A `parbegin` $b_1$; …; $b_n$ `parend` that runs the $n$ blocks concurrently.
- A library whose code runs concurrently, such as a database or graphics system.
- A distributed system.

Everything in this section applies to all of these, except for locks in a distributed system.

**Beyond shards and streams—struggle**

*Do I contradict myself? Very well then I contradict myself, (I am large, I contain multitudes.)* — Walt Whitman[Q96]

*I may be inconsistent. But not all the time.* —Anonymous

If you can't shard or stream, you will have to struggle. It helps to first show that a general nondeterministic program is correct, and then let performance constrain the choices: scheduling (including cache flushing, prefetching, timeouts, interleaving, losses), table sizes, etc. If the abstract state is not bulletproof (at least type and memory safe) you'll struggle more.

There are five kinds of concurrency; the first three provide *consistency* or *serializability*, which means that the concurrent system produces the same result as running the actions sequentially in some order, usually one that respects real-time: an action that starts after another action ends will be later in the order (this isn't the ACID consistency of transactions). If all the actions are on one object, it's called *linearizability*.
- **Really easy**: pure sharding or streaming. Either actions are *independent*, sharing no state except when you combine shards, or they communicate only by *producer-consumer* buffers.

- **Easy**: make a complex action *atomic* so that it behaves as if the entire action happened sequentially (serially), not interleaved with actions by other threads. To do this, group the actions into sets that don't commute (and hence break atomicity if they run concurrently), such as reads and writes of the same variable. Have a *lock* variable to protect each set, with the rules that:
  − Before running an action, a thread must acquire its lock.
  − Two locks in different threads *conflict* if their actions don't commute. For example, two reads commute, but writes of the same variable don't commute with reads or other writes.
  − A thread must wait to acquire a lock if another thread holds a conflicting lock.
  
  There are some complications, discussed below.
- **Hard**: anything else that's serializable, usually with small atomic actions. With hard concurrency you can choose: do a formal **proof** or have a **bug**. Stay away from it if you possibly can. If you can't, learn how to use a formal system like TLA+.[R57]
- **Eventual**: all updates commute, so you get the same eventual result regardless of the order they are applied, but you have to tolerate stale data. This is easy to code:
  − Make updates commute. The usual case is a blind write $v \coloneqq$ constant to a variable $v$, perhaps in a hierarchical name space. To make two writes to $v$ commute, timestamp them and let the last writer win. Collapse the update history into just keeping the timestamp of the last write with each variable, and when a write arrives, apply it if its timestamp is later. A deletion must leave a time-stamped tombstone. If you keep the whole update history, timestamps make arbitrary updates commute, but that's expensive.
  − Arrange the nodes in some structure that allows you to broadcast updates, such as a ring or a graph; check whether a node already has an update to keep it from getting into a loop.
  
  It's also highly available, since you can always run using only local data. The apps pay the piper: they must deal with stale data. The spec for eventual consistency is that a read sees an *arbitrary subset* of all the updates that have been done. Usually there's a *sync* operation; it guarantees that after it ends every read sees all the updates that precede the start of the sync. Three of many examples are name services like DNS (which has no sync), key-value stores like Dynamo, and "relaxed consistency" multiprocessor memory, in which sync is called "fence" or "barrier"[R4]. Most file systems buffer writes and don't guarantee that data is persistent until after an `fsync`, and this is somewhat similar after a failure: there are some updates that no one will ever see again.
- **Nuisance**: actions can run concurrently but produce strange results if they do, and some higher-level mechanism keeps this from happening. A familiar example is opening and renaming files. If the directories involved are themselves being renamed many strange things can happen, since a file system usually doesn't hold locks to serialize these actions. Applications use conventions to avoid depending on directories that might be rearranged while the app is running. A spec for nuisance concurrency (which you don't normally care about, because you're avoiding it) has the same flavor as eventual consistency: it collects all the changes that can occur during a non-atomic action, and the spec for the action chooses a subset of these changes non-deterministically to make the state that it acts on. The difference is that the client is not prepared to tolerate the inconsistency.

**More on easy concurrency**

The locking rules ensure that any concurrent action $c$ that happens between an action $a$ and the action $r_a$ that releases $a$'s lock commutes with $a$, since an action that doesn't commute must wait to acquire a conflicting lock. So $a; c; b; r_a$ is the same as $c; a; b; r_a$, where $b$ is the rest of the work that $a$'s thread is doing, and hence neither $a$'s thread nor $c$'s thread sees any concurrency until $r_a$ releases the lock. The atomic actions in the concurrent threads are *serialized*. Another way of saying this is that each atomic action is equivalent to a single action that occurs at a single instant called its *commit point*. You can reason about the code of such an atomic action as though it runs sequentially, with nothing going on concurrently. To reason about a collection of atomic actions, define a *lock invariant* on the state that holds except perhaps inside the code of an atomic action.

»Cluster locks: A lock is a resource, and like any resource it can become a bottleneck, so it needs to be instrumented. DEC built the first computing clusters, in which you could do more computing simply by adding machines. But one large cluster didn't scale, even though no physical resource was a bottleneck. It turned out there was a single lock that was 100% busy, but it took top engineers to figure this out because there was no way to measure lock utilization.

There are three tricky things about easy concurrency: enforcing the lock discipline, dealing with deadlock, and doing the right thing after releasing locks. Transaction processing systems (TPS) solve all these problems in draconian fashion; a custom app must worry about them.

*Discipline*: TPS interpose on the app's reads and writes of the shared state to acquire the necessary locks. A custom app must take responsibility for this; unfortunately locking is hard to debug, because the app will still work most of the time even if it touches variables without having locked them. Tools like Valgrind can detect most of these errors.

*Deadlock*: TPS detect deadlock and abort one of the transactions involved, undoing any state changes. A custom app usually doesn't have detection or abort, but avoids deadlock by defining a partial order on the locks and only acquiring a lock if it's bigger than all the ones it already holds.

*Lock invariant*: TPS don't allow an app to keep *any* private state after a transaction commits; the app has to re-read everything from the shared state. A custom app must follow a similar rule: have a lock invariant on the state, establish it before releasing any locks, and assume nothing about the state except the invariant when you acquire a lock. Another way of saying this: choose atomic actions wisely. For example, to increment $x$ atomically it's not enough to hold a read lock when fetching $x$ and a write lock when storing it; the lock must cover the whole atomic sequence.

An important special case of easy concurrency is *epochs*, a batching technique that maintains some invariant on the state except in between epochs. An epoch is a special case of locking that holds a global lock on certain changes throughout the epoch, so that they can only occur when the epoch ends and releases the lock. The code follows these rules by convention; there's no lock variable that's acquired and released, so epochs are cheap but easily subverted. Most often the change that is locked is deleting an object, so that objects won't disappear unexpectedly. For this to work well it has to be okay to defer the deletions. Sometimes the global lock prevents *any* changes to certain objects, keeping them immutable during the epoch.

Locks don't work well in a distributed system because they don't play nice with partial failures. Leases can be a workaround. The only meaningful content in an asynchronous message is facts that are *stable*: once they are true, they are true forever. For example, a lease implies "$P$ holds until time $t$," which is stable if you are careful about clock skew. "$P$ holds until $Q$" might be stable too, but a failure can make $Q$ inaccessible. A fact from an eventually consistent system is stable only if it has been synced.

There are many constructs other than locks and condition variables for coding easy concurrency. Each one handles some problems very elegantly, but unfortunately falls apart on others. My view is that there's only a limited amount of elegance to go around; don't spend it all in one place. Locks are an inelegant but all-purpose tool for atomicity; conditions do the same for scheduling.

**Commuting and atomicity**

When you don't have disjoint states (sharding) or strict dataflow (streaming), the essential idea is that commuting actions can be reordered into larger *atomic* actions to make it easier to reason about the code. Easy concurrency is the simple way to make actions commute: the locks *delay* any actions that don't commute. In hard concurrency there are more subtle reasons why actions commute, but the necessary proofs typically work by showing that certain actions are *movers* that commute with every other action in the system that can possibly run concurrently.

A good rule of thumb is the *scalable commutativity rule*: if the specs of two actions commute, then it's possible to write code in which they run concurrently, which is important for keeping all the cores busy on modern CPUs. For example, Posix file `open` returns the *smallest* unused file descriptor; if it returned an *arbitrary* unused descriptor, two `open`s could commute.[R21]

Locking is the standard way to make a complex action atomic; the commit point is some time between the last lock acquire and the first lock release. There are other ways, but they are used mostly in database systems because they need a lot of infrastructure. Two are worth knowing about: optimistic concurrency control (OCC) and multi-version concurrency control (MVCC).

The idea of OCC is to let the code of an atomic action touch variables fearlessly, but keep any changes in a write buffer private to the code. When the action commits, check that any variables that were read have not been written by someone else. If the check fails, abort: discard the changes and *retry* the action. If it succeeds, commit and install the changes. This check and install itself needs to be made atomic by locking, but it can be very fast and hence can be protected with a single global lock. The commit point is when this lock is held. Hardware transactional memory works this way, using the cache for the write buffer. OCC is good when conflicts are rare, since you only pay when there is a conflict. It has the drawback that you might see inconsistent data, but if you do you will always abort. Locking is better when conflicts are common, since waiting is better than the wasted work of many aborts and retries. OCC becomes less efficient as load increases, and under high load its performance can totally collapse.

MVCC remembers the state after every atomic action. An action can use its start time as its commit point, since it can always read the state as of that time. However, to write a variable that has already been read by a later action it must abort and move its commit point after the read; this works well for actions that only write private variables, most often after checking some property of the state such as whether the bank's books balance. Each version is immutable, so a cache entry tagged with the version it applies to is never invalid. Snapshot isolation is a limited form of MVCC. Multiple versions are good for version control of programs and documents, where they are an essential part of the user model, not just a coding technique; in these applications versions form a tree, and there's a semi-manual *merge* operation to reconcile two versions.

**Wait-free concurrency**

Multiple versions are also the basis for the general form of wait-free (non-blocking) computation: make a new version and then splice it in with a compare-and-swap instruction, retrying if necessary.[R55] Wait-free is good because a slow thread holding a lock can't force others to wait. If there is contention retry can livelock (keep retrying); the fix is for a conflicting thread to *help* with a failing update.[R95]

## 3.5 Adaptable

| | | |
|---|---|---|
| Fixed ↔ evolving, monolithic ↔ extensible [I] | { | The only constant is change. |
| | | Make it extensible. Flaky, springy parts. |
| Dynamic ↔ static | [E] | Stay loose. Pin it down. Shed load. Split resources. |
| Evolution ↔ revolution | | Stay calm. Ride the curve. Seize the moment. |
| Policy ↔ mechanism | | Change your mind. |

There are many things your system might need to adapt to during its lifetime:
- Changes in the **clients' needs**: new features or data formats, higher bandwidth, lower latency, better availability.
- Changes in the host **platform**: new interfaces or versions, better or worse performance.
- Changes in **regulation** or in security **threats**: privacy or other compliance requirements, data sovereignty, broken cryptography, new malware.
- Changes in **scale**, from 100 clients to 100 million or from storing text to storing video. If the change is temporary, in response to a bursty load, the system needs to be **elastic**, not just scalable; cloud computing provides an affordable elastic service by balancing the demand from many clients.

Such changes may force major rework, but usually a well-designed system can adapt less painfully. An old rule of thumb says that a 10 × change in scale requires a new design, but the Internet and the web are striking counterexamples. Ideally your system can be adapted to uses not anticipated by the designer.[R102]

The keys to adapting to functional changes are modularity and extension points in the design. The keys to adapting to scaling are modularity, automation, and concurrency.

Interface changes can be incompatible: unless the client and service specs change at the same time, there's a mismatch. This is okay if the new service spec is a superset of the old one. Ethernet, the Internet, many ISAs, some programming languages, and basic HTML have done this, and 40-year-old clients still work. The alternative is indirection: an *adapter* or *shim* that satisfies the old spec and is a client of the new one. When the new one is dramatically different this is *virtualization*. You're in trouble if you can't find all the places that need to be shimmed (too much complexity), or if the performance hit is too big (but translation can often keep this under control). A perfect adapter is hard, but even an imperfect one can be useful. If two parties connect dynamically each one should name its version of the interface so that it's clear what adapter is needed.

You can push this kind of adaptation a long way. It's a truism of computer architecture that the only mistake you can't recover from is making the address space too small, but in fact people have found many ways to recover, though all of them are slightly kludgy. The 32-bit IPv4 Internet address space is too small. IPv6 expands it cleanly, but the fix mostly used in practice is NAT, which works by making the 16-bit port number part of the address. Purists consider this to be heresy (which it is) and for years refused to standardize it, but the 2020 Internet depends on it. Early versions of Unix used separate processes connected by pipes, each with its own address space, to expand the PDP-11's 16-bit address space. This forced modularity and shell scripts, making a virtue of necessity.

Data has to adapt too, migrating when the representation or the hosting service changes. A simple schema makes this easier. Alternatively, use federation to keep old code around for old data while switching to new code for new data.

### 3.5.1 *Scaling*

Expanding on the catchwords above, scaling requires:
- Modularity for algorithms, so it's easy to change to one that scales better.
- Automating everything, both fault tolerance and operations, so that a human never touches just one machine (except to replace it if the hardware fails).
- Concurrency that scales with the load by sharding: different shards are independent because they don't share variables (of course it's okay to share immutable data) or resources: all communication is asynchronous.

The independent shards sometimes have to come back together. There are two aspects to this:
- Combining the independent outputs or synchronizing the shard states. Fancy versions of this are called *fusion*.
- Naming the shards, using big random numbers (which must be indexed) or path names.

If the shards already exist, use federation to put them into a single name space by making a new root with all of them as children.
- In a file system this is called *mounting*, and the shards stay independent.

- Global authentication is much like mounting: to accept identities that are registered with Intel, Microsoft adds to its directory a rule that the key for `groves.intel.microsoft.com` is whatever Intel says is the key for `groves.intel.com`.
- In a source code control system the shards are *branches* and synchronization is *merging*.
- Modules that satisfy the same spec can federate even if they use different techniques, as in SQL query processing, numerical computing, or mixtures of experts in machine learning. If the data's type determines the methods this is a form of subclassing. If not, you need to look at the data and select by hand which code to use.

A cloud's many independent clients make it easy to shard both computing and storage. A compute shard is either a virtual machine or a "serverless function," a short-lived chunk of computation that runs with some fixed OS, language-specific runtime, and initial state, but can read and update state in storage objects. A storage shard, called an object or blob, is a key-value pair where the value is just a sequence of bytes, possibly a few terabytes long. Clients can use objects to store sets of database rows or other structures. As usual, bigger shards perform better because there's a minimum cost to run a VM or access an object (and cloud providers throttle clients that access objects too fast), but the client has to manage the batching, which adds complexity.

The right algorithm can also help with scaling. Most important, the cost should not grow faster than $n \log n$, but constant factors matter too. For example, consistent hashing guarantees that resizing a hash table with $n$ keys and $m$ slots moves only $n/m$ keys to a different slot. This is especially important for a fault-tolerant or elastic distributed hash table, where the size changes frequently and load-balancing is important.[R100]

Many things scale according to Zipf's Law, which says that the $n$th most common or biggest thing is $1/n$ as common or big as the first. It's unintuitive, but this is very heavy-tailed: $\sum_1^n 1/k \approx \int_1^n 1/k \; dk = \ln n$, unbounded even though it grows very slowly, so the mean is undefined.[R2]

3.5.2 *Inflection points*—Seize the moment. Ride the curve.

*History never repeats itself, but it rhymes.* —John Robert Colombo[Q12]

Why do great new technologies often fail? They are great when compared with the current incarnation of the boring old technology, but during the 3 to 5 years that it takes to ship the new thing, the old one improves enough that it's no longer worthwhile to switch. This typically happens with new hardware storage technologies, such as magnetic bubbles, thin film memories, optical disks, and perhaps phase-change memory and memristors; for the last two the jury is still out.

The reverse happens when a new idea has some fundamental advantage that couldn't be fully exploited in yesterday's world, but conditions have changed so that it now pays off:
- In the 1960s virtual machine monitors were invented as a way to share expensive hardware without rewriting software, but time-sharing OSs like Unix replaced them. Around 2000 cheaper hardware made the system advantages of VMs affordable: decoupling an OS from hardware details, and isolating different users of a machine better than an OS can. Now VMs are the backbone of cloud computing.

- For many years multiprocessors were slower than spending the same amount to make a single processor faster. Capable single chip CPUs made it much cheaper to build a multiprocessor, and a workstation with several CPUs seemed like a good idea,[R105] but the advent of RISC delayed this. Eventually clock rate stopped increasing, the bag of tricks for increasing instructions per clock got empty, and multicore processors became the norm.
- Log structured file systems were invented around 1990, had marginal payoffs at best with the disks of the time, but are now very attractive because of the physical properties of SSDs and of high density disks—the hardware physics no longer permits random writes.
- Packets replaced circuits for communication when the computing needed to do the switching got cheap enough, and bandwidth got cheap enough for bursty data traffic to overwhelm voice.
- Ted Nelson invented the web in the 1960s (he called it hypertext), but it didn't catch on until the 1990s, when the Internet got big enough to make it worthwhile to build web pages.

Moore's Law provides exponential change, and there are inflection points when a curve crosses a human constraint. So you can capture a digital image with higher resolution than the human eye can perceive, store an entire music collection in your pocket, or stream a video in real time so you don't need lots of local copies. Likewise when an exponential crosses a boundary that makes something feasible, first technically and then economically, as with spreadsheets and word processors in the late 1970s, with email in the 1980s, and with machine learning around 2012. If you catch one of these moments you can be a hero.

The most fundamental constraints are three dimensions and the speed of light. This means that much of performance is about locality; less code operating on less data closer in space and time. Physics says that there will always be memory hierarchies, computation and communication, persistent storage and I/O. The relative bandwidth, capacity and latency change, but there are always tradeoffs among them.[R27]

»Three waves of computing. The development of electronics during World War II meant that around 1950 it became technically feasible to build computers that do *simulation*, first of physical processes like artillery trajectories and nuclear weapons and then of business processes like payroll and inventory, by running a model of the process in the machine. Simulation paid all the bills for the first 30 years, and it's by no means played out. Around 1980 hardware got cheap enough to use it for mediating *communication* between people, and this gave rise to the Internet, email, the web and social media, touching far more people than simulation did. This too still has plenty of room to evolve; for example, telepresence is still pretty bad. Around 2010 computing and sensing hardware got capable and cheap enough for a third great wave, which is about *interactions with the physical world*: robots, sensor networks, natural user interfaces, image recognition and many other things. This is just getting started, but it will have even more impact. The big system issues here are real time, uncertainty and safety.

## 3.6 Dependable

*The price of reliability is the pursuit of the utmost simplicity. It is a price which the very rich find most hard to pay.* —Tony Hoare[Q33]

*As a rule, software systems do not work well until they have been used, and have failed repeatedly, in real applications.* —David Parnas[Q61]

| | |
|---|---|
| Perfect ↔ adequate, exact ↔ tolerant   [ST] | { Good enough. Worse is better. End-to-end. |
| | Flaky, springy parts. Fail fast, fix fast. } |
| Precise vs. approximate software   [T] | Get it right. Make it cool. Shipping is a feature. |
| Centralized ↔ distributed, share ↔ copy [E] | Do it again. Make copies. Reach consensus. |
| Consistent ↔ available ↔ partition-tolerant | Safety first. Always ready. Good enough. |
| Generate ↔ check | Trust but verify. |
| Persistent ↔ volatile | Don't forget. Start clean. |

A system is dependable if it is:

– **Reliable**—it gives the right answers in spite of partial failures and doesn't lose data that it's supposed to preserve.
– **Available**—it delivers answers promptly in spite of partial failures.
– **Secure**—it's reliable and available in spite of malicious adversaries.

The secret of reliability and availability is fault tolerance by *redundancy*: doing things independently enough times that at least one succeeds. Redundancy can be in time or in space.

- Redundancy in **time** is *retry* or *redo*: doing the same thing again. You have to detect the need for a retry, deal with any partial state changes, make sure the inputs are still available, and avoid confusion if more than one try succeeds. The main design tool is end-to-end validation.
- Redundancy in **space** is *replication*: doing the same thing in several places. The challenges are giving all the places the same input and making the computation deterministic so that the outputs agree. The main tool is *consensus*.

It's very important for the redundancy to mostly use the *same* code as the normal case, since that code is tested and exercised much more, and hence has many fewer bugs. And of course redundancy won't do any good if a deterministic bug (a *Bohrbug*) caused the failure. On the other hand, many bugs are infrequent nondeterministic *Heisenbugs*, usually caused by concurrency.[R43]

Redundancy by itself is not enough; you also need *repair*. If one of two redundant copies fails the system continues to run, but it's no longer fault-tolerant. This is the reason to scrub mirrored disks, reading every block periodically and restoring unreadable data from the mirror. Similarly, if a component is failing half the time and a single retry costs three times as much as a success, the operation takes six times as long as it should.

The idea of redundancy is to have no *single points of failure*. Finding all such points is hard; it takes careful analysis, a good understanding of the environment and a clear notion of what risks you are going to accept. Cloud services, for example, worry about large-scale disasters like earthquakes and put data centers far enough apart to tolerate them. For security this analysis yields a threat model.

No single point of failure means a distributed system, which inherently is concurrent and has *partial failures*: some of the parts can fail, including the communication links, while the whole system keeps working. Hence there are many more rare states, which is why a distributed system is harder to get right than a centralized one, in which many errors just reset the whole system to a known state (perhaps passing through the Blue Screen of Death).

A Bohrbug is also a single point of failure, unless the redundancy includes different code. *Multi-version programming* does this, building more than one code to the same spec; so far it has not been successful, because it's quite expensive, it's easy for programmers to misinterpret the spec in the same way, and it's hard to get all the versions to track changes in the spec.

»Arpanet partitioning. On December 12, 1986, New England was cut off from the Arpanet for half a day. The map showed that there were seven connections to the rest of the network, but it didn't show that all seven of them went through the same fiber-optic cable between Newark and White Plains.[R48] In theory carriers can now guarantee that two connections share no physical structure.

»Cellphone disconnected. I tried to call a friend at the Microsoft campus on his office phone. It didn't work because it was a VOIP phone and his building's Internet connection was down. So I tried his cellphone, and that didn't work either because his building had a local cell base station, which used the building's Internet to connect to the carrier and was too stupid to shut itself off when it could no longer connect.

»Datacenter fans. An entire data center went down because of a bug in new firmware for the fans. Operations staff knew that you always do rolling updates for system software, converting only a fraction of the deployed machines at a time and only one at a time from a replication group. But nobody realized that the fan firmware was system software.

### 3.6.1 *Correctness*

The best way to get your code to be correct is to keep it simple, and the best way to do that is to structure your system so that the most critical parts of the spec depend only on a small, well-isolated part of the code. This is the *trusted computing base* (TCB), invented to keep computer systems secure but applicable much more broadly. It's a good idea, but there are some difficulties:

– Keeping the TCB isolated from bad behavior in the rest of the system.
– Keeping the "most critical" parts of the spec from growing to be all of it (mission creep).
– Maintaining the structure as spec and code change.

The single best tool for making a TCB is the *end-to-end* principle;[R90] its underlying idea is that the client is in control. Specifically, if the client can easily *check* whether an answer is correct and has a *backup* procedure, then the code that *generates* the answer isn't in the TCB, and indeed doesn't need to be reliable at all. To use this idea you need a check for failure; if you're just sending a message this is a strong checksum of the contents, and a timeout in case the message never arrives. The checksum also works for storage, but since you can't retry, storage needs replication.

You probably don't want to give up if the check fails, so you need the backup; end-to-end says that this decision is up to the client, not the abstraction. You need to undo any visible state change caused by the failure, or ensure that it doesn't confuse the next steps. After that, if the failure is nondeterministic retrying is a good backup. The canonical example is TCP, which makes the flaky best-efforts packet service of the raw Internet into a reliable flow- and congestion-controlled byte stream. Other possibilities are trying something more expensive, especially if it was a hint that failed, or running in a degraded mode such as eventual consistency (with or without notice to the client). There may be no backup; encryption, for example, can't prevent a denial of service attack, though it can guarantee secrecy and integrity.

Fault tolerance means that the code doesn't run sequentially, because it can be redirected at any point by a fault. Instead you should think of it as a collection of atomic actions, each one enabled by some predicate on the state that is not just "PC = x," much like a concurrent program.

In fact, a fault tolerant program *is* a concurrent program, in which you don't have much control over the concurrency. An example is a crash-tolerant file system, where every chunk of code that ends with a write to the disk is an atomic action, after which recovery might run instead of the next sequential action.

If your program is going to be reliable, of course it has to be correct, right? And if you detect something wrong, for example a failing assert, you should report an error and abort, right? Well, not exactly. For one thing, if you're running a reliable service you should never abort; you should always log the failure, undo the failing action and retry. If you had a Heisenbug, the retry will work. More drastic retries are more disruptive, but also more likely to work because they reset more of the state.

It may not be important to guarantee a correct answer, and you can do surprisingly well with *failure-oblivious computing*: fix up the error in some straightforward way and press on. For example, if the program dereferences null, catch the segfault and return 0.[R86] Of course this idea can be pushed too far, but it works surprisingly often.

»Unverified build script. IronFleet verifies code for distributed systems, using the methodology described earlier. It has a tool for building the system (verifying, compiling, linking, etc.) which turned out to have a bug: it would sometimes think that the verification succeeded even though it had actually failed.[R37] Lesson: Bugs can show up anywhere; the toolchain is not exempt.

### 3.6.2 *Retry—Do it again*

If you can tell whether something worked, and after it fails there's a good chance that it will work better the second time, then retry is the redundancy you want. This applies especially to networking, where often you don't have good control of the communication, and even if you do it's much cheaper to tolerate some errors. Retry is based on the end-to-end principle, and in most applications you expect it to succeed eventually unless the network is partitioned or the party you are talking to has failed. Retry is a form of slow path: success on the first try is the fast path, with cost $f$, and if $p$ is the chance of failure and $r$ is the cost for one retry (the time it takes to detect a failure—usually a timeout but perhaps a negative acknowledgment—and try again), the cost of the slow path is $s = r(1 + p + p^2 + \cdots) = r/(1-p)$. As usual, the slowdown caused by retries is $1 + p(s/f)$. To make this small you need $p \ll f/r$. For example, if a retry costs 10 × a success ($r = 10f$), then you need $p \ll 10\%$ to make the slowdown from retries small. If instead $p \approx f/r$, for example, then the slowdown is $1 + 1/(1-p) \approx 2 \times$.

Too many retries can overload the system. Use exponential backoff to limit the load.

If $p$ is too big (perhaps because the chance of corrupting a message bit is too great), you can make it smaller with forward error correction (an error-correcting code). Or make $r$ smaller by fragmenting: breaking the work into smaller chunks that fail and retry independently.

A retry that succeeds is supposed to yield the same final state as a single try (as long as there are no concurrent actions that don't commute with this one); this is *idempotence*. Some actions are intrinsically idempotent, notably a *blind write* of the form $x := $ constant. To make an arbitrary action such as $x := x + 1$ idempotent, make it *testable*: give it a unique ID, remember the ID of a completed action (often as the version of a variable, encoded by the log sequence number of the operation that last changed it), and discard any redundant retries. In communication this is discarding duplicate messages at the receiver; it's called *at-most-once* messaging, "at most" rather than "exactly" because the message is lost if the sender fails or gives up. A common case is TCP, whose code is tricky because a persistent ID would be too expensive, so each connection has to do a "handshake" to set it up. (TCP has other complications because it sets up a connection rather than sending a single message, and because of the window mechanism for flow control and exponential backoff for congestion control.) If the messages and IDs are persistent it's called a *message queue* and it *is* exactly once. The reason that the payment pages of online commerce often say "don't hit the back button and retry" is that they do this wrong.

Another form of retry is redo recovery from a log after a crash. If every pair of actions $a$ and $b$ in the log either commute ($a; b = b; a$) or absorb ($a; b = b$), then redoing prefixes of the log repeatedly (which happens if there are crashes during recovery), followed by redoing the whole log, is equivalent to redoing the whole log once. This is *log idempotence*. A blind write absorbs an earlier write to $x$ and commutes with a write to any other variable. A testable action absorbs itself.

Retries can happen at different levels, resetting more and more of the state. For something like a web service where you control the whole system, they are the 5 Re's:[R14]

- Retry, the level discussed here; if it succeeds the system runs normally.
- Restart a transaction, an application, a thread or whatever, and then retry.
- Reboot the OS; this resets a lot of state that may be bad. The client will likely see a hiccup.
- Reinstall/reimage the application or the entire OS.
- Replace the hardware, since it must be broken (unless the application has a Bohrbug).

3.6.3 *Replication*—Make copies

The simplest kind of replication is several copies of the bits that represent the state, but it's very tricky to make this work when there are failures because you can't update all the copies atomically. The most powerful kind of replication is a log that records the sequence of operations that produced the current state. With this and a checkpoint of some past state, you can reconstruct the current state by redoing the operations. There are many variations on this idea.

The strongest variation provides uninterrupted service even when there are failures. It is a *replicated state machine* (RSM), a way to do a fully general fault-tolerant computation using the ideas of being and becoming. You make several replicas of the host, all running the same code, start them in the same state, and feed them the same sequence of deterministic commands, either in real time or from a log. Then they will produce the same outputs and end up in the same state. Any of the outputs will do as the output of the RSM, or the replicas can vote if there are at least three of them and a minority might be Byzantine.

Of course there are some complications:
- The replicas must all see the same sequence: they must all agree about the first command, the second command, etc. The *Paxos* algorithm for distributed asynchronous consensus does this, by getting a set of nodes to agree on a value; it guarantees that replicas will never disagree about commands, and it makes progress as long as a suitable quorum of replicas can communicate for long enough. The idea behind Paxos is to collect a quorum for some value, and if failures keep you from seeing the quorum, to try again. The tricky part is ensuring that if there *was* a quorum, a retry is forced to choose the *same* value.

- The commands must be deterministic; this requires some care. It means no randomness (pseudo-randomness is okay if they all use the same seed) and no real time (putting the current time into a command is okay). Concurrency requires great care, because it's inherently nondeterministic. Each replica must run the same code, or code that implements the same deterministic spec. This can be tricky; for example, two codes for sorting might order items with equal keys differently if the sort is not stable.
- To restore a failed replica, you can redo the whole sequence of commands from scratch, or copy the state of some other replica and redo recent commands. If the commands are log idempotent you can make a fuzzy copy and then redo all the commands since copying started; this is also how a replica with persistent state recovers from a failure.
- You can use the RSM to add or delete replicas; this is also a bit tricky.

Reads must go through the RSM as well, which is expensive. To avoid this cost, use the fact that physics provides a reliable communication channel called real time. One replica takes out a time-limited lock called a *lease* on part of the state through the RSM; this stops anyone else from changing that state. Drawbacks are that clocks must be synchronized, the leaseholder can be a bottleneck, and if it fails everyone must wait for the lease to expire (including the maximum clock skew) unless you can reliably detect the failure.

The usual way to do replication is as *primary-backup*: one replica is the primary, chosen by the RSM, and it has a lease on the whole state so that it can do fast reads and batch many writes into one RSM command. The backups see all the writes because of the RSM, and they update their state to be ready in case the primary fails. The RSM needs three replicas, but they only need to store the commands; only two have to store the entire state. A variation is *chain replication*, which arranges a set of replicas in a sequence, with reads done at the tail and updates started at the head and passed down the chain, but acknowledged only by the tail. It maintains the invariant that each replica has seen every update that its successor has seen. You can shard the chains, and as long as an object stays in a single chain, access to it is serialized.

Replication can make things faster as well as fault tolerant, since you can read from any replica that you know is up to date, such as a cache. This only helps if there are a lot more reads than writes from different writers, since a replicated write costs more.

If all you need is fault-tolerant *storage*, an *error-correcting code* (not to be confused with the code that satisfies a spec) uses much less storage than replication. ECC comes in many flavors; cloud providers want storage to be as reliable as simple three-way replication, and *local reconstruction codes* can do this using $1.33 \times$ storage, spending about $2 \times$ the latency and $5 \times$ the bandwidth of an error-free read to reconstruct a bad block. Rewriting data in place is expensive with LRC; the cloud systems use three-way replication for writes, bundling up data for LRC only after it's settled down.[R49] The first major effort along these lines was RAID (Redundant Arrays of Inexpensive Disks), which was quite popular for a while but eventually faded because doing a good job on every workload and in every failure condition was too complicated.

»Tandem mirrored disks. Tandem Computers built one of the first fault-tolerant disk systems, writing each sector onto two drives. Several years later, data from the field revealed that if one of the drives in a pair failed, the other was about 40 times more likely to fail than a drive in a pair without any failures. The reason was that the technician trying to replace a failed drive often pulled out the working one instead. The solution was a green light on the back of each drive, and instructions never to pull a drive with the green light lit. Lesson: independence is tricky.

»Ariane 5. The first flight of the European Space Agency's Ariane 5 rocket self-destructed 40 seconds into the flight because both inertial reference system computers failed, delivering a diagnostic bit pattern to the control computer instead of correct flight data. The computers shut down because of an uncaught exception from an overflow in a floating point to integer conversion. It was a deliberate design decision not to protect this conversion, made because the protection is not free, and (incorrect) analysis had shown that overflow was impossible. Shutdown seemed reasonable to engineers familiar with random hardware failures rather than software Bohrbugs.[R12] Lesson: independence is tricky.

»Diesel generators. Once there was a bank with a mission-critical data center. They had a diesel generator to provide backup power, and they tested the backup regularly by switching to it for 15 minutes once a week. Eventually there was a real power failure, and after 30 minutes the generator seized up for lack of lubricating oil.[R41] Lesson: Every part of the system, including the failure handling, should be used routinely in normal operation.

3.6.4 *Detecting failures: real time*

Real time is not just for leases. It's the only conclusive way to detect that a service is not merely slow but has failed—it hasn't responded for too long. (Another way is for the service to tell you about it, but it might be wrong or dead.) How to decide how long is too long? Choose a timeout, and when it expires either retry or declare a failure and run recovery; in both cases report the problem. For a client device the report goes to the human user, who can decide to keep trying or give up. For a service it ultimately goes to the operations staff. In a *synchronous* system a part that doesn't respond promptly has failed; a *heartbeat* checks for this at regular intervals.

How do you choose a timeout? If it's too short there will be unnecessary retries, failovers or whatever. If it's too long the overall system latency will be too long. If the service reports the progress it's making, that might help you to choose well. This story applies to a *fail-stop* system, which either satisfies its spec or does nothing. After a *Byzantine* failure the system might do anything. These are trickier to handle, and out of scope here.

»Thrust reversers. The two major suppliers of commercial airplanes have somewhat different philosophies about the role of computers. Boeing believes that when push comes to shove, the pilot is in control. Airbus believes that sometimes the computer knows best. Thirty years of experience have not been enough to reveal which is right, but they have provided some good stories. One dark and stormy night an Airbus plane was landing in a rainstorm with a tail wind and wind shear. The plane touched down and applied the brakes, but it just aquaplaned. The pilot pulled the lever to engage the thrust reversers. They didn't come on. Cycling the lever again didn't help. Eventually the plane ran off the end of the runway and caught fire, and two people were killed. Why? The computer knows that if the thrust reversers come on when the plane is in the air, it will fall like a stone, but how do you know it's on the ground? The answer that was designed in: the wheels are turning fast enough, or shock absorbers are compressed at both main landing gears. But they weren't.[R113] Something similar seems to have happened with Boeing's MCAS failures in 2018.

### 3.6.5 *Recovery and repair*

It's common to describe availability by counting nines: 5 nines is 99.999% available, which is five minutes of downtime per year (there are $\approx 2^{25}$ or $\pi \times 10^7$ seconds in a year). A good approximation is $MTTR/MTTF$, mean time to repair over mean time to failure (how long the system runs before it fails to serve its clients promptly enough). When part of a fault-tolerant system fails, $MTTR$ is the time to fail over to a redundant component, not the time to fix the failing part. In a well-engineered system failover is less than the specified response time, so the *system* doesn't fail at all; this is why it's important to make failover fast. Repair is also important.

»Memory errors. At Xerox Parc in 1971 we built a medium-sized computer called Maxc, using the new Intel 1103 1024-bit dynamic RAM chip (2020 chips are more than 50 million times bigger). We didn't really know whether this chip worked, but with single bit error correction we could tolerate a lot of failures, and indeed we never *saw* any failures in the running system for quite a while. So we used the same chips 18 months later in the Alto, but error correction was a much larger fraction of the cost in a much smaller machine and we decided to just have parity. Everything was fine until we ran the first serious application, the Bravo full-screen editor, and we started to get parity errors. Why? It turned out that 1103's are pattern-sensitive (sometimes a bit will flip when the values of surrounding bits are just so) with a very long tail. Although Maxc hardware reported a corrected error, there was no software to read the reports, and there were quite a few of them. Lesson: Measure failures and do repairs.

With some effort we got the problem under control using a random memory test program that ran whenever a machine was idle and reported errors over the ethernet. Two years later we built the Alto 2, using 4k RAM chips and error correction. The machine seemed to work flawlessly, but after another two years we found that in one quarter of the memory neither error correction nor parity worked at all, because of a design error. Why did it take us two years to notice? The 4k chips were much better than 1103's, and most bits in RAM don't matter much, so most single-bit errors don't actually cause software to fail. This is why consumer PCs don't have parity: chips are pretty reliable, parity adds cost, and parity errors make the PC manufacturer look bad, but if random things happen Microsoft gets blamed. Lesson: Different parties may have different interests.

»Ethernet slowdown. The ethernet on one of the Altos at Xerox was transferring data at 1/10 the speed of the others. This turned out to be because its ethernet interface was on the verge of failing and was dropping a lot of packets. The Pup Internet reliable transmission protocol was getting the data through, but with much lower bandwidth. Lesson: fault tolerance works, but you have to measure failures and do repairs.

### 3.6.6 *Transactions*—Make it atomic

*In bacon and eggs, the chicken is involved, the pig is committed.* —Anonymous

If a complex action is atomic (either happens or doesn't), it's much easier to reason about. The slogan for this is ACID: Atomic, Consistent, Isolated, Durable.

- Atomic: Redo recovery makes it atomic with respect to *crashes*: after a crash either the whole action has happened, or none of it. Persisting uncommitted writes requires an undo log as well.
- Consistent: If each transaction leaves the system in a good state (maintains an invariant) when running sequentially, then the whole system does so in spite of concurrency or failures. In addition, the transaction can decide to *abort* before committing, which undoes any state changes and so makes it atomic with respect to its *own* work. So it only needs to leave the system in a good state (consistent) if it commits. It just aborts if it runs into internal trouble and can't complete. Don't confuse this with a consistent distributed system.
- Isolated: The locks of easy concurrency make it atomic with respect to *concurrent* actions. Optimistic concurrency control has the same effect.

- Durable: A committed transaction writes its changes to persistent storage, usually in several copies, so that they survive anything short of a truly catastrophic failure.

Transaction processing systems ensure all these properties by draconian control over the transaction's application code.

An essential aspect of this control is that a transaction has no private persistent state, so that it is always doing an atomic action against the global system state. This means that it's easy to update the transaction's code even if it's running— the current transactions keep running with the old code, and any new ones start with the new code.

A transaction commits by first *preparing*: writing a persistent record of everything that's needed to make the transaction's updates durable. Then it writes a *commit record*. All these writes are usually done into a log. Once committed, it does whatever is necessary to make sure that others will see the updates; after a failure, recovery will redo this work. Finally, it releases the locks that were keeping it isolated. Prepare is usually called *write-ahead log* for a transaction on one system. If there are updates on several systems it's a *distributed transaction* (even if all the systems, usually called *resource managers*, are on one machine) and they must all prepare, and then wait to learn whether the transaction committed. For fault tolerance, use a consensus protocol to write the commit record.[R96]

Atomic transactions don't scale across organizations, because once you have prepared they can force you to hold locks until another agent's work is done, and usually an organization won't give up that much control. OCC doesn't solve this problem, because you still have to give up control at commit time. The alternative is *compensating transactions*, which undo steps that are taken as part of a bigger action that turns out to abort; this is necessarily ad hoc.

Isolation by locking works badly for long-running transactions such as balancing a bank's books, because concurrent transactions either are blocked for a long time or force the long transaction to abort. The solution to this is a form of multi-version called snapshot isolation.

»Pixie dust. Transaction processing systems are the pixie dust of computing. They take an application that understands nothing about fault tolerance, concurrency, undo, storage or load-balancing, and magically make it atomic, abortable, immune to crashes, and easy to distribute across a cluster of machines.

3.6.7 *Security*

*But who will watch the watchers? She'll begin with them and buy their silence.* —Juvenal[Q38]

*If you want security, you must be prepared for inconvenience.* —Gen. Benjamin Chidlaw[Q11]

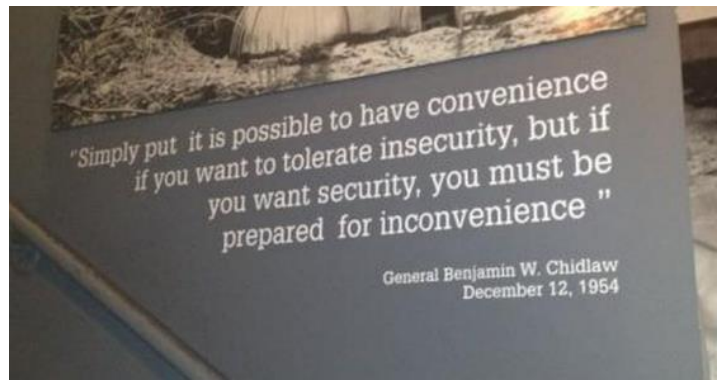

Computer security is hard because of the conflict between *isolation* and *sharing*. People don't want outsiders to mess with their computing, but they do want to share data, programs and resources. In the early days isolation was physical and there was no sharing except by reading paper tape, punch cards or magtape. Today there's a lot more valuable stuff in your computers, and the Internet enables sharing with people all over the world. The job of security is to say "No," and people like to hear "Yes," so naturally they weaken the security until they actually get into trouble.

Here are the most important things to do for security (which all add inconvenience):

- **Focus**: figure out what you really need to protect.
- **Lower aspirations**: secure only things so important that you'll tolerate the inconvenience.
- **Isolation**: sanitize outside stuff to keep it from hurting you, or don't share dangerous stuff.
- **Whitelisting**: decide what you do trust, rather than blacklisting what you don't. This is a bounded, not an unbounded task; you don't have to defend against all the bad guys' tricks.
- **Respond:** recover from security breaches rather than trying to prevent all of them. This means backing up data and punishing adversaries. The result is a little like eventual consistency: after recovery there are several states you might see.

There are basically two approaches to security: *high assurance* and fixing bugs. The former tries to build a TCB that is simple enough to be formally verified or thoroughly tested. This has proved easier to say than to do; the closest approximations that are widely deployed are hypervisors. Everyone practices the latter for want of anything better, but decades of experience tell you that there are always more bugs. Defense in depth can help.

It's traditional to describe the goals of security as *confidentiality* (secrecy), *integrity* and *availability*; the acronym is CIA. Integrity means that only authorized agents can change the state. In practice, systems that keep track of money or other critical data do use authorization, but they rely on detecting and undoing bad changes, rather than always preventing them, because even authorized agents sometimes make mistakes or do bad things; this is an example of the end-to-end principle, and the data is only eventually consistent. If you can't undo something, such as a wire to Russia, you must be much more careful in allowing it. Long-lived systems have levels of undo, ending with bankruptcy court.

The mechanisms of security are *isolation* and the gold standard of *authentication* (who is making a request), *authorization* (who can access a resource), and *auditing* (what happened). A decentralized system also has to establish *trust*, which you do by indirection: you come to trust someone by asking someone else that you already trust. Thus to answer questions like, "What is the public key for `billg@microsoft.com`," you trust a statement from `microsoft.com` that says, "The public key for `billg@microsoft.com` is $K$, valid through 3/15/2020."[R62]

An authorization policy is a function *permissions*(*agent*, *resource*), usually thought of as a matrix and stored as a set of triples. A small policy can be centralized, but usually it's stored

- at the resource, listing the agents with permissions for it in an *access control list* (*ACL*), or
- at the agent, listing the resources for which it has permissions in a *capability list*.

Only ACLs work for managing the policy, because the manager's question is, "Who has access to this resource?" It's okay to make short-term copies of parts of it into capabilities (usually called *file descriptors*), which are faster to check. Stating the policy using named sets of both agents and resources makes the manager's job feasible.

What are the points of failure? For security they are called a *threat model*, especially important because there are so many possible attacks (hardware, operating system, browser, insiders, phishing, …) and because security is fractal: there's always a more subtle attack. For example, how do you know that your adversary hasn't hacked the BIOS on your PC, or installed a Trojan Horse in the hardware?[R115] So you need to be very clear about what you are defending against and what you are not worrying about. The TCB is the dual of the threat model; it's just what you need to defend against the threats. The end-to-end principle makes the TCB smaller: encryption can make a secure channel between the two ends, so that the stuff in the middle is not a threat to secrecy or integrity.

Code for security is often tricky; don't roll your own. For secure channels, use TLS. For parsing text input to complex modules like SQL or the shell that accept programs in their input, use standard libraries to block SQL injection and similar attacks. Similarly for encrypting data; it's easy to make mistakes in coding crypto algorithms, managing keys, and blocking side channels.

Isolation can protect a system from an untrusted network (usually the host is trusted), or it can protect the host from an untrusted application, browser extension, webpage with JavaScript, etc. The latter is called *sandboxing*. Isolation is hard, because the adversary only has to find one flaw. As with so many other things, simpler is better. Physical isolation is simpler than software, and virtual machines are simpler than operating systems, because the interfaces are less complex and better specified. A library OS can drastically simplify the interface that isolation depends on.[R50] Side channels make isolation harder.

» Outrunning a bear. Two hunters run into a grizzly bear in the woods. One says, "We'd better run!" The other objects, "You can't outrun a grizzly." The first replies, "But I only need to outrun *you*." Lesson: Be a harder target than someone else who's just as rich.

»Bitcoin proof of work. The blockchain ledger underlying the Bitcoin cryptocurrency needs a distributed consensus protocol to make sure that all the copies of the ledger are the same, and the one it uses is based on proof of work, known as *mining* here. The idea is that in order to corrupt the ledger you would need to control more than half of all the mining capacity in the world. If mining is being done on millions of PCs, this seems impossible. Unfortunately, specialized hardware can do mining so much more efficiently than a general purpose machine that by 2018 a few miners in China in fact did control the ledger.[R51] This was not a surprise to people who had thought about using proof of work to control spam. Every proof of work system that I know about has this problem.

»Orange Book. The DoD's Orange Book is a famous security failure.

## 3.7 Yummy

*The Mac is the first personal computer good enough to be criticized.* —Alan Kay[Q42]

Simple ↔ rich, general ↔ specialized   [S]  KISS: Keep It Simple, Stupid. Do one thing well

A system is much easier to sell if it's yummy, that is, if customers are enthusiastic about it. There are some good examples:

- Apple makes consumer products that people love to use, sacrificing functionality for completeness (within the overall goals for the product), coherence and elegance. The Macintosh, the iPod and the iPhone are well known.
- Amazon's mission statement is, "To be Earth's most customer-centric company," and they approach a project by "working backwards": first write the press release, then the FAQ, then the customer scenarios, then the user manual.[R110] So they make Prime, Kindle, Alexa, AWS.
- People use and love the web as soon as they see it. Writing for it is less yummy, though.
- Facebook must be yummy, judging by the number of users it has, though I don't get it. This goes double for Snapchat.
- Spreadsheets are loved (especially by accountants and list-makers); VisiCalc, and later Lotus 1-2-3, are what made PCs take off.
- Porsches, Corvettes and Teslas are yummy.

By contrast, Microsoft Word, SharePoint, dishwashers and the Honda Accord are good products, but not yummy. Linux is yummy for developers, but not for users.

So what—is it important for your system to be yummy? If it's a consumer product it certainly helps a lot, and it might be crucial. For an enterprise product, staying power is more important. Clearly there's a lot of noise, but to cheaply boost your chances of making a yummy system, Amazon's approach is best. Much more expensive, but even better, is to study the users deeply. This is much easier if the designers are also users; this isn't always possible, but when it is the resulting system is much more likely to succeed. Unix, Bravo and the Internet are obvious examples.

### 3.7.1 *User interfaces*

*And the users exclaimed with a snarl and a taunt, "It's just what we asked for but not what we want."* —Anonymous[Q100]

*There are only two industries that refer to their customers as "users".* —Edward Tufte[Q85]

*Design principle: Don't make the user feel stupid.* —Alan Cooper[Q14]

*EMACS could not have been reached by careful design, because that arrives only at goals visible at the outset. No one visualized an extensible editor.* (paraphrased) —Richard Stallman[Q76]

People think that good user interfaces are all about dialog boxes, animations, pretty colors and so forth. Two things are much more important:

- The *user model* of the system: is there a way for the user to think about what the system is doing that makes sense, is faithful to what it actually does, and is easy to remember?
- *Completeness and coherence* of the interface: can the user see clearly how to get their whole job done, rather than just some piece of it? Are there generic operations like copy and paste that tell the user what operations are possible? Do the parts look and feel like a coherent design?

User models and coherence are hard because it's hard to find out what the users really need. You can't just ask them, because they are paid to *do* their jobs, not to *explain* them—no user would have asked for the iPhone. The only way is to watch them at their work or play for a long time to learn the use cases that are really important. A much cheaper substitute is to make up scenarios or use cases, but it's hard to ensure that they are both common and complete.

Here are some examples of good user models:
- Files and folders on the desktop.
- The web, with links that you click on to navigate.
- Web search, which pretty often finds what you're looking for.
- Spreadsheets, which can do complex calculations without any notion of successive steps.

And here are some less good examples:
- Microsoft Word, with styles, sections, pages, and other things interacting confusingly.
- The user interface to security—there's no intelligible story about what's going on and how a user should think about it.
- System administration, where the sound idea that the user should describe the desired state by a few parameters is badly compromised by poor engineering of the components.

Programmers often think that the data structures and routines they deal with are the real system, and the user interface just a façade. The truth is closer to the opposite: the UI is what the customers are buying, and it won't bother them at all if you replace all the underlying code. The germ of truth in the misconception is that if the UI really is a sloppily done façade, it's easy to replace it with something else that's just as sloppy.

It's good to separate the internal state of a system from the details of the UI. The usual way to do this is called model-view-controller (MVC). The model is the internal state, the view is the way it's displayed to the user, and the controller is the way the user changes the model.[R84]

In a world of communication and collaboration, often asynchronous, some form of eventual consistency is very important. Changes trickle in and must be integrated with what's currently being displayed, and perhaps changed by the user, in a way that's not too distracting. To be responsive a system must be able to run with only local data, and catch up later.

When you have input such as audio or mouse coordinates where timing is important, you can't slow it down, and you may not be able to keep up, make a log of time-stamped input events and process it at your leisure. Of course this isn't as good as real-time response, but at least you're not losing anything.

»Bravo and Gypsy. The most successful application on the Alto was the Bravo editor (the ancestor of Microsoft Word), the first What You See Is What You Get editor, built to exploit the Alto's display and processing power. When Charles Simonyi and I designed it, we made a deliberate decision not to work seriously on the user interface, not because we thought it was unimportant but because we knew it was hard and we didn't have the resources to both build an editing engine and invent a new UI. We were very lucky that Larry Tesler and Tim Mott came along with their Gypsy system for the book editors at Ginn, a publishing company that Xerox had bought. Their first step was to spend several weeks watching their customers at their daily work, learning what they actually spend time on. They used the Bravo engine but completely replaced our UI, and they invented modeless commands and copy/paste, the basis of all modern UIs.[R104] Later versions of Bravo adopted their work.

»Lotus Notes. It's nice when the user is the customer. Often, however, the IT department is the customer. Why did Lotus Notes and Microsoft SharePoint succeed? Ordinary people find them almost unusable, but IT departments really like them because their customization is a good match for the skills of IT staff. And of course once IT has done some customization, they are hooked. Lesson: know your customer. Lotus Notes is also one of the many examples of a strange fact about email: for about the first 20 years, an email product that you had to pay for was bound to be junk. Lotus Notes, Digital All-In-One, and Microsoft Outlook are just three examples. I think the reason for this is that none of these systems was built to be an email client, but rather to solve the much grander problem of office automation (which they all failed to do). Only Outlook has succeeded in transcending its origins. Lesson: do one thing well.

## 3.8  Incremental

| | |
|---|---|
| Being ↔ becoming | How did we get here? Don't copy, share. |
| Indirect ↔ inline | [E] Take a detour, see the world. Use what you know. |
| Fixed ↔ evolving, monolithic ↔ extensible | [A] The only constant is change. Make it extensible. |
| Iterative ↔ recursive, array ↔ tree | Keep doing it. A part is like the whole. |
| Recompute ↔ adjust | Take small steps. |

There are three aspects to incremental:
- *small* steps—otherwise it wouldn't be incremental,
- *useful* steps—you make some progress each time, and
- steps *proportionate* to the size of the change—you don't have to start over.

Incremental steps are easier than big steps to understand, easier to get right, less disruptive, and more likely to be useful building blocks. But it's important to start with a good idea; Alan Kay says, "It's hard to tinker a great sculpture from malleable clay just by debugging."[Q43]

Increments can be qualitative or quantitative. Qualitative ones are being and becoming, many forms of indirection, relaxation, evaluating small changes, subclassing, path names and many other techniques. Quantitative ones add elements without changing the structure:
- Nodes to the Internet or a LAN (and you don't even have to take it down).
- Peripherals to a computer.
- Applications to an OS installation or extensions to a browser.
- Features to a language, often as syntactic sugar.

### 3.8.1 *Being and becoming*

This is an opposition: being is a **map** that tells you the values of the variables, becoming a **log** of the actions that got you here. Some examples:
- A bitmap can represent an image directly, but a "display list" of drawing commands can produce the image; this generalizes to an arbitrary program, as in PostScript.
- There are many ways to represent a *store*, a map from variable names called addresses to variable values. Virtual memory uses page tables; an address is a virtual page number. File systems do it with a variety of schemes, among them the hierarchical index blocks and buffer cache of Unix version 6, the delta-based structures of copy-on-write systems, and the striped and redundant layout of high performance network file systems such as GFS; an address is a byte position in a file. Some text editors use a "piece table" (see fig. 2g); an address is a character position in the document, and a single replacement requires touching at most three pieces.

- A log-structured file system uses the log to store the data bytes, with an index just like the one in an ordinary file system except that the leaf nodes are in the log, which is enough to reconstruct the index. Amazon's Aurora pushes this to a limit.
- Checkpoints and deltas can compress a long sequence of states, such as the frames of a video or successive versions of a file. The checkpoints are a few complete states (called key frames for MPEG videos), and the deltas are actions that take one state to the next.
- Becoming lets you do time travel, since you can recover any previous state by replaying the log. Checkpoints make this faster. Undo's in the log also makes it faster because you can run time backward, but there'll be more bugs because the undo's aren't used in normal operation.
- Source code control systems like GitHub make it easy to access multiple versions. They usually work by storing just the edits (actions, deltas) for each new version, creating a complete version by redo only when it's needed for display or for feeding to a compiler. Content management systems for large sets of documents that are frequently revised also use this scheme.
- This idea generalizes to any computation with a result that's expensive to store. If it's deterministic it's okay to discard old results, remember the program that produced them, and recompute them transparently on demand, as in the version of map-reduce in Nectar.
- *Memoizing* is the opposite: caching the result of a function call in case it's needed again. Database systems call this a *materialized view*, remembering the result of a query. If there are some inputs to the function that change, you can generate a *slice* that just computes the corresponding changes in the results.[R34] *Dynamic programming* breaks down a problem into subproblems and memoizes solutions to avoid redoing work.
- The standard way to recover from failures in a data storage system is to apply a redo log that produces the current state from a persistent state that reflects only some prefix of the actions. If the system has transactions, the persistent state may contain updates from uncommitted transactions, and applying an *undo log* backs out the effects of a transaction that aborts.[R96 §19]
- A more general approach to fault tolerance uses a replicated state machine, which applies the same log to several identical copies of the state.
- Any kind of indirection is a move from being toward becoming, because it introduces some amount of computing into the map's job of returning the value of a variable.
- Becoming gives you a history that you can audit as well as replay. You can also use it to keep track of the *provenance* of your data: what agent with what inputs made each change.

How do you find the value of a variable $v$ (that is, construct a bit of the map) from the log? Read the log backward, asking for each logged action $u$ how it relates to the action $r$ that reads $v$.

If $u$ is a blind write $m(v') \coloneqq x$ then either $u$ and $r$ commute (if $v \neq v'$) or $u$ determines that $v = x$ and you don't need to look farther back in the log. Other kinds of $u$ need ad hoc treatment.

»Bravo undo. How do you undo some actions to get back to a previous version $v$? Simply replay the log up through the last action that made $v$. We did this in Bravo, logging the user commands, although our original motivation was not undo but reproducing bugs, so the replay command was called BravoBug. I've never understood why later systems didn't copy this; perhaps they didn't want to admit that they had bugs.[R64]

**Optimizations**

There are many variations on these ideas. To keep a log from growing indefinitely (which increases both the space to store it and the time to find the value of a variable), you can take a *checkpoint*, which is a map as of some point in the log. You can *share* parts that don't change among multiple versions; a copy-on-write file system does this, as does a library for immutable data like immutablejs. So does a multiprocessor that implements total store order: all the processors share the RAM, and each processor has a private write buffer holding its writes that have not yet made it to RAM.

There's a common idea behind all these optimizations: deconstruct the map, moving it closer to a log, by putting it together out of independent parts. The base case that the hardware provides is a fixed-size finite array of bytes in RAM, pages on disk or whatever; here the variables are integers called addresses $A$. Call this a store $S: A_S \to V$ and represent it abstractly by a hierarchical structure $S = A_S \to (V \text{ or } (T, A_T))$, where $A_T$ is an address in a lower level store $T$. Each level takes an address and either produces the desired value or returns a lower level store and address. Index blocks in file systems are an obvious example, but often you can think of this as a way to compress or index a log of updates, as in log structured memory or copy on write file systems.

To efficiently build a store $S$ on top of lower-level stores $T_1, T_2, ...$, build an index from (ranges of) $S$ addresses $[a_S, a_S + \Delta]$ to pairs $(T_i, a_{T_i})$; each entry in this index is a *piece*. Then the value of $S(a)$ for $a_S \leq a \leq a_S + \Delta$ is $T_i(a_T + (a - a_S))$ (fig. 2a). A write changes the index for the range of addresses being written (fig. 2b). There are many data structures that can hold the index: a sorted array, a hash table, a balanced tree of some kind.

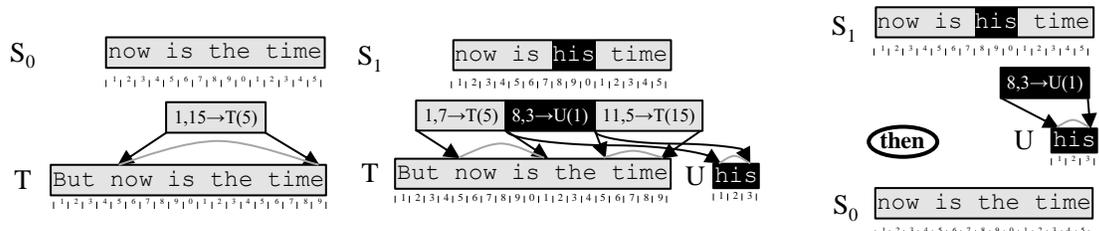

**Fig. 2a**: A single range **Fig. 2b**: Writing "his" in place **Fig. 2c**: Reusing $S_0$ in the write

Since the $T_i$ are stores themselves, this idea works recursively. And the indexes can be partial overlays, with a sequence of stores $S_n, S_{n-1}, ... S_0$; if $a$ is undefined in $S_n, ..., S_i$ then you look in $S_{i-1}$ (fig. 2c, with just $S_1$ and $S_0$). In the extreme each $S_i$ holds just one byte and corresponds to a single log entry and a write of a single byte. Several successive writes can appear explicitly (fig.

2d, with $S_2, S_1$ and $S_0$), or you can collapse them to a single level (fig. 2e, with just $S_2$ and $S_0$, like CPU write buffers), or all the way to an index that maps every address (fig. 2f, like a copy-on-write file system optimized for speed, with a full index for each version). Keeping $n$ versions as in fig. 2d means a worst-case cost $O(n)$ to find the value of an address.

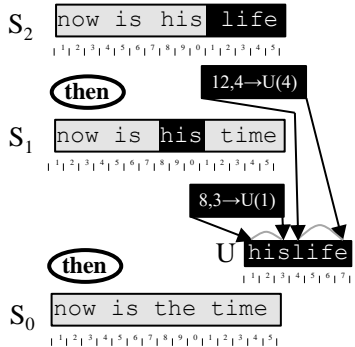
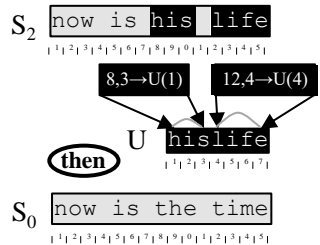
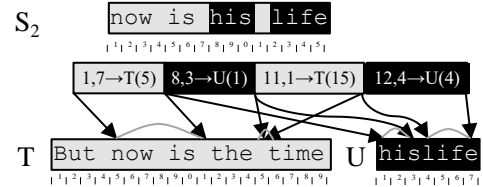

**Fig. 2d**: Two writes, three versions    **Fig. 2e**: A single discontinuous write    **Fig. 2f**: Back to a full index for $S_2$

These structures work in almost the same way for editing text, where the addresses can change. The difference is that you must adjust the $a_S$ addresses in the ranges that follow the edit (fig. 2g, corresponding to fig. 2e). When reusing $S_0$ you don't need the index range preceding the edit because those addresses don't need to be adjusted.

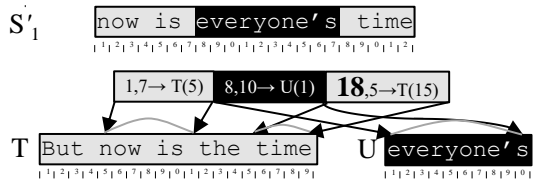
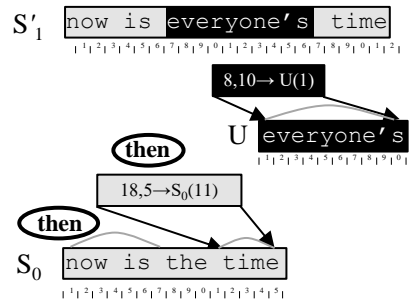

**Fig. 2g**: Writing "everyone's" in place. Compare fig. 2b; note that " time" is now at 18-22    **Fig. 2h**: Reusing $S_0$ twice in the write; compare fig. 2c

An extension of these ideas is to apply a function lazily to a range of addresses by attaching the function to each piece that is part of the range. To actually get the value of an address, compose all the functions on the path from the root to the stored value $x$, and apply the result to $x$. To preserve ordering give them sequence numbers. This is the way Bravo represents character and paragraph properties like bold and centered.

The interface to a piece just maps an index to a value, so the code can be an arbitrary function instead of an array or tree. Some examples: a compressed or encrypted sequence of values, the integers in $[i..j]$, the digits of $\pi$, an audio recording sped up by $2\times$.

Amazon Aurora applies many of these techniques to a cloud database, separating storage completely from the database code. It treats the redo records that contain database writes as the truth; when the database reads a page, storage reconstructs it from the redo records. If there are many of them, it takes a checkpoint just for that page. This drastically reduces write bandwidth, especially taking the cost of replication into account, since you only write the new data, not whole pages.[R109]

**Multi-version state**

The log, checkpoints, and shared index structure can make it cheap to get access to *any* old version of the state, but there are some important cases where you don't need this generality. One of them is a *wait-free* tree structure, which you update by constructing a new subtree and then splicing it in with a *compare-and-swap* instruction `CAS(linkAddress, oldLink, newLink)` that atomically stores `newLink` into `linkAddress` if its old contents is `oldLink`. This is a highly specialized form of optimistic concurrency control where the hardware directly provides the needed atomicity.

Another is the use of epochs to keep a state (or some properties of it, such as the existence of objects) immutable until the end of an epoch. This makes it much easier for concurrent computations to work on the state. Of course the price is that any updates have to be held aside until they can be applied at the end of the epoch, but as long as they commute with other updates, the log is a good place to keep them.

*Snapshot isolation* for long-running transactions keeps one extra version of the state, as of the transaction's start. The code does the inverse of copy-on-write, saving in the snapshot the *old* value of a variable that's written. The transaction is still atomic if it writes only to private variables.

3.8.2 *Indirection*—Take a detour, see the world.

Indirection is in opposition to inlining, but there are many other examples, often having to do with binding a client resource less tightly to the code or objects that implement it. Indirection replaces the direct connection between a variable and its value, $v \to x$, with an indirect connection or link, $v \to u \to x$. This means that you go through $u$ to get to the object, and $u$ can do all kinds of things. It can *multiplex x* onto some bigger object or *federate* it with $y$ so that its own identity becomes invisible. It can *encapsulate x*, giving it a different interface to make it more portable or more secure. It can *virtualize x*, giving it properties its creators never dreamt of. It can *interpose* between $v$ and $x$ to instrument the connection. It can act as a *name* for $x$, decoupling $x$ from its clients and making it easy to switch $v$ to a different $x$.

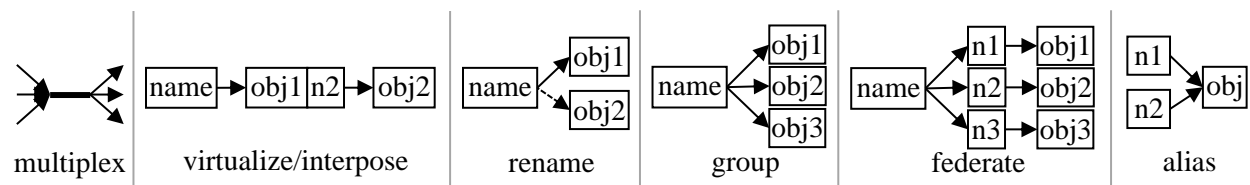

**Fig. 3**: Forms of indirection

**Multiplexing** divides up a resource into parts. The classic example is dividing a communication channel into subchannels, either statically by time, frequency, or code division multiplexing, or dynamically by packet switching. An OS multiplexes files onto a disk or processes onto a CPU. When combined with naming, this makes it easier to move a resource. Routing does this repeatedly; Internet packets, email messages and web page requests (through proxies) all go through several indirections.

**Federation** is almost the opposite, combining several resources into a single one: several disks into one volume, several filesystems into a bigger one by mounting, a sea of networks into the Internet. Load-balancing federates servers: each client sees a single resource, but there are many clients and the balancer spreads the load across many servers.

**Encapsulation** isolates a resource from its host, as a *secure enclave* that keeps the resource safe from the host or a *sandbox* that keeps the host safe from an app. This can be for security, or just to control the environment that the resource sees, as containers like Docker do.

**Virtualization** converts a "physical" host resource into a "logical" guest one that is less limited (virtual memory much bigger than physical memory, missing instructions trapped and done in software) and easier to move (virtual machines not bound to hardware). Usually this is paired with multiplexing, as in an operating system or hypervisor. It can also change the interface, for example with a different ISA on the guest so you can run old programs (*emulation*) or for portability, as with the Java Virtual Machine (JVM). An interpreter can run the guest ISA by executing instructions of the host, or a compiler can translate guest programs to the host ISA either statically, or dynamically using JIT. Other examples: virtual hard disks, overlay networks, NAT, the C library. An adapter can handle a smaller interface change.

Virtualization allows more than one code for the same spec, making it easy to experiment with new codes and to migrate clients from one code to another without any changes and even without downtime. For this to work smoothly, of course, all the codes must implement exactly the same spec. Thus when you assemble a big value out of pieces (for example, by mounting one file system on another) the different pieces can use different code. If the pieces are immutable this is easier.

**Interposing** splices more or less arbitrary code between a client and a service, often to log audit records or to collect information about performance. It's easy to do this for a class, but it's always possible by patching, even at the level of machine instructions. Proxies and content distribution networks such as Akamai do this on a larger scale to distribute load and improve locality.

**Naming** decouples a service such as Twitter from the physical machines that implement it. Such a service uses several levels of indirection: DNS maps `twitter.com` to an IP address, and the Internet delivers packets with that address to a machine, possibly passing through NAT translators and load-balancers on the way. The mapping from file names to physical disk addresses is similar. A copy-on-write file system uses indirection to assemble a big thing out of pieces that are shared or reused; more on this here. It may make sense to cache parts of these mappings. In other cases the mappings are done statically, as between variable names and CPU register numbers in a compiled program, but then the CPU uses renaming to map register numbers dynamically to physical registers. You can name a *group*: a style in a word processor names a group of character or paragraph properties, decoupling the markup from the appearance, and a mailing list, security group or role names a group of people, decoupling the structure of an organization from the current membership. An index makes name lookup or search cheaper. Indirection makes it easier to have *aliasing*: several different $v$'s that map to the same $x$.

A namespace is just a key-value store, with `read` and `write` usually called `lookup` and `set`. Some namespaces lack `enumerate`, because it's constant, impractical to implement (as with DNS), or overlooked by the designer. Likewise, `search` is often missing, or implemented by an external index and a crawler.

**Certificates** use indirection to establish trust.

Sometimes indirection yields not a single result but a *set* from which you choose an element. This is useful for fault tolerance and for load-balancing. BitTorrent is a multi-level example with sets, where you go server → torrent file → tracker → peers/chunks. You fetch the rarest chunk first and become a seed for it, to balance the number of copies of each chunk.

# 4. Process

*The most important single aspect of software development is to be clear about what you are trying to build.* —Bjarne Stroustrup[Q78]

*Systems resemble the organizations that produce them* (paraphrased). —Melvin Conway[Q13]

*If you can't be a good example, then you'll just have to be a horrible warning.* —Catherine Aird[Q1]

*SOFTWARE IS HARD. ... Good software ... requires a longer attention span than other intellectual tasks.* —Donald Knuth[Q46]

*There are only two ways open to man for attaining a certain knowledge of truth: clear intuition and necessary deduction.* —René Descartes[Q17]

The summary is STEADY by AID with ART: **A**rchitecture, **A**utomation, **R**eview, **T**echniques, and **T**esting are the essentials of process. I don't have much personal experience with this; I've never run a team. But I have watched a lot of systems being developed, with teams that range in size from six to several thousand people. If you find yourself working on a team that breaks the rules in this section, look for another job.

You can build a small system with willpower: one person keeps the whole design in their head and controls all the changes. You can even do without a spec. But a system that's bigger (or lives for a long time) needs process. Otherwise it's broken code and broken schedules. Process means:

- Architecture: Design that really gets done, and documented so that everyone can learn it.
- Automation: Code analysis tools (very cheap for the errors they can catch) and build tools.
- Review: Design review—manual, but a much cheaper way to catch errors than testing.
- Review: Code review—manual, but still cheaper than testing.
- Testing: Unit and component tests; stress and performance tests; end-to-end scenarios.[R15]

Here's another take on the same story, where I haven't labeled the components from the acronym, but I have italicized the ones that don't seem to fit under it:

- Process: have a spec; use source control, build with one command, build daily, track bugs in a database; fix bugs first; *have a truthful schedule*.
- Coding: *minimize distractions*; buy the best tools; design for test, build tests for the code.[R97]

None of this will help, though, if the goal is badly conceived. If your system isn't going to be yummy, it had better at least be useful. If it's entering a crowded field, it needs to be a *lot* better than the market leaders. If there's a strong ecosystem of languages and applications in place, build on it rather than fighting it. And usually simplicity is key: if your system does one thing well, it's easier to sell and easier to build. If it's successful it will expand later. Some well-known examples:

- Dropbox just syncs a subtree of the file system, unlike Groove and Microsoft Windows Live Mesh, which do a lot more very confusingly and less reliably.
- The C language stays as close to the machine as possible.
- The original HTML gives you links, text with simple formatting, and bitmap images.
- Unix gives you files as byte strings, path names, processes linked by pipes, and the shell.
- Internet email gives you string names rooted in the DNS as addresses, and plain text or HTML with extensible attachments for the contents.
- Twitter gives you short tweets that can go to millions of followers.

The symbiotic relationship between a platform and its applications can take one of two forms:

- **Controlled**: The platform only accepts applications that fit its self-image, with the goal of coherence and predictability for the whole ecosystem. Apple does it this way, and specialized systems like SQL databases or Mathematica get a similar result technically by making it much easier to build the kind of applications they want than to do anything else.
- **Wild and free**: The platform accepts anything, and it's up to the market to provide whatever coherence there is. Windows does it this way. Android is in the middle.

Successful systems last, and you want your system to succeed, right? You don't get to rewrite it from scratch; that's not compatible with agile development and shipping frequently. And the shipping code reflects lots of hard-won knowledge, much of which isn't written down and has slipped out of the team's heads (or the team has changed). This is why it pays to think through the initial design, and to put as much code as possible into modules with clean interfaces, especially performance-critical code. It also pays to clean up messy code when you need to change it; IDE tools can help. If the system is too slow, first measure and then work on the few modules need to be fast and predictable. Your system doesn't have that structure? Then you have incurred *technical debt*.[R66] The solution is to change it until it does; those changes are expensive, but they have enduring value. Then keep it that way. And keep shipping.[R98]

»Intel Itanium. When Intel made a big bet on a VLIW (Very Long Instruction Word) design for its 64 bit Itanium architecture to replace the x86, the performance predictions were apparently based on a single hand-coded inner loop, 30 instructions long, since they didn't have the optimizing compiler working.[R23] Most real programs turned out to be much less amenable. Usually chip designs are based on extensive simulation of real workloads.

»Windows phone. Microsoft was early to the smartphone market, but was completely blindsided by the iPhone and Android, and ended up abandoning the Windows phone after spending billions of dollars. It failed because it didn't have apps, and that was because a good phone came to market too late to compete with the incumbents. Earlier

Windows phones were not good enough to catch on, and this was because they were built by a weak team. The team was weak for two reasons:
- First, there was no business model; it doesn't work for a large company to license software at $10 per phone into a market of 50 million phones. Apple's business model was selling hardware, which Microsoft didn't do at the time, and Android's business model was ads, which Microsoft didn't understand.
- Second, Microsoft is a horizontal company. Since the cellphone market was controlled by the major carriers, this meant that they all had to be satisfied by the design. But carriers have no clue about design or about software. The only reason that AT&T allowed Apple to control the iPhone design is that they were desperate, losing badly to Verizon in wireless, and they were not afraid of Apple. They never would have given Microsoft this control.

First-class engineers didn't want to work in a team facing these two problems.

Later it became clear that a smartphone was strategically important, and management assigned an A-team, but by then it was too late to create a healthy app ecosystem in competition with Apple and Android. It's ironic that for several years Microsoft refused to recognize this, since the main reason that MS-DOS and Windows were so dominant was the carefully cultivated app ecosystem.

## 4.1 Legacy

A successful system attracts lots of clients, and when the clients are software this causes problems for the next version. Of course the clients' developers don't like it when the spec changes, but in addition they often don't pay attention when the spec improves or adds features. The reason is that there are lots of installations of the system, and a client that *depends* on a new version won't work on an old one. But people are slow to upgrade their systems because it's often a hassle (though cloud-based automatic update systems help) and they fear the new bugs of the new version. So improvements to a platform pay off a lot more slowly than you might expect. User interface changes don't have this problem, as long as people really like them.

## 4.2 When tools are weak

It's much easier to build a system on a platform that has the functions, performance, scaling and fault tolerance you need, with good tools such as a language specific editor, a static analyzer, source code control, a build system, a debugger, a profiler, test and deployment automation, and telemetry to collect and analyze operational data at scale. If you don't have good tools at hand, try to acquire them. If that fails, think seriously about building basic ones. This seems expensive, but handwork is even more expensive. Don't be too ambitious, though; you're probably not being paid for tool building. The biggest problem is almost always to make the performance good enough that the tool can be used as serious building material.[R54]

# 5. Oppositions

Finally, here is a brief discussion of each opposition. These are not alternatives but extremes; the text explores the range of possibilities between the extremes. The brackets refer to relevant goals.

**Simple ↔ rich, fine ↔ features, general ↔ specialized [S Y]**

   —KISS: Keep It Simple, Stupid. Do one thing well. Don't generalize.
   —Don't hide power. Leave it to the client. Make it fast. Use brute force.

*If in doubt, leave it out.* —Anonymous

*The cost of adding a feature isn't just the time it takes to code it, [it's the] obstacle to future expansion. ... Pick the features that don't fight each other.* —John Carmack[Q10]

*Exterminate features.* —Chuck Thacker[Q82]

*Any intelligent fool can make things bigger and more complex. It takes a touch of genius—and a lot of courage—to move in the opposite direction.* —E. F. Schumacher[Q73]

*'Free' features are rarely free. Any increase in generality that does not contribute to reliability, modularity, maintainability, and robustness should be suspected.* —Boris Beizer[Q5]

Systems are complicated because it's hard work to make them simple, and because people want them to do many different things. You can read a lot about software bloat, the proliferation of features in browsers and in rich applications like Word and Excel. But almost every feature has hundreds of thousands of users at least. The tension between keeping things simple and doing a lot is real, and there is no single right answer, especially for applications that interact with users.

 Still, it's best to add features and generality slowly, because:
– You're assuming that you know the customers' long-term needs, and you're probably wrong. It's hard enough to learn and meet their immediate needs.
– It takes time to get it right, but once it's shipped legacy customers make it hard to change.
– More features mean more to test, and more for a bad guy to attack.

So why do systems get overambitious? Because there are no clear boundaries,[Q9] as there are with bridges for example, and programmers are creative and eager to tackle the next challenge.

 But features that have a lot in common can add power without adding too much complexity; the best design is a single mechanism that takes different parameters for the different features. So a search engine can index many different data types, a webpage can include text, images and video, or an email program can keep a calendar. A user interface feature that just invokes a sequence of existing features is less dangerous because it only complicates the UI, not the rest of the system.

 For software whose clients are other programs, the solution is building programs on components. A single component should do one thing, and its code should do it well and predictably so that clients can confidently treat it as a primitive building block; beware of components that don't have these properties. Building one of these components is a lot of work. It's worth doing if the component is critical for your system, or if it's part of a platform like an operating system, a browser or a library where it will have lots of clients.

 Even better is a complete set of such components, with both the functionality and the performance you need to write programs for a significant application domain. Then a client can do a lot without writing much code and without much cleverness. Some examples:
– key-value stores;

- Unix shell programming on top of primitives like `diff`, `sort`, `grep`;
- graphics on top of BitBlt, spline curves, and compositing;
- mathematics systems like Mathematica and Julia.

This takes both lots of work and deep insight into the application domain, but the payoff is big.

The opposite of doing one thing well is doing a lot of things badly. Assuming reasonably competent engineering, this usually happens because of excessive aspirations. Some striking examples:

- The original Arpanet spec mandated that a packet handed to the network should be delivered reliably (rather than the best-efforts delivery of the Internet, which is free to drop packets when the going gets tough). This turned out to make the network much slower than expected.
- Many systems that rely on remote procedure calls have run into trouble because RPCs look like ordinary procedure calls, but in fact are much more expensive, have very unpredictable latency, and can fail completely when the network or the target system fails.
- A crucial feature of the web is that it makes no attempt to prevent broken links. They are annoying, but even *trying* to prevent them would make things much more complicated.
- »Air traffic control. The FAA's air traffic control modernization system, the Advanced Automation System, begun in 1981, was abandoned a decade later, after several billion dollars had been spent. The leading causes of this failure were a wildly unreasonable requirement for 99.99999% reliability (3 seconds per year of downtime) and the assumption that the contractor was an enemy to be regarded with suspicion, rather than a partner.[R106,R17]
- »Orange Book. The DoD's Orange Book program to deploy multilevel secure systems failed miserably. Vendors developed several such systems, but essentially no one in the DoD bought them in spite of formal requirements to do so, because they were more expensive, slower, and less functional than alternative, less secure systems. It was much more expedient to get a waiver from the requirement.[R67]

**Perfect ↔ adequate, exact ↔ tolerant [S T D]** —Good enough. Flaky, springy parts.

*Worse is better.* —Richard Gabriel[Q29]

*The best is the enemy of the good.* —Voltaire[Q88]

*It's impossible to make anything foolproof, because fools are so ingenious.* —Anonymous

This is not about whether there is a precise spec, but about how close the answer needs to be to an ideal result. "Close" can take different forms: a tolerance or a probability of being right, results that may just be wrong in some difficult cases, or a system that behaves well as long as its environment does. Some examples:

*Tolerance or probability:*

- Available 99.5% of the time (down no more than one hour per week), rather than 100%.
- Response time less than 200 ms with 99% probability, rather than always.
- A 98% hit rate in the cache on the Spec benchmark, rather than 100%.
- A schedule with completion time within 10% of optimal, rather than optimal.

Such properties usually come as statistics derived from measuring a running system, or from a randomized algorithm.

*Wrong in difficult cases:*
- Words are hyphenated if they appear in a hyphenation dictionary, rather than always.
- Internet packets are discarded when there's too much congestion.
- Changes to DNS may not appear immediately at all servers (because it uses eventual consistency for high availability).
- A database system may fail, but it recovers without losing any committed work.
- An operating system usually just tries to avoid disaster, rather than using the hardware resources optimally.

*Friendly environment:*

Every system at least depends on its host to execute its instructions correctly, but often the system can be simpler or cheaper by assuming more about its environment:
- Data is not lost as long as the power doesn't fail.
- Your files are available if you have a connection to the Internet.
- Faces are recognized reliably if the lighting is good enough.
- The routine works if its precondition is satisfied; otherwise all bets are off.

The environment is not just the host you depend on; it's also your clients. If they are not too demanding, your system may be adequate even if it doesn't satisfy an ideal spec.

**Spec ↔ code [P S]**

—Keep secrets. Good fences make good neighbors. Free the implementer.
—Embrace nondeterminism. Abstractions leak.

*Don't tie the hands of the implementer.* —Martin Rinard[Q68]

*Writing is nature's way of letting you know how sloppy your thinking is.* —Richard Guindon[Q31]

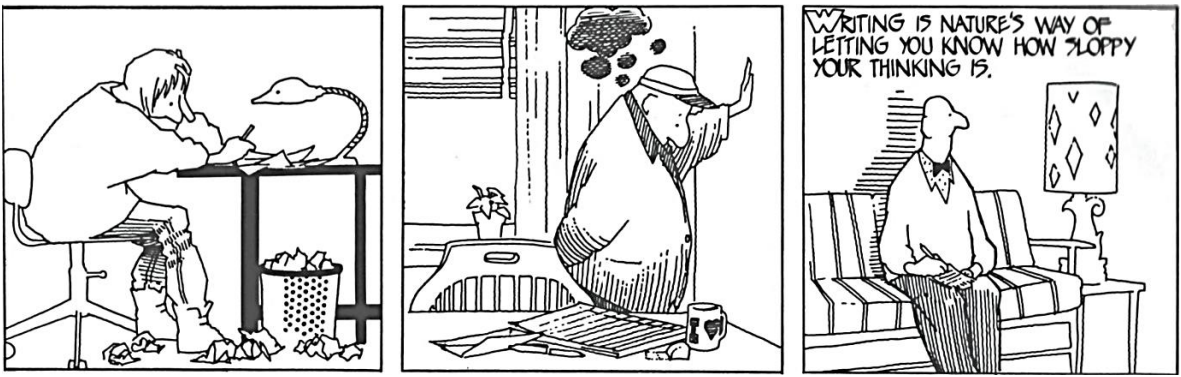

*We're much better at building software systems than we are at predicting what they will do.* —Alan Kay[Q40]

A spec tells you *what* a system is supposed to do, and the code tells you *how*. Both are described by actions; how do they differ? A spec constrains the visible behavior of the system by saying what behaviors (sequences of steps) are acceptable or required. A spec is not a program, and the right language for writing it is either English (if the design ideas are still too vague to be expressed precisely) or mathematics.

The code is executable, but it still may not be a program you can run; it may be an algorithm such as Quicksort or Paxos, described in pseudocode that abstracts from the details of how the machine represents and acts on data. Pseudocode can have a precise definition and a toolchain.[R58] Code is less likely to be nondeterministic, except that concurrency leaves open the ordering of actions in different threads: at the top level a concurrent program keeps freely choosing a ready thread's action to run: **while true do** var $i\ |\ thread_i$ is ready **in** $thread_i.pc.action$.

Sometimes you can give a safety property for the code rather than demand liveness in the spec. For example, absence of deadlock is safety (the bad thing that mustn't happen is a cycle of processes waiting for each other) but continued progress is (stronger) liveness property, and keeping public outputs untainted by secrets is safety but non-interference is liveness.

**Imperative ↔ functional ↔ declarative [S E]** —Make it atomic. Use math. Say what you *want*.

The many styles of programming can be grouped into three broad classes: imperative (with object-oriented as a subclass), functional and declarative. These three styles vary along the axis of expressing detail. The other important axis is modularity: how easy it is to change one part of the program without affecting the rest. The essential idea for modularity is abstraction. Object-oriented programming is one way to code abstraction; usually the objects have state and the code is imperative, though it needn't be.

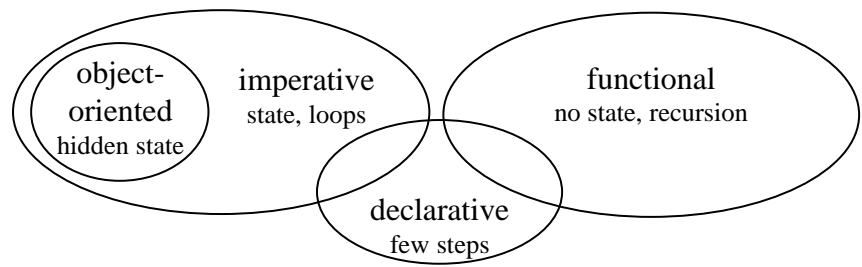

**Fig. 4**: Styles of programming

An imperative program (for example, one written in Java or C) has a sequence of steps and a program counter, as well as named variables that the program can read or write. Interesting programs take lots of steps thanks to loops or recursion. If big chunks are atomic (can be viewed as a single step), the code is easier to understand and more likely to be right. Most computing hardware is imperative.

A functional program (perhaps written in pure Lisp or in the functional subset of ML or Haskell) has function calls instead of steps, and immutable values bound to function parameters or returned from the calls instead of state variables; it also has implicit state that keeps track of how the nested function calls are evaluated. Another way to say this: its steps simplify terms rather than changing state, though programmers don't think about it that way. Interesting programs have recursive functions, so they can make lots of calls. Real languages aren't purely functional because when the values are big and changes are small, it's hard to map a functional program efficiently to the hardware's step-by-step execution; you can guide this process by writing part of the program imperatively. Inversely, you can embed immutable data structures in an imperative language, and a library like `immutablejs` can make this efficient using the techniques of becoming. The most widely used programming languages are functional: spreadsheets and database query systems. However, they are special-purpose and have escapes to imperative code. Single-assignment imperative programs are also functional.

»Haiku. Jim Morris explained the problem with functional languages by analogy with haiku and karate. All three are "unnatural", like all disciplines; that's not the issue. Haiku have great aesthetic appeal, but they need an appreciative audience. Karate will work in a barroom brawl even if no one else knows it. Programmers care about results.[R75]

I agreed to write a piece for Alan Kay's 70th birthday celebration,[R60] and recklessly provided the title "Declarative Programming"; this seemed safe, since everyone knows that declarative is good. When it came time to write the paper I realized that I didn't actually know what declarative programming is, and searching the literature didn't help much. I finally concluded that a program is declarative if it has few steps; this makes it easier to understand (as long as each step is understandable), since people are bad at understanding long sequences of steps. Often it's also easier to optimize, since it doesn't commit to the sequence of steps the machine should take. The *open* action of the file system spec is a small example.

Powerful primitives help to make a program declarative; for example, code to compute a transitive closure has lots of steps, but a transitive closure primitive is a single easy step. The SQL query language for relational databases has many such primitives, as does HTML as an abstract description of a desired webpage.

Another declarative style is to write the code as a constrained optimization problem such as a linear program, or as the solution to an equation: $x \mid p(x)$. For example, the median of a multiset $s$ of reals is

$$x \mid (\exists l, h \mid l \cup h = s \text{ and } l.\text{size} = h.\text{size} \text{ and } l.\max \leq h.\min \text{ and } x = (l.\max + h.\min)/2)$$

*Program synthesis*

It's obvious how to run a declarative program written with powerful primitives; the only big question is how much to optimize. SQL is an example, and query optimizers make it clear that you can get good performance for many workloads. It's less clear how to get a reasonably efficient executable program from an equation or optimization; exhaustive search for a solution works in principle if the solution space is finite, but it's too slow. Program synthesis is sometimes the answer; for imperative or functional programs it's called *inductive programming*, and in practice usually requires exhaustive search in some space of possible *programs*, checking each one against the declarative spec.[R44] This works when there's a good program that's not too long, so it's important to have powerful but efficient primitives that keep the programs short. The spec can be extremely partial, perhaps just a few input-output examples as in Excel's FlashFill feature. An important application is automating system management by asking the manager just to give a desired state, leaving the system to figure out a sequence of changes that gets there, and report if it can't.

A different kind of program synthesis is *machine learning*, in which the program is a function expressed as a multilevel neural net, and learning consists of adjusting the weights by stochastic gradient descent to bring the output as close as possible to the training data. This works well surprisingly often, but it doesn't give any guarantees about the results.

The paradigm for program synthesis is "signal + search = program". The signal is some sort of partial spec such as I/O examples in FlashFill, demonstrations of desired behavior, natural language, or training data. The search is optimizing over some sort of DSL (which could be a neural net), guided by the partial spec, past successes on similar problems, user hints about how to subdivide the problem, or inverting a probabilistic program that generates the training data to recover the model that drives it, as in Alexandria.[R114] If the program is differentiable, gradient descent can make the search much faster.

**Immutable ↔ append-only ↔ mutable [S]** —Make it stay put.

Data is much easier to deal with if it doesn't change:
- Concurrency is easier: the data doesn't move around under you, and your local copy doesn't need to be updated.
- Caching data (or functions of it) is easier: the cache entry can't become invalid.
- Reasoning is easier: an expression always has the same value, as it does in mathematics.

Functional programs are simpler and more reliable than imperative ones because the variables are immutable.

So it's a no-brainer? No, because if you want to change something (say, your sister's salary) you have to make a copy of the entire HR database, and if it's big that's expensive. The only way around this in general is to keep multiple versions with some form of MVCC, moving from the being to the becoming view of the state the way a library like `immutablejs` does; if you want the value of a variable, you have to specify the version. Now you can think in terms of big transactions that make meaningful state changes. The main design challenge is the contention when many of these run concurrently; often some form of eventual consistency can help with this. Usually you optimize the data structure so that getting the most recent version is fast.

Sometimes it's enough to make the data append-only, writing only to the end of a sequence rather than updating in place; the length of the sequence determines the version. Log-structured file systems do this.

An important special case is to keep changes in a buffer, consult it on every read, and merge it into the main data structure when it's convenient (often at the end of an epoch) or when a transaction commits (with OCC). CPU write buffers do this, and the Bw-tree elaborates the idea by keeping the buffers local to the deepest tree node being changed, out of the way of most reads.[R68] Another important case is a data warehouse: you load lots of data into a relational database, but don't update the data once it's been loaded. A third is audio or video: rather than processing it in real time, record it as a sequence of samples indexed by time and use the techniques of becoming to manipulate it; this is how audio/video editing is done.

Other aspects of immutability are in the dynamic-static and imperative-functional oppositions.

**Precise ↔ approximate software [T D]** —Get it right. Make it cool. Shipping is a feature.[Q57]

*Unless in communicating with [a computer] one says exactly what one means, trouble is bound to result.* —Alan Turing[Q87]

*It is better to be vaguely right than precisely wrong.* —Leonard Lodish[Q50]

*There's no sense being exact about something if you don't even know what you're talking about.* —John von Neumann[Q90]

Broadly speaking, there are two kinds of software, precise and approximate, with the contrasting goals "Get it right" and "Get it soon and make it cool."

Precise software has a specification (even if it's not written down very precisely), and the customer is unhappy if the software doesn't satisfy its spec. Obviously software for controlling airplanes or nuclear reactors is precise, but so are word processors, spreadsheets, software for handling money, and the Internet packet protocol. The spec might be nondeterministic (the Internet might drop packets), partial (Excel should evaluate its formulas correctly) or opaque (Word should generate the same paragraph numbers today that it did 10 years ago), but that doesn't make it imprecise.

Approximate software, on the other hand, has a very loose spec, or none at all; the slogan is "Good enough." Web search, retail shopping, face recognition, and social media are approximate. This kind of software does have a (weak) spec, but the customers don't think in those terms.

Approximate software is not better or worse than precise, but they are very different, and it's important to know which kind you are writing. If you wrongly think it's precise, you'll do extra work that the customers won't value and it will take too long. If you wrongly think it's approximate, the customers will be angry when code doesn't satisfy the (unwritten) spec they counted on.

**Dynamic ↔ static [E A]** —Stay loose. Pin it down. Shed load. Split resources.

A computer is infinitely flexible, but a program is not; both what it does (the spec) and how (the code) are more specialized. Yet the code can be more or less flexible, more or less able to adapt to changes in itself or in the environment. Flexibility costs because you have to check more things at runtime, but it can save if the checks let you skip some work. Code that takes advantage of things that stay constant is more efficient if they really are constant, and static checking automatically proves theorems about your code before you ship it. To some extent you can have both with *just-in-time* (JIT): make a static system based on the current code and environment, and remake it if there are changes. Like retry, this performs well if changes are rare.

There are (at least) four aspects of this opposition: interpret vs. compile, indirect vs. inline, scalable vs. fixed, and online vs. preplanned resource allocation. Another name for it is late vs. early binding.

Compiling commits the code to running on a host that is usually closer to the hardware. The compiler chooses how data is represented, and often it infers properties of the code (examples: at this point $v = 3$ always; the address of $a[i,j]$ increases by $n$ each time around this loop) and uses them to optimize. It may do *trace scheduling*, using information from past runs or heuristics about common code patterns to predict code properties (in this JavaScript program, $i$ is usually an integer).[R39] These predictions must be treated as hints and checked at runtime, with fallback to slower code when they are wrong. Together with JIT, trace scheduling can adapt a very general program to run efficiently in common cases.

A different aspect of the dynamic-static opposition is resource allocation, and scheduling in particular. CPUs and operating systems can allocate resources online to a sequence of tasks that's not known in advance (using caches, branch prediction, asynchronous concurrency, etc.), but if you know the sequence you can do this work just once. Examples: resources reserved for a real-time application; the graphics pipeline of a GPU; a *systolic* array in which work items pass through a sequence of processors, taking the same amount of time in each one with no queuing.[R56] Storage allocation is similar; you can do it dynamically, but static allocation (splitting up the storage) is cheaper if you know the sizes in advance or can guess them well. And when it fails, it's much easier to figure out why.

If scheduling is dynamic, continuity (in the form of many small jobs of about the same size) makes it work better, especially when the resources are not identical and it's hard to predict how long a job will take. Individual scheduling decisions are less critical, and stragglers hurt the overall performance less. Examples: the Cilk parallel computing system, virtual nodes for DHTs, parallel sorting. Continuity helps with resource allocation in general: it's easier to do a good job when you have lots of similar storage blocks, network packets, or whatever, because aggregates of complex things can be simple. Dynamic allocation can have a surprisingly long tail of bad performance, though, even if it has good average performance.[R31]

Being dynamic is an aspect of adaptability, and scaling and extensibility are two other ways in which systems can be dynamic.

**Indirect ↔ inline [E I]** —Take a detour, see the world. Use what you know.

*Any problem in computing can be solved by another level of indirection.* —David Wheeler[Q91]
*Any performance problem can be solved by removing a level of indirection.* —M. Haertel[Q32]

**Indirection** is a special case of abstraction that replaces the direct connection between a variable and its value, $v \to x$, with an *indirect* connection $v \to u \to x$, often called a *link*; the idea is that ordinary lookups to find the value of $v$ don't see $u$, so that clients of $v$ don't see the indirection. You can change the value of $v$ by changing $u$, without changing $x$. Often $u$ is some sort of service, for example the code of a function, reached indirectly by jumping to the code for the function, or even more indirectly by obtaining the address of the code from a "virtual function table" if the function is a method of a class; this gives the most flexibility, since you can run arbitrary code in the service. The link doesn't have to be explicit; it could be an *overlay* that maps only some of the possible $v$'s, like a TLB or a cache; the code for this detects the overlaid $v$'s with a hash table or content addressable memory, or a trap on a direct link like a page table entry. Often indirection is lazy: if you never need the variable's value you don't pay for it. Simple examples of indirection are the hard links, indirect blocks, symbolic links and mount points of a file system. Virtualization is an important subset of indirection.

**Inlining** replaces a variable $v$ with its value $x$. This saves the cost of looking up $v$, and the code can exploit knowing $x$. For example, if $x = 3$ then $x + 1 = 4$; this saves an addition at runtime. If $v$ is a function you can inline its code, avoiding the control transfer and argument passing, and now you can specialize to this particular argument. But inlining takes more space and makes it hard to change the function's code. The tradeoff is the usual one between dynamic and static: flexibility versus being able to do static analysis and optimization; doing it dynamically is a special case of JIT. Inlining generalizes to *currying*, which replaces $f(x, y)$ with $f(x)(y)$. At first this looks like a step backward because making a new function $f(x)$ is costly, but it saves the cost of passing the argument $x$, and this can pay if there are many calls to $f$ with $x$ as the first argument, for example in a loop. Also, the code of $f(x)$ can exploit the known value of $x$; this is program *specialization*.

Another way to look at it is that inlining is evaluating a function composition: $v \to u$ composed with $u \to x$ yields $v \to x$. If you do this dynamically, saving the composition is a form of caching; it's often called *snapping* the link. To make it work you need a way to get control when following an unsnapped link; this is easiest when it's a control link (a jump), but it could be something that traps such as a page table entry. Examples: TLBs, shadow page tables, dynamic linking, JIT compiling of interpreted code, content distribution networks. The snapped link could be a hint that has to be checked, such as the compiled code address for a method call in Smalltalk, where the check is that this code was used for the object's class last time it was called.[R32] To invalidate the snapped link, use the usual techniques for caches. Soft state in networking is similar.

**Time ↔ space [E]** —Cache answers. Keep data small and close.

*Time is nature's way of keeping everything from happening at once.* —Ray Cummings[Q16]

There's often a tradeoff between execution time and storage space: *precompute* or save something to speed up a later computation. The reasons are to do less total work (a form of speculation) or to reduce latency. Some examples:

- An index into a set of data items makes it much faster to retrieve the items that have some properties, but the size of the index grows linearly with the size of the set and the typical number of properties. The Google search engine is an extreme example; it indexes hundreds of billions of webpages and took up more than $10^{17}$ bytes in April 2019.[R42]
- Caching the result of a function call (memoizing) saves time if the call is repeated, but takes space for the cache.

- Precomputing a "rainbow table" of size $R$ speeds up the search to invert a hashed password and find its plaintext from $N$ (the number of possible passwords) to $N/R$ (but takes time $N$ as well).[R79]
- Many other algorithms run faster with precomputation. For example, binary space partitioning speeds up graphics rendering by precomputing information about the ordering of overlapping polygons in a scene.[R77]

On the other hand, getting data from storage needs communication, which can't be faster than the speed of light, so the latency to access $n$ bits grows at least like $\sqrt[3]{n}$ in our 3-D world. Put another way, fast computing needs locality, and it's easier to make small things local, so saving space can also save time. To get locality as a system scales up, you have to shard: divide the state and the work into independent tasks.

**Lazy ↔ eager ↔ speculative [E]** —Put it off. Take a flyer.

*When you come to a fork in the road, take it.* —Fort Gibson New Era[Q99]

The common theme is to improve efficiency by reordering work. The base case is *eager* execution, which does work just when the sequential flow of the program demands it; this is the simplest to program. **Lazy** execution *defers* work until it must be done to produce an output, gambling that it will never be needed. It can pay off in lower latency because it first does the work that produces output, and in less work if a result turns out not to be needed at all. Laziness has been studied mainly in programming languages such as Haskell, but it's used much more widely in systems.

Indirection is lazy as well as dynamic—if you never need the value of the name, you never pay the cost of following the link. Other examples of laziness are write buffers, which defer writes from a cache to its backing store; redo logging, which replays the log only after a crash; eventual consistency, which applies updates lazily and in an arbitrary order until there's a need for a consistent result; and computing in the background, using resources that would otherwise be idle. Often representing the state as becoming is lazy.

If a function is a predicate (the result is `true` or `false`) then it defines a set lazily: the values for which it returns `true`. You can list the members of the set explicitly (an *extensional* definition), or represent the predicate as code that enumerates the set or tests for membership (an *intensional* definition). It's the difference between "{Cain, Abel, Seth}" and "the children of Adam and Eve." The code for a predicate is often called a pattern (as in `grep`) or a query (as in SQL). To avoid storing the set, package the code in an *iterator*, a routine that returns an element of the set each time it's called, or that passes each element of the set to a client function.

More generally, it's lazy to represent a function by code rather than as a set of ordered pairs. Of course if the set is infinite then code is the only option. Pushing this idea farther, to defer the execution of some code, wrap it in a function and don't invoke it until the result is needed.

Sometimes lazy execution postpones work until it's convenient, rather than necessary; this is a form of batching. Read-copy-update and generational garbage collection use the end of an epoch to trigger execution. B-link trees and write buffers accumulate small changes until there are enough of them to make cleanup worthwhile. Sloppy counters keep a per-core fragment of a counter to make incrementing fast, and use a demand for a true total value to trigger consolidation; if all you need is a monotonic counter you can even read the total without blocking any incrementors.

**Speculative** execution does work *in advance*, gambling that it will be useful. This makes sense if you have resources that are otherwise idle, or to reduce latency in the future. Prediction is the most common form of speculation, for example when a storage system prefetches data from memory to cache or from disk to memory, or when a CPU predicts which way a branch instruction will go. Caching speculates that an entry will be used before it has to be replaced. Exponential backoff in networks and optimistic concurrency control in databases speculate that there will be little contention. Precomputing speculates that there will be enough later tasks that will run faster.

If the gamble fails and the result isn't needed, the speculative work is wasted. The result might also be stale: no longer valid because some inputs have changed since it was computed. If you detect this at the time of the change, you can discard the speculative result or update it. The other possibility is to treat it as a hint and check that it's still valid before using it. Memoizing a function call is the same idea.

A twist on speculation is *work stealing*, where one agent takes up work from another agent's queue. Usually the agent is a thread, as in the Cilk system for fine-grained concurrency or in helpers for wait-free data structures, but it might be recovery code processing a redo log after a crash. This idea also works for cleanup operations that don't change the abstract state: rather than doing the cleanup right away or in the background, a client that notices a need for it can just do it.

Usually laziness or speculation keeps the program's results unchanged. This is simplest if the parts being reordered commute. They do in a functional program, but code with side effects may not. Sometimes you settle for sloppy results, for example with eventual consistency.

**Centralized ↔ distributed, share ↔ copy [E D]** —Do it again. Make copies. Reach consensus.

*A distributed system is one in which the failure of a computer you didn't even know existed can render your own computer unusable.* —Leslie Lamport[Q48]

If you have a choice, it's better to be centralized. Distributed systems are more complicated because they have inherent concurrency and partial failures, and they have to pay for communication. But they are essential for serious fault tolerance, and for scaling beyond what you can get in a single box. A well-engineered distributed system has much better availability and much more total performance than any centralized system that is likely to be built, because it can be built out of commodity components. Any big or highly fault-tolerant centralized system will have a small market; witness supercomputers, and fault-tolerant systems since the decline of Tandem and Stratus. Hence it will be expensive—there are only a few sales to pay for all the engineering.

A distributed system needs fault tolerance because it has to deal with *partial failures*; you don't want to crash the whole system when one part fails, and in a big distributed system there are always some failed parts. This means that there has to be *redundancy*: retry for communication and replication for the data. For the latter the essential primitive is fault-tolerant consensus. But even a very large system can be centrally managed (in a fault-tolerant way) because management doesn't require that much computing or data; this is how big cloud systems like AWS and Azure and big search engines like Google and Bing work. The jargon for this originated in networking: separating the control plane from the data plane. Sometimes politics prevents centralization, as in the Internet. Sometimes it happens anyway, as with DNS.

**Fixed ↔ evolving, monolithic ↔ extensible [A I]**

>—The only constant is change. Make it extensible. Flaky, springy parts.

*No matter how far down the wrong road you have gone, turn back now.* —Turkish proverb

*Always design your program as a member of a whole family of programs, including those that are likely to succeed it.* —Edsger Dijkstra[Q20]

*It's cheaper to replace software than to change it.* —Phil Neches[Q56]

*Success is never final.* —George Starr White[Q92]

*It is a bad plan that admits of no modification.* —Publilius Syrus[Q80]

*If you want truly to understand something, try to change it.* —Kurt Lewin[Q49]

The classic way to develop software, known as the waterfall model, is sequential: figure out what's needed (requirements), design something that meets the needs, write code to implement the design, test it, and deploy it. This implicitly assumes (among other things) that the needs are known at the outset and don't change, but this is seldom true. Often the customer's needs are unclear, and successful systems live for a long time, during which needs change. Just thinking hard is usually not enough to make unclear needs clear, because you aren't smart enough and don't know enough about the customer. It's better to build a prototype, try it out, improve it. Agile and spiral models for software development do this, trying to converge on a system that does what the customers want by taking fairly small steps, either down a gradient toward the goal or repeating the sequence of the waterfall steps.[R38]

A successful system must do more—it must adapt and evolve, because needs change as people see ways to make it do more, as the number of users grows, as the underlying technology changes, and as it works with other systems that perhaps didn't even exist originally. Evolution requires modularity, so that you can change parts of the system without having to rebuild it completely. Interfaces allow clients and code to evolve independently. These are aspects of divide and conquer.

Evolution is easier with *extensibility*, a well-defined way to add certain kinds of functionality. This is a special form of modularity, and it needs a lot of care to keep from exposing secrets of the code that you might want to change. Examples:

- You can add new tags to HTML, even complicated ones, and old code will just ignore them.
- Most operating systems can incorporate any number of I/O drivers that know about the details of a particular scanner, printer, disk, or network.

- Inheritance and overloading in programming languages like Smalltalk and Python make it convenient (if dangerous) to add functionality to an existing abstraction.
- Defining policy for a system specializes its mechanism for particular needs.
- Backward compatible changes enable new things without breaking old ones. The evolution of ethernet, C++ and HTML show how far this can be taken. It's much easier if interfaces and persistent data have explicit version identifiers that are always visible.
- A platform that supports applications lets you build whole new things on top of it.

Another way to extend a component is to let the client pass in a (suitably constrained) program as an argument; for example, a search engine can take a parser for an unfamiliar format, and an SQL query can take a user-defined function as a selection predicate. You can do this without pre-planning by patching, but it's tricky to maintain all the code's invariants.

**Evolution ↔ revolution [A]** —Stay calm. Ride the curve. Seize the moment.

Usually things change slowly, by small improvements on what's there already, and dramatic innovations fail; expensive examples are phase change memory, photographic data storage, the Intel iAPX 432 and Itanium, the DEC Alpha, Windows Vista. But sometimes the environment changes enough that a big change can succeed, and this happens more often in computing than in most fields since Moore's Law means that the cost of computing drops fast. The result has been a "virtuous cycle" in which cheaper computing opens new applications, expands the market dramatically and justifies more investment to make it cheaper still. Obvious examples are personal computers, the Internet, the web and cloud computing. Very few companies have ridden these waves and thrived in computing for more than a couple of decades.

Often someone has an idea that is new to them, but was tried 30 years ago and failed. Things are different now, so perhaps it can succeed, but you need to understand why it failed earlier. Striking examples are virtual machines and the web (as the successor to hypertext).

**Policy ↔ mechanism [A]** —Change your mind.

*When the facts change, I change my mind. What do you do, sir?* —Paul Samuelson[Q71]

The mechanism is what the system *can* do, determined by its specs and code, and the policy is what the system *should* do: the control system for the mechanism. Policy is different for each installation and typically changes much faster than the code. Administrators, rather than engineers, set policy, and they think of it as part of the spec. It should give them as much control over the mechanism as possible, bearing in mind that policy is more likely to be wrong because most administrators are not good programmers.

The most elaborate example of the distinction is in security, where the mechanism is access control and the policy is what principals should have access to what resources. Other examples: policy establishes quotas, says how much replication there should be, or decides what software updates should be applied. Policy is an aspect of system configuration, which also includes the hardware and software elements that make up the system and the way they are interconnected. Historically all these things were managed by hand, but cloud computing has forced automation.

*Application resources* are constants that sit between policy and mechanism: fonts, colors, search paths etc. You can think of them as part of the code that you can change without changing any of the executable instructions, or as parameters to the code supplied when it is installed or updated.

**Consistent ↔ available ↔ partition-tolerant [D]** —Safety first. Always ready. Good enough.

If you want a system to be *consistent* (that is, all the parts of it see the same state; not the same as ACID consistency) and highly *available* (very unlikely to fail, because it's replicated in different places), then the replicas need to communicate so that all the parts can see all the changes. But if the replicas are *partitioned* then they can't communicate. So you can't have all three; this is the CAP "theorem". The way to get around it in practice is to make partitioning very unlikely. Because that's expensive, usually you have to choose, most often to sacrifice consistency in the hope that clients will notice inconsistency less than unavailability. A partial mitigation is *leases*, which are locks that time out, using the passage of real time for uninterruptible communication. If you spend enough money on a well-engineered network, as major cloud providers do, you can make partitioning so unlikely that it's negligible next to other failures that make a system unavailable.[R16]

**Generate ↔ check [D]** —Trust but verify.

A problem is in complexity class NP if finding a solution is hard (takes work $O(2^n)$), but checking it is easy (work $O(n^k)$). In code most checks are in an `assert`—the code has done something complicated, and the check confirms that it hasn't gone too far off the rails. But other examples are closer to the NP paradigm, mostly from randomized algorithms or program verification:

- If a randomized algorithm gives a result that's wrong with probability $p$ but that you can check, then retrying it until the check succeeds gives a correct result in expected time $1/(1-p)^2$.
- *Proof-carrying code* exploits the fact that it's much easier to check a proof than to generate it. If a client wants to safely run some code from an untrusted source, the source can provide a proof that the code satisfies its spec, and the client only needs to check the proof. Pushing this further, the code might carry only enough hints about how to generate the proof that it's cheap for the client to do so, for example, the loop invariants. Or write the code in a stylized way that's easy for the client to check, as in the Informer.[R33]
- *Validation* just checks a *result*, not the *program* that finds it; for example, a complicated sort routine, but a check that the result is ordered and the same size as the input (checking that it's actually a permutation is too expensive, but you could check that $Op_{x \in s} hash(x)$ is the same for some commutative and associative $Op$ such as × mod $2^{64}$).
- *Translation validation*, instead of verifying that a compiler is correct, just checks that the behavior of the compiled code is a subset of the behavior of the source code.[R94]

The general idea, however, is much broader: keep a hint that might be wrong, but is easy to check. This is a narrower meaning of "hints" than in the title of this paper, but there are many examples of it throughout. The end-to-end principle is closely related.

**Persistent ↔ volatile [D]** —Don't forget. Start clean.

Persistent state remains the same in spite of failures, unless code changes it explicitly; the canonical examples are file systems and databases. A crash resets volatile state such as RAM; more generally, a failure may change it or a bug may corrupt it (crashes and failures are events that the code does not control). Persistent state is more expensive because

- fault tolerance requires redundancy,
- physical storage that survives power failures is slow (though only 50-500 times slower than RAM when it's flash memory, rather than 10,000 times slower when it's disk), and
- to ensure that state survives bugs and changes in the code, you must represent it conservatively rather than efficiently.

But a dependable system almost always needs persistent state.

There are two ways to make complex state persistent: checkpoints and transactions. A checkpoint writes the volatile state to persistent storage, preferably using a simple representation; an example is saving a document from an editor to a file. This is fairly expensive. In contrast, a transaction makes state changes, often just a few, and then *commits*. The system maintains a volatile cache of the state, uses a persistent redo log to make sure that the changes are not lost, and sometimes takes checkpoints to keep the log short. It's the difference between being and becoming.

A simpler form of volatile state is a write-through cache, which just contains a subset of some more persistent state. Discarding the cache may slow down the program, but won't change its meaning. These caches speed up reads, but don't help with writes. A related technique is soft state, an unreliable cache of hints that time out if not refreshed.

If data changes frequently but needs to persist for months or years, it must have a very simple form. Otherwise the inevitable bugs in the code that is changing it will corrupt the data. Two such forms have stood the test of time: a text file without complicated internal structure, and a relational database with a simple schema. If the text file gets messed up, you can edit it with `ed`. The schema keeps the database from being messed up too badly. You can read one of these simple data structures and make fancier volatile state such as an index, a complicated data structure and one that you might want to change, but anything that needs to last should be expressed in text or tuples. An interface that lets the client both read out and restore the persistent state is a must, so that you can take a backup that will still work next year or move the data out of the system entirely. This is a more fragile alternative to preserving the internal state.

»Persistent objects. In 1981 I heard about a novel idea: extend the **new** operator to create a *persistent object* that works just like an ordinary object, with updatable links to other persistent objects. Of course you need transactions so that a change involving several objects can't be interrupted in the middle by a crash. I thought it wouldn't work, because even if every bit is preserved, millions of objects connected by arbitrary links that are created over many months by code that is changing will end up as a rubble of objects, rather than anything useful. And so it has proved.

»Alto file system. To make the file system reliable even though there is lots of experimental software with unprotected access to the physical disk, a disk block has an extra *label* field, block $x$ of file $f$ has the label $(f, x)$, and a disk write first checks that each block has the expected label. In case of trouble, a scavenger program reads and sorts all the labels to rebuild the index information that the file system normally uses.

**Being ↔ becoming [I]** —How did we get here? Don't copy, share.

There are two ways to represent the state of a system:
- *Being*: the values of the variables—a *map* $v \to x$
- *Becoming*: a sequence of actions that gets the state to where it is—a *log* of actions.

Different operations are efficient in different representations. If you're only interested in a single point in time, you want the map. If you care about several different versions (to recover the current state from a checkpoint, undo some actions, or merge several versions), you want the log. There are ways to convert one representation into the other, and points between the extremes: applying the actions gets you the values, a `diff` produces a *delta* (a sequence of actions that gets you from one state to another), *checkpoints* shorten the log. Ordinary programs use being; fault-tolerant programs use both, getting redundancy by writing the actions into a log and replaying the log from a checkpoint after a crash. More on this in § 3.8.1.

Another way to look at it is that becoming (the log of the actions) is fundamental: after all, that's how you got here. Being (the values of the variables) is just a cache. Becoming is also the essence of lazy computing: you don't run the program until you really need the result.

**Iterative ↔ recursive, array ↔ tree [I]** — Keep doing it. A part is like the whole.

*To iterate is human, to recurse divine.* —Peter Deutsch[Q18]

*The basic principle of recursive design is to make the parts have the same power as the whole.* — Bob Barton[Q4]

Iteration and recursion are both Turing-complete. You can write an iteration recursively using tail-recursion (which is easy: the last step in the loop is the only recursive call), and you can write a recursion iteratively using a data structure to simulate a call stack (which is a pain). A simple example is these two versions of factorial:

Iteration: $x := n; f := 1; \textbf{while } x > 1 \textbf{ do } f := x \times f; x := x - 1 \textbf{ end}$

Recursion: $f(x) \equiv \textbf{if } x = 1 \textbf{ then } 1 \textbf{ else } x \times f(x - 1)$

But iteration is more natural when there's a list or array of unstructured items to process, and recursion is more natural when the items have subparts, especially when the parts can be as general as the whole.

Thus recursion is what you want to process a tree (or a graph, taking care to visit each node only once) where the description of the structure is itself recursive. You don't need recursion to treat a sequence of different items differently, though, because you can make them into objects that carry their own methods. Here are examples that illustrate both points:
- A hierarchical file system can have different code at each directory node. Some nodes can be local, others on the Internet, yet others the result of a search: `bwl/docs/?author=smith`.[R40]

- Internet routing is hierarchical, using BGP at the top level, other protocols locally in an AS.

These examples also show how a *path name* (a sequence of simple names) identifies a path in a graph with labeled edges and provides decentralized naming. Just as any tree node can be the root of an entire subtree, a path name can grow longer without conflicting with any other names. This even works when there's no tree; if you have a sequence of items labeled with path names and sorted by them, you can always insert another one by extending the label of its predecessor.

If all the nodes have the same interface for looking up the next name in the path, the path name is just a sequence of names. If nodes have different interfaces, you may need syntax in the pathname to tell you which interface to use. File system path names are a familiar example of the first case, URLs of the second.

A different kind of recursion codes an interface using the same interface, usually to get better portability, security, or quality of service. Thus a hypervisor makes several virtual machines on top of a single one that can be either physical or virtual itself, an overlay network can provide specialized but efficient multicast or better routing on top of the public Internet, and software fault isolation sandboxes arbitrary user-mode code within a user process. Sometimes there are intermediate levels; thus a modem provides a 56 kb datastream on top of an analog voice-grade circuit that in turn uses a 56 kb datastream for everything except the last mile.

**Recompute ↔ adjust [I]** —Take small steps.

If you have computed $f(x)$ the simplest way to compute $f(x \oplus \Delta x)$ is to start from scratch and apply $f$ to the new argument $x \oplus \Delta x$. Trickier, but often more efficient, is to take advantage of knowing $f(x)$. For example, if $x$ is an array of 1000 numbers, $f(x) = \sum x_i$, and $\Delta x$ a new value for $x_i$, that is, a pair $(i, v)$, then $f(x \oplus \Delta x)$ is $f(x) - x_i + v$. Adding an entry to a B-tree usually requires no more than a lookup plus an insertion into a leaf node. When the leaf node overflows it has to be split, but that is rare. If insertions are concentrated in one part of the key space, the tree may have to be rebalanced, but that's even more rare. If an approximate answer is good enough and you can compute a derivative $f'$ for $f$, then $f(x \oplus \Delta x) \approx f(x) \oplus f'(x)\Delta x$.

# 6. Conclusion

I don't know how to sum up this paper briefly, but here are the most important points:
- Keep it simple. Complexity kills.
- Write a spec. At least, write down the abstract state.
- Build with modules, parts of the system that people can work on independently
- Exploit the ABCs of efficiency: algorithms, approximate, batch, cache, concurrency.
- Treat the state as both being and becoming: map vs. log, pieces, checkpoints, indexes.
- Use eventual consistency to keep data available locally.

In addition to the papers I've referenced, there are some good books about building systems: Lamport[R57,R58] and Ousterhout[R81] on how to write specs, Hennessy and Patterson[R47] on hardware architecture, Cormen, Leiserson, Rivest and Stein[R24] on algorithms, Bentley[R8] on efficiency, Hellerstein, Stonebraker and Hamilton[R46] on databases, Tanenbaum and Wetherall[R103] on networking, and Anderson[R6] and Schneier[R92] on security.

# Acknowledgments

I am grateful to Terry Crowley, Peter Denning, Frans Kaashoek, Rebecca Isaacs, Alan Kay, John Ousterhout, Fred Schneider, Charles Simonyi and John Wilkes for comments on many drafts.

# Appendix: Program semantics

The usual way to describe an action in code is by a statement in an imperative program, and this also works well for many specs. Since a statement is an action, its meaning is a relation (a predicate on *pairs* of states $s$ and $s'$), built up from smaller statements (except for the basic `x := y`; evaluating an expression is not an action).[R57,R63 §9,R78]

- For `a; b`, it's the composition of the actions: $a \circ b$. In a deterministic program the actions are functions, there's exactly one next state $a(s)$ after $a$, and this is just function composition, one statement after another: $s' = b(a(s))$. More generally, $a \circ b$ relates $s$ to $s'$ if there's some intermediate state $s_i$ that $a$ can produce from $s$ and that $b$ can turn into $s'$: $\exists s_i \mid a(s, s_i) \text{ and } b(s_i, s')$. If $a$ is deterministic, $s_i = a(s)$.

In the predicates below $p$ is short for $p(s)$, $p'$ for $p(s')$, action $a$ for $a(s, s')$, and $skip$ is $s' = s$.

- For `if p then a else b`, the relation is $(p \text{ and } a) \text{ or } (\text{not } p \text{ and } b)$.
- For `f(x)`, where $y$ is $f$'s formal parameter, it's $(y = x) \text{ and } f$.
- For `while p do a`, it's $(KleeneClosure(p \text{ and } a) \text{ and not } p')$;

You usually can't have a nondeterministic action in an imperative program, but you can certainly have one in a spec or in pseudocode, and a relation still gives its meaning:

- For `a □ b` (nondeterministic choice) the relation is to do a or do b: $a \text{ or } b$.
- For `if p then a` (if p is true do a, otherwise block) it's $p \text{ and } a$. Another way to write this is `await p; a`, where the meaning of `await p` is just $p \text{ and } skip$.
- For `a else b` (do a, but if it's blocked do b) it's $a \text{ or } (blocked(a) \text{ and } b)$.
- For `var x | p(x) in a(x)` (find a value for x that makes p true and do a, like `let x = e in a(x)`, which says explicitly how to compute the value) it's $\exists v \mid (p(v) \text{ and } a(v))$, which is somewhat surprising: the **var** statement relates $s$ to $s'$ if there's some value $v$ such that $p(v)$ is true in $s$

and $a(v)$ relates $s$ to $s'$. This is general nondeterminism; there's a small example in § 2.2 above.

In a concurrent program these composition rules only apply to the sub-actions within a single atomic action in a thread; the thread program counters determine which atomic actions are enabled. Lamport's PlusCal is similar to the language sketched above, and it has a toolchain: a translator to the first-order logic of TLA+ and a model checker.[R58]

If $A$ is the set of atomic actions, $Next(s, s') = \bigvee_{a \in A} a(s, s')$ defines the possible steps: any action that's enabled could take the next step. A behavior $b = s_1, s_2, \ldots$ is allowed if $Next$ holds for each step: $\Box Next$ is true of $b$ iff $\forall i \mid Next(s_i, s_{i+1})$. Here $\Box$ is the "henceforth" operator of temporal logic: $Next$ is true of every state. The initial state $s$ must satisfy a predicate $Init(s)$. So the spec for the whole system is a predicate on behaviors:

$$Init \text{ and } \Box Next$$

It's customary to call this kind of state-actions spec the "meaning" of the system, or (especially for programming languages) the "semantics", since it depends only on mathematics (in fact, only on set theory and first-order logic), so that you don't need anything else to reason about the system.

Another popular scheme for reasoning about programs is Hoare triples. $\{P\}\ a\ \{Q\}$ means that if $a$ relates $s$ to $s'$, then $P(s) \Rightarrow Q(s')$. $P$ is called the precondition, $Q$ the postcondition, so that the precondition implies the postcondition as you would expect. Thus $a$ as a relation between states determines $a$ as a relation between predicates.

A system satisfies a spec if its visible behavior is a subset of the spec's visible behavior, but for reasoning you want a logical formula rather than a set, and you just saw it: $JS = Init \land \Box Next$. To get a formula that refers to only the visible state, say that any values for the internal variables will do: $S = \exists\ v_1, v_2, \ldots, v_n \mid JS$. Then $S_2$ satisfies $S_1$ is just $S_2 \Rightarrow S_1$: every behavior that satisfies $S_2$ also satisfies $S_1$. So implementation is implication. This story works for any system.

*Logic*

Here is a way to remember the rules of logic. Propositions form a lattice, a partially ordered set with `sup` and `inf` (generalized `max` and `min`). Ordering ($\geq$) is implication ($\Rightarrow$, the arrows in the figure), `max` is **and** ($\land$), `min` is **or** ($\lor$), and $\exists$ / $\forall$ repeat **or** / **and** over a set just as $\Sigma$ / $\prod$ repeat $+$ / $\times$. The largest element is F (`false`, the strongest, the hardest to prove) and the smallest is T: $F \Rightarrow p \Rightarrow T$ for any $p$. Negation, **not** $p$ ($\overline{p}$), reverses everything as usual, but also $p$ and $\overline{p}$ are as distinct as possible: $\overline{p} \land p = F$, $\overline{p} \lor p = T$, and hence $(p \Rightarrow q) = (\overline{p} \lor q)$. This story also works for the sets defined by predicates, though it's confusing that $\geq$ is $\subseteq$ and the top element is the empty set. Or you can turn the whole picture upside down so that $\geq$ is $\supseteq$; unfortunately then $\Rightarrow$ is $\leq$.

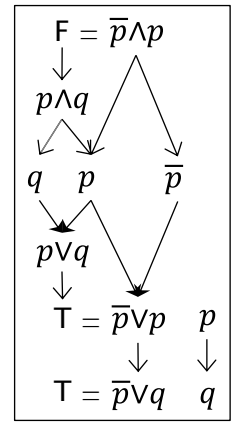

$F = \overline{p} \land p$
↓
$p \land q$
↙  ↘
$q$  $p$  $\overline{p}$
↘  ↙
$p \lor q$
↓
$T = \overline{p} \lor p$  $p$
↓  ↓
$T = \overline{p} \lor q$  $q$

# Quotations

I've tried to find attributions for all the quotations; some were unexpected, and it's disappointing that some of the best ones lack citations and may even be apocryphal. References of the form [Author99] are to PDF files that might not be at any link I've given; you'll find them here.

Q1. Catherine Aird, *His Burial Too*, Collins, 1973. [Aird73; last ¶]

Q2. Gene Amdahl, Gerritt Blaauw and Fred Brooks, Architecture of the IBM System/360, *IBM Journal of Research and Development* **8**, 2 April 1964, 87-101. The original is, "The term *architecture* is used here to describe the attributes of a system as seen by the programmer, i.e., the conceptual structure and functional behavior, as distinct from the organization of the data flow and controls, the logical design, and the physical implementation." Link [Amdahl64; footnote *]

Q3. Dan Ariely, You are what you measure, *Harvard Business Review* **88**, 6, June 2010, pp 38-41. Link [Ariely10; ¶ 5]

Q4. Bob Barton, quoted by Alan Kay in The Early History of Smalltalk, *ACM Conf. History of Programming Languages II*, *SIGPLAN Notices* **28**, 3, March 1993, pp 69-95; § I, ¶ -2. Link

Q5. Boris Beizer, *Software Testing Techniques*, Van Nostrand Reinhold, 2 ed., 1990, chapter 2, § 3.2.2.

Q6. Gordon Bell; he told me in January 2019, "I'm pretty sure I made the statement, but I can't find it in writing."

Q7. Bernard of Chartres, exhaustively documented in Robert Merton, *On the Shoulders of Giants*, Free Press, NY, 1965. Reprinted Chicago, 1993. Link. John of Salisbury says in *Metalogicon* (1159) bk. 3, ch. 4, "Dicebat Bernardus Carnotensis nos esse quasi nanos gigantium humeris insidentes, ut possimus plura eis et remotiora videre, non utique proprii visus acumine, aut eminentia corporis, sed quia in altum subvehimur et extollimur magnitudine gigantea." Cambridge, Corpus Christi College, MS 046, f. 217r; link. Translation from Henry Osborn Taylor, *The Mediaeval Mind*, Macmillan, London, 1911, vol. 2, p. 133: "Bernard of Chartres used to say that we were like dwarfs seated on the shoulders of giants. If we see more and further than they, it is not due to our own clear eyes or tall bodies, but because we are raised on high and upborne by their gigantic bigness." Link. Misattributed to Newton.

Q8. Yogi Berra, *When You Come to a Fork in the Road, Take It! Inspiration and Wisdom from One of Baseball's Greatest Heroes*, Hyperion, 2002, p. 53. For the title, see Q99.

Q9. Fred Brooks, No silver bullet, *IEEE Computer* **20**, 4 (April 1987), pp 10-19. Link [Brooks87; p 11, ¶ 3]

Q10. John Carmack, Archive - .plan (1997), July 7, 1997, p 41. Link [Carmack97; p 41, ¶ -1]

Q11. General Benjamin W. Chidlaw, Commander in Chief, Continental Air Defense Command, 1954. Link

Q12. John Robert Colombo, A Said Poem, in *Neo Poems*, The Sono Nis Press, Department of Creative Writing, University of British Columbia, 1970, p 46. Attributed to Mark Twain without evidence by Colombo and many others. Link [Colombo70]

Q13. Melvin Conway, How do committees invent?, *Datamation* **14**, 5, April 1968, 28-31. The original is, "Organizations which design systems ... are constrained to produce designs which are copies of the communication structures of these organizations." Link [Conway68; conclusion, ¶ 1]

Q14. Alan Cooper, *About Face 3: The Essentials of Interaction Design*, Wiley, 2007, p 97, ¶1. [Cooper07]

Q15. Terry Crowley, What to do when things get complicated, *Hacker Noon*, Sep. 27, 2017. Link [Crowley17-9-27; p 4, ¶1]

Q16. Ray Cummings, *The Girl in the Golden Atom*, in *All-Story Magazine*, 15 March 1919, ch. 5. Expanded to a novel, 1922. Reprinted by the University of Nebraska Press, 2005, ISBN 978-0-8032-6457-1. Misattributed to Einstein or Wheeler; see Link, Link [Cummings22; p 46, ¶ 10]

Q17. René Descartes, *Regulae ad Directionem Ingenii*, 1701; Rules for the direction of the mind, in Elizabeth Anscombe and Peter Thomas Geach, *Descartes: Philosophical Writings*, Nelson, 1954. "... nempe nullas vias hominibus patere ad cognitionem certam veritatis praeter evidentem intuitum, et necessariam deductionem;" translated as "There are no ways of attaining truth open to man except self-evident intuition and necessary inference." Link, Link; p 40, ¶ 3.

Q18. Peter Deutsch, quoted in James O. Coplien, *C++ Report* **10,** 7, July/August 1998, pp 43-51; title. Sometimes attributed to Robert Heller. Link. Also quoted in Bjarne Stroustrup, *The C++ Programming Language*, Special Edition (3rd Edition), Addison-Wesley, 2000, ch. 7, p 143.

Q89. John von Neumann, quoted by Edsger Dijkstra, EWD 563, Formal techniques and sizeable programs, in *Selected Writings on Computing: A Personal Perspective*, 1982, pp 205-214. I did not find a direct citation to von Neumann. Link [Dijkstra82; p 8, (6)]

Q90. John von Neumann. Alas, this quote appears to be apocryphal; see Link.

Q91. David Wheeler, but I don't know a direct citation for this (it has been widely but wrongly attributed to me). It appears in Bjarne Stroustrup, *The C++ Programming Language* (4th Edition), Addison-Wesley, 2013, p v. Link [Stroustrup13]. Stroustrup was Wheeler's PhD student. But, "It is easier to move a problem around (for example, by moving the problem to a different part of the overall network architecture) than it is to solve it. (corollary) It is always possible to add another level of indirection." In *The Twelve Networking Truths*, RFC 1925, April 1996, truth 6. Link [RFC1925]

Q92. George Starr White, *Think; Side Lights, What Others Say, Clinical Cases, Etc.*, Phillips Printing Co., Los Angeles, 1920, p 73. Link [White1920]. See also Link. Misattributed to Winston Churchill (often as "Success is not final."), according to Langworth, *Churchill by Himself*, Random House, 2008, p 579.

Q93. A.N. Whitehead, *Dialogues of Alfred North Whitehead* as recorded by Lucien Price, Little Brown, 1954, Ch. 43, November 11, 1947, p 369. Link [Whitehead54; p 10, ¶ 3]

Q94. A.N. Whitehead, *An Introduction to Mathematics*, Holt, 1911, ch. 5, p 43. Link [Whitehead11]

Q95. A.N. Whitehead, *The Concept of Nature*, Cambridge, 1920, p 163. The context is: "The aim of science is to seek the simplest explanations of complex facts. We are apt to fall into the error of thinking that the facts are simple because simplicity is the goal of our quest. The guiding motto in the life of every natural philosopher should be, Seek simplicity and distrust it." Link, Link [Whitehead20; p 163, last sentence]

Q96. Walt Whitman, *Song of Myself*, part 51. Link

Q97. Ludwig Wittgenstein, *Tractatus Logico-Philosophicus*, Harcourt, Brace, 1922, §5.6, p 149. Link [Wittgenstein22]

Q98. Frank Zappa, "Packard Goose", *Joe's Garage*, Act III, track 2, Zappa Records, 1979. Link [Zappa79]

Q99. Fort Gibson New Era, *Wise Directions* (filler item), p 2, col 6, July 31, 1913, Fort Gibson, OK. The original adds, "I will, if it's a silver one." Misattributed to Yogi Berra.[Q8] Link

Q100. From "'Twas the night before release date," in many places on the Internet.

# References

Each reference includes a link to the ACM Digital Library if I could find it. If the Library has a PDF for an item I assume it will always be there. If it doesn't, I give both a link to another source and (in case that link stops working) a citation of the form [Author99]; in my archive here there's a PDF file whose name starts with Author99. Some documents have been OCRed from ugly originals, which you can find in the eponymous folder.

# Index

# Players

# Stories